\documentclass{aa}  

\usepackage{graphicx}
\usepackage{txfonts}
\usepackage{lipsum}
\usepackage{subcaption} 
\usepackage{lscape} 
\usepackage{placeins} 
\usepackage[colorlinks = true,
            linkcolor = blue,
            urlcolor  = blue,
            citecolor = blue,
            anchorcolor = blue]{hyperref}
\usepackage{natbib}
\usepackage{multirow}
\usepackage{comment}
\bibpunct{(}{)}{;}{a}{}{,} 

\newcommand{\orcid}[1]{\unskip\protect\href{https://orcid.org/#1}{\protect\includegraphics[width=8pt,clip]{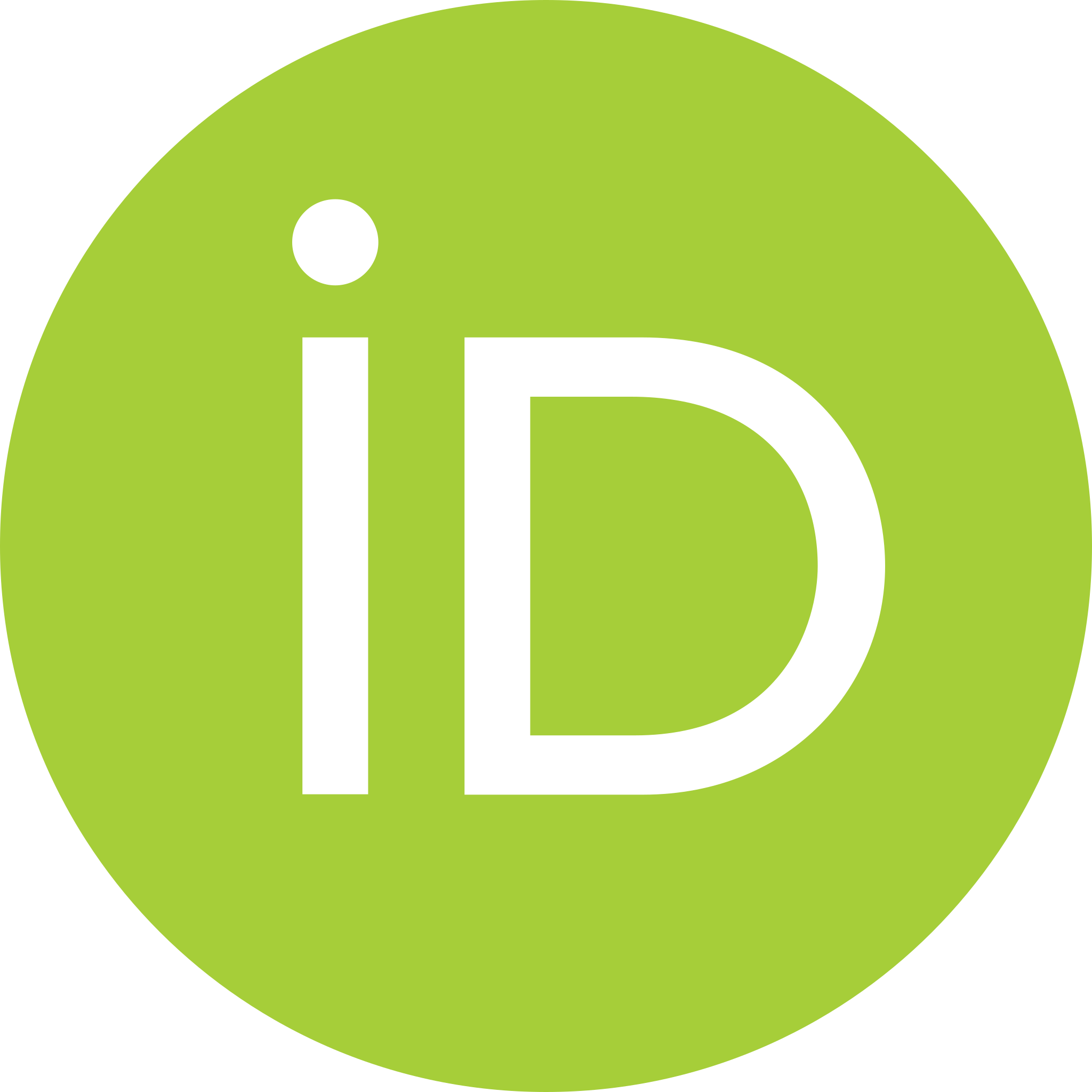}}}

\begin{document}

\title{Dust characterization of the HD 163296 disk with high-resolution multi-wavelength ALMA observations}

\author{
Kiyoaki Doi \orcid{0000-0003-1958-6673}\inst{\ref{mpia}}\thanks{\email{doi.kiyoaki.astro@gmail.com}}
\and
Myriam Benisty \orcid{0000-0002-7695-7605}\inst{\ref{mpia}}
\and
Akimasa Kataoka \orcid{0000-0003-4562-4119}\inst{\ref{naoj},\ref{sokendai}}
\and
Haochang Jiang \orcid{0000-0003-2948-5614}\inst{\ref{mpia}}
\and
Francesco Zagaria \orcid{0000-0001-6417-7380}\inst{\ref{mpia}}
\and
Hauyu Baobab Liu \orcid{0000-0003-2300-2626}\inst{\ref{kaohsiung},\ref{taiwan}}
\and
Carsten Dominik \orcid{0000-0002-3393-2459}\inst{\ref{amsterdam}}
\and
Ryo Tazaki \orcid{0000-0003-1451-6836}\inst{\ref{komaba}}
\and
Takahiro Ueda \orcid{0000-0003-4902-222X}\inst{\ref{naoj}}
\and
Tomohiro C. Yoshida \orcid{0000-0001-8002-8473}\inst{\ref{milan}}
\and
Takashi Tsukagoshi \orcid{0000-0002-6034-2892}\inst{\ref{ashikaga}}
\and
Yoshihide Yamato \orcid{0000-0003-4099-6941}\inst{\ref{riken}}
\and
Ryuta Orihara \orcid{0000-0003-4039-8933}\inst{\ref{ScienceTokyo}}
}

\institute{
Max-Planck Institute for Astronomy, Königstuhl 17, D-69117 Heidelberg, Germany\label{mpia}
\and 
National Astronomical Observatory of Japan, 2-21-1 Osawa, Mitaka, Tokyo 181-8588, Japan\label{naoj}
\and 
Astronomical Science Program, Graduate Institute for Advanced Studies, SOKENDAI, 2-21-1 Osawa, Mitaka, Tokyo 181-8588, Japan\label{sokendai}
\and 
Department of Physics, National Sun Yat-Sen University, No. 70, Lien-Hai Road, Kaohsiung City 80424, Taiwan, R.O.C. \label{kaohsiung}
\and
Center of Astronomy and Gravitation, National Taiwan Normal University, Taipei 116, Taiwan, R.O.C. \label{taiwan}
\and 
Anton Pannekoek Institute for Astronomy, University of Amsterdam, Science Park 904, 1098 XH Amsterdam, The Netherlands \label{amsterdam}
\and
Department of Earth Science and Astronomy, The University of Tokyo, Tokyo 153-8902, Japan \label{komaba}
\and
Dipartimento di Fisica, Università degli Studi di Milano, Via Celoria 16, 20133 Milano, Italy \label{milan}
\and 
Faculty of Engineering, Ashikaga University, Ohmae 268-1, Ashikaga, Tochigi, 326-8558, Japan \label{ashikaga}
\and
RIKEN Pioneering Research Institute, 2-1 Hirosawa, Wako, Saitama 351-0198, Japan\label{riken}
\and
Department of Earth and Planetary Sciences, Institute of Science Tokyo, 2-12-1 Ookayama, Meguro, Tokyo 152-8551, Japan\label{ScienceTokyo}
}
\date{Received April 19, 2026; accepted August 20, 2026}
\authorrunning{K. Doi et al.}

\abstract
{Planets form through the growth and accumulation of dust grains in protoplanetary disks. 
Observational characterization of dust properties such as dust size, surface density, and temperature is key to understanding the planet formation process.}
{We characterize the dust properties of the protoplanetary disk around HD 163296 by performing spectral energy distribution (SED) fitting of multi-wavelength high-resolution observations.}
{We present new high-resolution ALMA Band 9 (0.45 mm) observations, which are sensitive to the temperature.
We performed SED fitting at a common resolution of $0\farcs066$ using these new Band 9 observations along with archival ALMA Band 3, 4, 6, and 7 observations. We also compared the fitted results with VLA observations. 
We explored multiple dust models with different optical constants and porosities.}
{The Band 9 image shows the central disk, two rings at $0\farcs67$ and $1\farcs00$, and outer extended emission previously seen at other wavelengths. 
At higher frequencies, the rings appear wider, the gaps appear shallower, and the extended emission appears brighter, which can be explained by optical-depth effects and/or size segregation.
We characterized the dust properties, including temperature, surface density, and dust size, for each dust model.
However, the inferred dust properties are dependent on the dust model, and the ALMA data alone do not allow us to determine which dust model is preferred.
We identified the DSHARP Zubko (porous) model as the preferred model based on comparison with the VLA profiles and its consistency with other physical and observational constraints.
The estimated temperature in the outer ring is lower than predicted by a passively irradiated disk model, suggesting shadowing by the inner ring. 
Although the estimated surface density and dust size are dependent on the dust model, the preferred model indicates that the central disk, both major rings, and the extended disk each contains more than a few Earth masses of dust.}
{}

\maketitle
\nolinenumbers

\section{Introduction}

Planets form through the growth and accumulation of dust grains in protoplanetary disks.
In the core accretion scenario \citep{Safronov1972,Hayashi1981}, micron-sized dust grains initially grow into planetesimals through collisional sticking.
However, several barriers exist during this initial growth stage. 
The radial drift barrier is the problem that dust grains rapidly drift inward toward the central star due to gas drag, ultimately falling onto the star and thereby losing the solid material for forming planets \citep{Whipple1972,Weidenschilling1977mnras,Brauer2008}.
The fragmentation/bouncing barrier is the problem that the collisional velocity of dust grains can become too high for sticking, leading to fragmentation or bouncing \citep{Ormel2007,Birnstiel2010_fragmentation,Wada2009,Zsom2010,Blum2008ARAA,Dominik2024}.
Rapid planetesimal formation via streaming instability \citep{Youdin2005,Johansen2007} and gravitational instability \citep{Sekiya1983,Youdin2002} has been proposed as a way to overcome these barriers, but the details of where and when these processes operate in protoplanetary disks remain unclear.
Thus, dust growth in protoplanetary disks is still not fully understood.

Observational characterization of dust grains in protoplanetary disks provides direct constraints on the dust growth process.
High-resolution dust continuum observations with the Atacama Large Millimeter/Submillimeter Array (ALMA) have revealed substructures in many disks such as rings, spirals, and crescents \citep[e.g.,][]{alma2015,Andrews2018,Long2018}.
These structures in the dust emission distribution indicate that dust grains are locally concentrated, which may be a clue to overcoming the barriers to dust growth.
The radial drift of dust grains may be halted by local concentration \citep{Pinilla2012a,Zhu2012}, and locally concentrated dust may promote planetesimal formation by triggering instabilities.
Thus, disk observations play an important role in understanding the dust growth process.

It is necessary to quantitatively constrain dust properties such as the dust surface density and dust size to understand planet formation.
The total dust mass indicates how much material is available for planet formation, and the surface density distribution indicates where dust is concentrated and where planet formation may occur.
The dust size indicates how much dust growth has progressed or is suppressed.
Moreover, the surface density and dust size determine the conditions for instabilities such as the streaming instability and gravitational instability \citep{Li2021,Lim2024}.

These dust properties can be constrained from multi-wavelength observations \citep[e.g.,][]{Testi2014review,Carrasco-Gonzalez2019,Macias2021,Sierra2021,Ueda2022,Ueda2025,Guidi2022,Zhang2023,Zagaria2025,Guerra-Alvarado2024}.
The continuum intensity depends on the dust surface density, the dust size distribution, and the temperature.
However, these parameters often have degeneracies, particularly between temperature and the other parameters.
In the optically thin limit and Rayleigh-Jeans limit, the continuum intensity is proportional to both the dust surface density and the temperature, resulting in a strong degeneracy.
To break this degeneracy, previous studies have inferred the surface density and dust size by assuming a temperature structure or by imposing priors on the temperature.
However, the temperature structure of disks is non-trivial, and if shadows exist, as suggested by infrared observations \citep{Benisty2023}, the temperature may vary significantly on local scales.
Hence, dust characterization including temperature based on multi-wavelength observations is necessary.

High-frequency observations with ALMA Bands 9/10 are key to breaking the degeneracy between these parameters.
As shown in Fig. \ref{fig:planck_RJ}, the blackbody spectrum deviates significantly from the Rayleigh-Jeans approximation at these bands at typical disk temperatures and retains a strong dependence on temperature \citep{Planck1901,Rybicki1979}.
This behavior enables us to break the degeneracy and characterize the dust properties, including temperature, from multi-wavelength spectra \citep{Kim2019}.

In this study, we focus on the protoplanetary disk around HD 163296.
HD 163296 is a Herbig Ae star located at a distance of $101.0 \pm 0.4$ pc \citep{Gaia2023}.
The disk is very bright at submillimeter and millimeter wavelengths, and extensive observations have been conducted with ALMA and the Karl G. Jansky Very Large Array (JVLA) \citep{Isella2016,Isella2018,Guidi2016,Guidi2022}.
High-resolution continuum observations have revealed the central disk, two clear rings at radii of 67 and 100 au, and extended emission (also referred to as the 155 au ring), as well as the crescent-like structure within the gap \citep{Huang2018_ring,Isella2018}.
Molecular line observations have also been used to characterize the gas disk of HD 163296. 
The disk gas mass has been estimated from CO isotopologue observations and dynamical mass measurements based on molecular-line kinematics \citep{Booth2019,Zhang2021,Martire2024,Pezzotta2025}. 
In addition, localized gas kinematics have revealed rings and gaps in the gas corresponding to the dust substructures \citep{Teague2018_HD163296,Rosotti2020}, as well as vertical flows possibly associated with embedded planets within the gaps \citep{Teague2019_Nat,Benisty2026}.
These structures have been discussed as potentially arising from embedded planets \citep{Liu2018}.
In addition, the presence of planets at separations of 94 and 260 au has been suggested from localized kinks in the CO channel maps and chemical signatures \citep{Pinte2018_planet,Izquierdo2022,Izquierdo2026}.

Optical and infrared observations have also been carried out to investigate scattered light from the disk and to search for planets.
Infrared polarimetric observations have detected a ring-like structure in scattered light corresponding to the 67 au ring seen by ALMA, while the 100 au ring is not detected, suggesting shadowing in the outer disk \citep{Garufi2014,Garufi2017,Monnier2017,Muro-Arena2018,Rich2019,Ren2023}. 
In addition, the inner ring shows time variation that may be associated with shadowing \citep{Mullin2026}.
Hubble Space Telescope (HST) observations have detected faint scattered light on larger spatial scales up to 440 au\footnote{The radial extent is updated using the Gaia DR3 distance \citep{Gaia2023} instead of the Hipparcos distance used in these references.}, and its time variation is also associated with shadowing \citep{Grady2000,Wisniewski2008}.
High-contrast imaging campaigns aimed at the direct detection of planets have not detected planets \citep{Guidi2018,Mesa2019,Uyama2025}.

In this study, we characterize the dust in HD 163296 by combining new ALMA Band 9 observations with archival ALMA Band 3, 4, 6, and 7 observations.
These new Band 9 observations enable us to characterize the dust properties by breaking the degeneracy between temperature and the other parameters.
This paper is organized as follows.
In Section \ref{sec:observations}, we describe the observations and data reduction.
In Section \ref{sec:obs_result}, we present the observational results.
In Section \ref{sec:SED_fitting}, we perform spectral fitting to characterize the dust properties.
In Section \ref{sec:discussion}, we discuss the dust properties of HD 163296 and their implications for planet formation.
Finally, we summarize our conclusions in Section \ref{sec:conclusion}.

\section{Observations and data reduction} \label{sec:observations}

\begin{figure*}[thbp]
    \begin{center}
        \includegraphics[width=18cm]{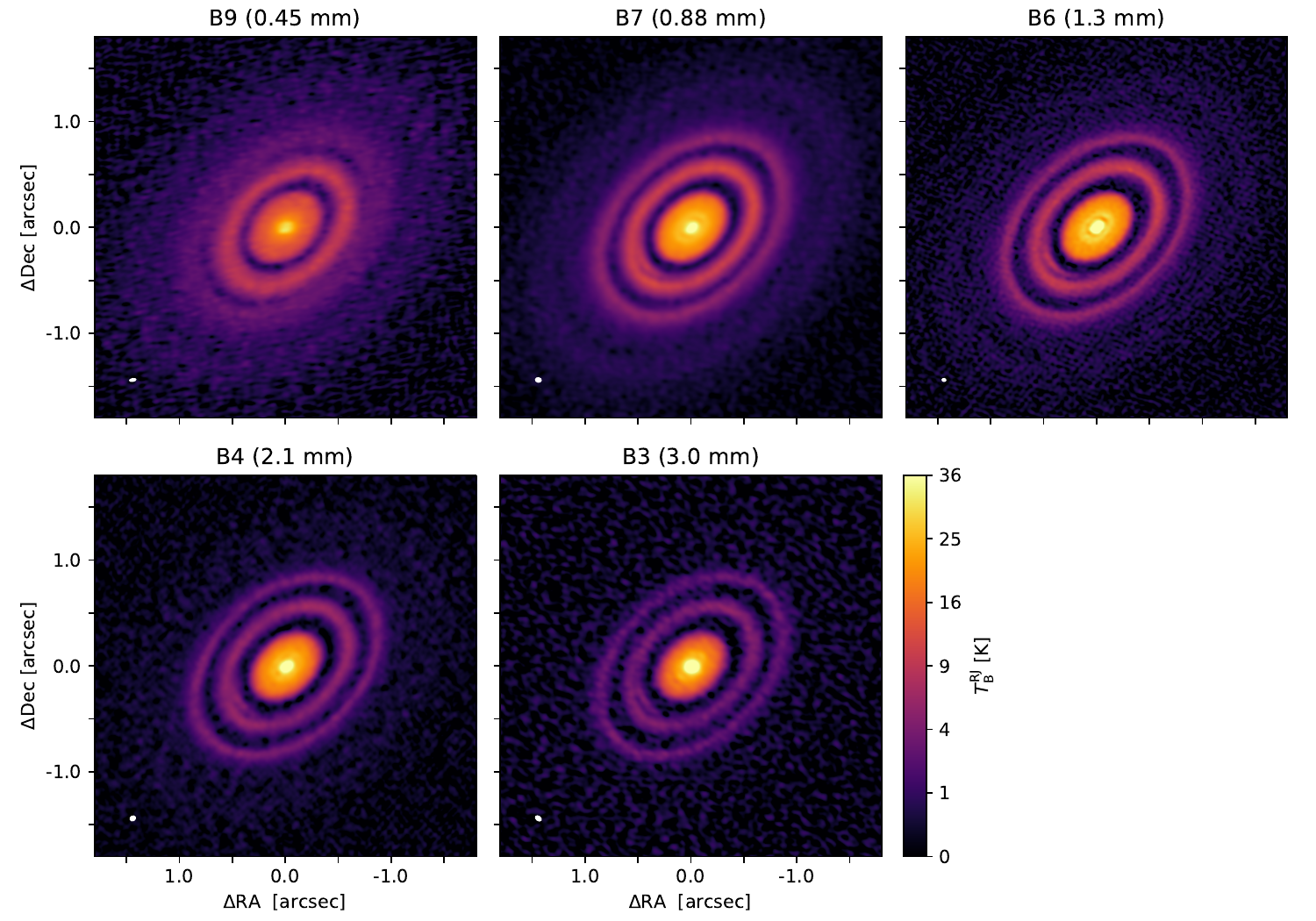}
    \end{center}
    \caption{
    Continuum images of the protoplanetary disk around HD 163296.
    The imaging settings, rms noise, and intensity are summarized in Table \ref{tab:imaging_results}.
    The intensities are shown as brightness temperatures converted using the Rayleigh-Jeans approximation, which preserves linearity with the observed intensity.
    The white ellipses at the bottom left of each panel indicate the beam size.
    }
    \label{fig:image_5panels}
\end{figure*}

\begin{table*}[ht!]
\caption{Imaging results.}
\centering
\begin{tabular}{lcccccccc}
\hline\hline
Band & Frequency & Robust & Beam size & Beam PA & rms noise & Peak intensity & SNR & Integrated flux \\
& [GHz] & & [arcsec] & [deg] & [$\mathrm{\mu Jy beam^{-1}}$] & [mJy beam$^{-1}$] & & [mJy] \\
\hline
Band 9 & $670.98$ & $-0.5$ & $0.073 \times 0.037$ & $-80.9$ & $308.0$ & $31.75$ & $103.1$ & $6138.3$ \\
Band 7 & $339.94$ & $-0.5$ & $0.066 \times 0.054$ & $79.9$ & $41.7$ & $14.15$ & $339.4$ & $1645.1$ \\
Band 6 & $238.98$ & $-0.5$ & $0.048 \times 0.038$ & $81.7$ & $22.6$ & $4.26$ & $189.0$ & $715.1$  \\
Band 4 & $140.02$ & $0.0$ & $0.065 \times 0.055$ & $-62.6$ & $11.1$ & $3.02$ & $271.2$ & $174.0$ \\
Band 3 & $100.44$ & $-0.5$ & $0.073 \times 0.052$ & $54.3$ & $12.4$ & $1.96$ & $158.5$ & $71.6$ \\
\hline
\end{tabular}
\label{tab:imaging_results}
\end{table*}

We obtained new ALMA Band 9 observations in Cycle 10 (project code: 2023.1.00578.S, PI: K. Doi) and combined them with archival ALMA data in Bands 3, 4, 6, and 7.
The archival Band 6 data were taken from the DSHARP program \citep{Andrews2018}, and the Band 3, 4, and 7 data were the same data as those used in \citet{Guidi2022}.
The details of these observational data are summarized in Table \ref{tab:observations}.

We used the image product from the DSHARP project for the Band 6 data.
For the new Band 9 observations and the Band 3, 4, and 7 archival data, we performed calibration and imaging of the ALMA data.
We executed the standard ALMA pipeline calibration for Band 9 Configuration 6 data in CASA v6.6.1.17 (pipeline version 2024.1.0.8) \citep{CASA2022}.
For the other data, we used the calibrated measurement sets provided by the observatory.

For the pipeline-calibrated measurement sets, we performed calibration following the strategy of the exoALMA collaboration \citep{Loomis2025}.
We used CASA modular version 6.7.0.31.
First, we flagged the channels with molecular line emission and averaged over channels to create a continuum measurement set, and then performed one round of phase self-calibration for each EB.
We then corrected the center offset by minimizing the imaginary part of the visibilities for each Execution Block (EB), and performed relative alignment by minimizing the difference in visibility amplitude with respect to the reference measurement set.
After that, we measured the flux difference from the ratio of visibility amplitude as a function of deprojected baseline and performed flux rescaling.
We then performed self-calibration on the concatenated Atacama Compact Array (ACA) data if they exist, then concatenated the Short Baseline (SB) data and performed self-calibration, and finally concatenated the Long Baseline (LB) data and performed self-calibration.
After phase calibration, we checked the flux difference again and repeated flux rescaling and self-calibration as needed.
We also applied one round of amplitude and phase calibration.

For the imaging, we used the \texttt{tclean} task with the \texttt{mtmfs} deconvolver with \texttt{nterms=2} \citep{Rau2011}.
For the Band 9 image including multiple antenna sizes, we used \texttt{gridder=mosaic}, while for the other images, we used \texttt{gridder=standard}.
We applied the primary beam correction to the images.
The subsequent analyses, including the radial profiles and the SED fitting, are based on the primary-beam-corrected images.
We made two sets of images with different resolutions.
For the high-resolution images, we adjusted the robust parameter to achieve a similar spatial resolution of around $0\farcs06$, and we adopted these images as the fiducial set for the following discussion.
Fig. \ref{fig:image_5panels} shows the fiducial images, and the robust parameters, resolutions, sensitivities, and SNRs of the final images are summarized in Table \ref{tab:imaging_results}.
The rms noise levels and SNRs in Table \ref{tab:imaging_results} were measured on the images before the primary beam correction, while the integrated fluxes were measured on the primary-beam-corrected images.
The local noise level at each position was then estimated by dividing the measured rms noise by the primary beam response.
We also made low-resolution but more sensitive images with a robust parameter of 0.5, which are shown in Fig. \ref{fig:image_5panels_low} of Appendix \ref{app:low_res_image}.

We also used the image products of VLA Ka-band at 8.6 mm (upper sideband) and 9.7 mm (lower sideband) from \citet{Guidi2022} for comparison.
The beam size for the upper sideband is $0\farcs102 \times 0\farcs051$ with a position angle of $5.3^\circ$, and the sensitivity is $3.76\ \mathrm{\mu Jy\ beam^{-1}}$, and the beam size for the lower sideband is $0\farcs113 \times 0\farcs057$ with a position angle of $4.3^\circ$, and the sensitivity is $3.39\ \mathrm{\mu Jy\ beam^{-1}}$.
We did not use the VLA images for the SED fitting in Section \ref{sec:SED_fitting_method}, as these images have lower resolution and would reduce the common resolution required for the SED fitting.
Moreover, we will analyze the radial profiles along the major axis for the ALMA data, whereas the sensitivity of the VLA images is not sufficient to obtain adequate SNR without averaging over the entire azimuthal direction. Therefore, the ALMA and VLA data cannot be analyzed under the same conditions.
In addition, the VLA images have characteristic extended sidelobes in their point spread functions (PSFs), and if the SNR is too low for sufficiently deep cleaning, the deconvolution model cannot be sufficiently built, so sidelobe contamination from bright regions can significantly affect the intensities in the gaps.
For these reasons, we decided not to use the VLA images for the SED fitting, but instead to use them for comparison with the intensities predicted at VLA wavelengths from the dust properties obtained from the fitting to the ALMA data.

\section{Observational results} \label{sec:obs_result}

\begin{figure*}[thbp]
    \begin{center}
        \includegraphics[width=18cm]{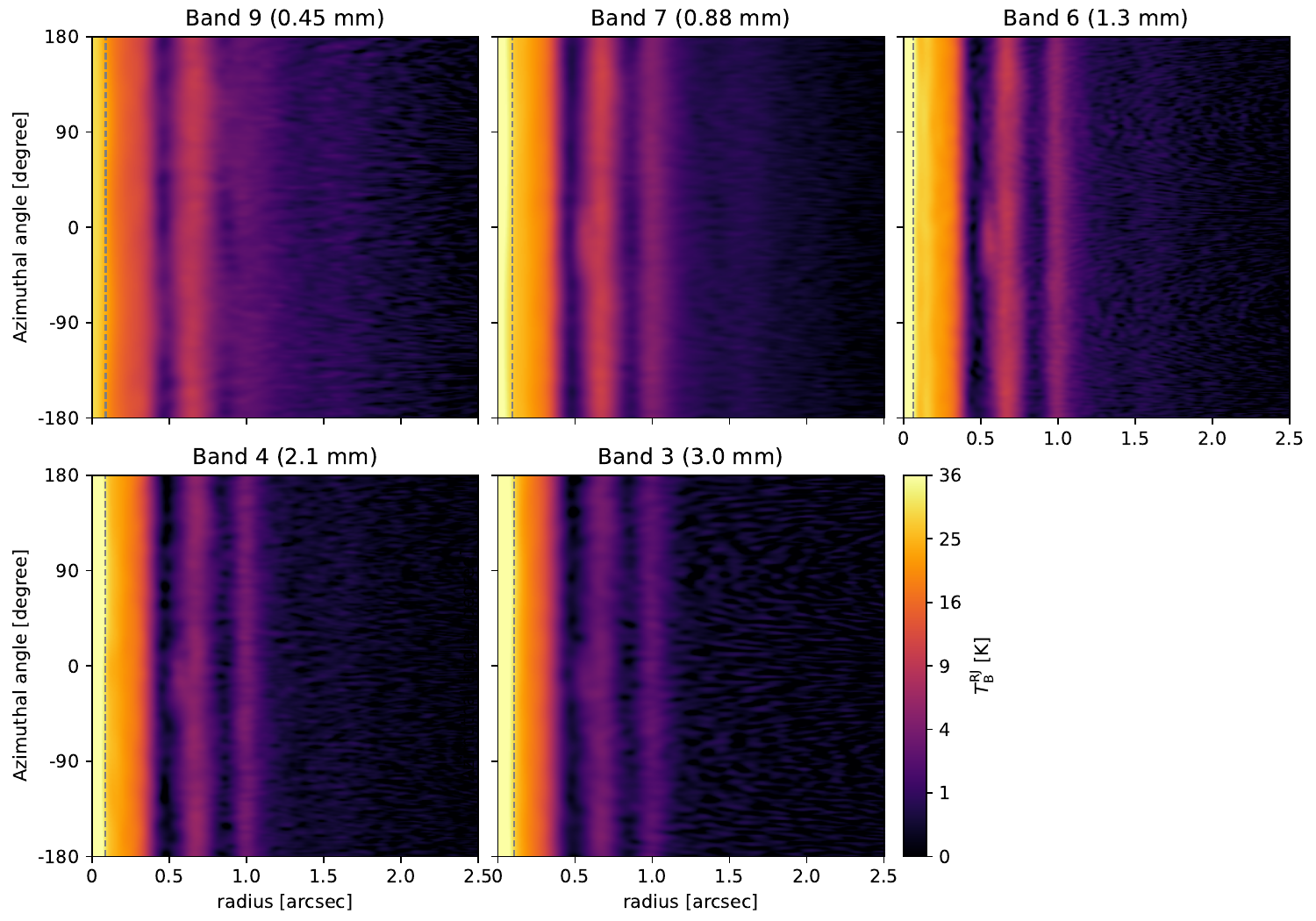}
    \end{center}
    \caption{
Polar plots of the fiducial continuum images of the protoplanetary disk around HD 163296.
The images were smoothed so that the beam is circular in the deprojected plane.
The reference for the azimuthal angle is defined such that the major axis in the southeast direction corresponds to $0^\circ$ and increases counterclockwise.
The gray dashed lines on the left side of each panel indicate the beam FWHM.
    }
    \label{fig:polar_5panels}
\end{figure*}

\begin{figure}[htbp]
    \begin{center}
        \includegraphics[width=8cm]{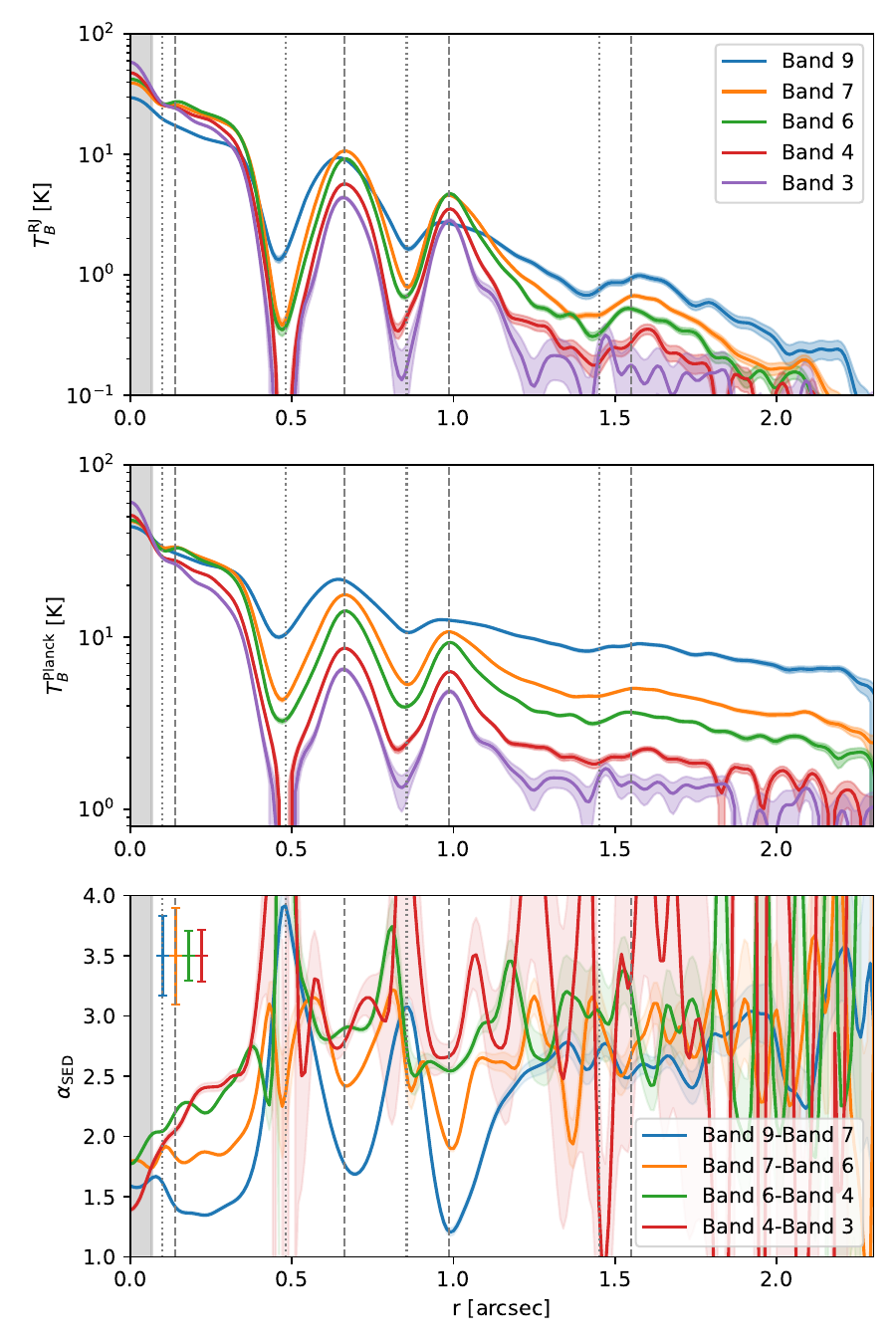}
    \end{center}
    \caption{
Radial profiles and spectral index along the major axis on the northwest side, averaged over a range of $\phi \pm 22.5^\circ$.
The top panel shows the radial profiles in brightness temperature using the Rayleigh-Jeans approximation, while the middle panel shows those in brightness temperature computed using the Planck function.
The bottom panel shows the spectral index.
The shaded areas in the top and middle panels indicate the $\pm 1\sigma$ range of the rms noise, and those in the bottom panel indicate the $\pm 1\sigma$ uncertainties propagated from the rms noise of the intensities.
The bar at the top left of the bottom panel indicates the $\pm 1\sigma$ uncertainty of the spectral index due to the ALMA flux calibration uncertainties.
The vertical gray dashed and dotted lines indicate the ring and gap radii, respectively, reported by \citet{Huang2018_ring}.
    }
    \label{fig:radial_obs}
\end{figure}

\begin{figure}[htbp]
    \begin{center}
        \includegraphics[width=9cm]{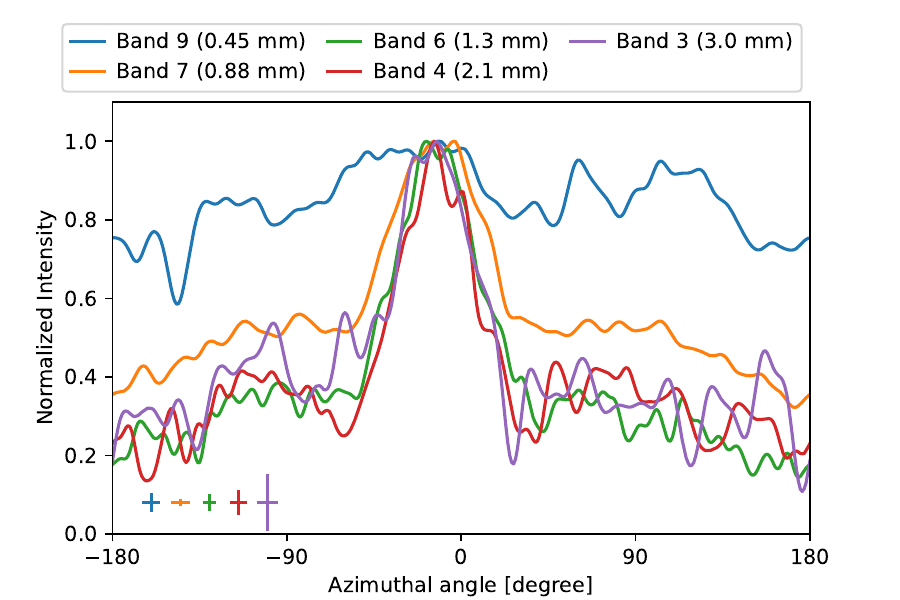}
    \end{center}
    \caption{
Normalized azimuthal profiles of HD 163296 extracted at a radius of $r=0\farcs55$ along the crescent.
The crosses at the bottom left of each panel indicate the beam size in the azimuthal direction and the $\pm 1\sigma$ noise level.
The solid lines are the azimuthal profiles from images with circular beams in the deprojected plane, as in Fig. \ref{fig:polar_5panels}.}
    \label{fig:azimuthal}
\end{figure}

The images of the HD 163296 disk at ALMA Band 3, 4, 6, 7, and 9 are shown in Fig. \ref{fig:image_5panels}.
These images are aligned by correcting for center offsets by minimizing the residuals between the original images and their $180^\circ$ rotated images, as described in Appendix \ref{app:geometry}.
The intensity is converted to brightness temperature using the Rayleigh-Jeans law to preserve the linearity with the observed intensity.
The new Band 9 image recovers the structures found in previous observations, such as the central disk, the two rings, and the extended disk.
However, the structures appear smoother than in previous observations, with wider rings and shallower gaps.

On the other hand, the crescent structure at $r=0\farcs55$ and the innermost asymmetry at $r=0\farcs04$, previously detected at other wavelengths \citep{Isella2018}, are less obvious in Band 9.
Fig. \ref{fig:image_5panels_rotated_difference} shows the residual images obtained by subtracting the images rotated by $180^\circ$ around the center from the original images.
In these residual images, both features are visible at all wavelengths, although the crescent remains less prominent in Band 9 than at the other wavelengths. 
This may be because the surrounding emission becomes more optically thick, thereby reducing the contrast of the crescent.
The innermost asymmetry appears brighter on the far side regardless of the observation date.
This may be due to a geometric effect in which the irradiated inner edge of a vertically puffed-up inner disk appears brighter on the far side \citep{Ribas2024,Guerra-Alvarado2024}.

Fig. \ref{fig:polar_5panels} shows the polar plots of the images.
We adopt the inclination of $i=46.48^\circ$ and the position angle of $\mathrm{PA}=133.27^\circ$ by minimizing the residuals between the original images and their $90^\circ$ rotated images, as described in Appendix \ref{app:geometry}.
To reduce the effect of asymmetry created by the beam in the polar plots, we smoothed the images so that the beam shape becomes circular in the deprojected plane.
If we focus on the azimuthal intensity variations in the Band 9 image, the gaps before and after the $0\farcs67$ ring appear shallower along the minor axis than along the major axis.
Since the dust is vertically puffed up at the 67 au ring \citep{Doi2021,Doi2023,Liu_Yao2022}, the gaps along the minor axis may appear shallower than the intrinsic radial profile due to the projection effect of the inclined line of sight crossing the vertically extended structure \citep{Pinte2016,Villenave2025review}.
Therefore, it is desirable to consider the profile along the major axis, where such effects are not expected, when examining the radial profile in the disk.

Fig. \ref{fig:radial_obs} shows the radial profiles and the spectral index profile, sampled at radial intervals of $0\farcs01$.
Since the minor axis may be affected by the projection effect of the vertical thickness, and there is a crescent structure in the southeast direction, we show the profiles averaged over a range of $\pm 22.5^\circ$ from the major axis on the northwest side, i.e., over a range of $45^\circ$.
Here, we used the images smoothed to have the smallest common beam at all wavelengths.
The beam size along the major axis is $0\farcs066$ and is indicated by the gray shaded area on the left side of the figure.
The uncertainty in the radial profile is calculated as $\sigma_{\mathrm{rms,radial}} = \sigma_{\mathrm{rms}} / \sqrt{N_{\mathrm{indep}}}$, where $\sigma_{\mathrm{rms}}$ is the rms noise level, and $N_{\mathrm{indep}} = l_{\mathrm{arc}} / (2 \sqrt{\pi} \sigma_{\mathrm{beam}})$ is the number of independent beams, where $l_{\mathrm{arc}}$ is the length of the arc corresponding to the range of azimuthal averaging, and $\sigma_{\mathrm{beam}}$ is the beam size in standard deviation.
The spectral index in the bottom panel is calculated from the radial profiles in the top panel.
The uncertainties of the spectral index are propagated from the rms noise of the intensity profiles.
We adopt the nominal ALMA flux calibration uncertainty of 20\% for Band 9, 10\% for Band 7 and Band 6, and 5\% for Band 4 and Band 3.
The flux calibration errors are systematic errors that uniformly scale the intensities across the disk and thus shift the spectral index uniformly at all radii, and therefore we show them as the bar in the bottom panel, separately from the thermal noise, which is independent at each radius.
Since the high-resolution profiles are noisy in the outer regions, we also show in Fig. \ref{fig:radial_obs_low} of Appendix \ref{app:low_res_image} the radial profiles from the low-resolution but more sensitive images, averaged over the full $360^\circ$ azimuthal range, which better recover the outer regions.

The top panel of Fig. \ref{fig:radial_obs} shows the radial profiles of brightness temperature using the Rayleigh-Jeans approximation, while the middle panel shows the radial profiles of brightness temperature computed using the Planck function.
From the top panel, we can see that the ring widths are wider and the gaps are shallower at shorter wavelengths.
Also, emission at shorter wavelengths, which is more optically thick, is brighter and more extended up to $2\farcs3$.
In the top panel, the brightness temperature of Band 9 is lower than those of the other wavelengths at the central disk and the ring locations, especially the outer ring, despite being more optically thick.
However, in the middle panel, the brightness temperature is higher at shorter wavelengths in all regions except for the central beam.
This indicates that the low brightness temperature at high frequencies seen in the top panel is due to the deviation from the Rayleigh-Jeans approximation.
On the other hand, in the innermost region within one beam radius, the brightness temperature is higher at lower frequencies, indicating a significant contribution from free-free emission \citep{Guidi2022}.

The bottom panel of Fig. \ref{fig:radial_obs} shows the spectral index.
At the locations of the two rings, the spectral index shows local minima, which may be due to the rings being more optically thick or having a smaller dust opacity index.
Also, the spectral index tends to be smaller at higher frequencies. In the outer regions beyond $1\arcsec$, the uncertainties are large and this tendency is difficult to discern in the high-resolution images, but it can be seen more clearly in the radial profiles of the low-resolution images in Fig. \ref{fig:radial_obs_low}.
This may be due to the emission being more optically thick at higher frequencies, or having a smaller dust opacity slope, or having a lower temperature and thus a larger deviation from the Rayleigh-Jeans approximation.
In particular, the locally small value of the spectral index in Band 9 at the outer ring at $1\arcsec$ indicates a large deviation from the Rayleigh-Jeans approximation.

Fig. \ref{fig:azimuthal} shows the azimuthal profiles at the crescent radius, $r = 0\farcs55$, normalized by their peak intensities. For a similar presentation, see \citet{Guidi2026}. 
The profiles shown here were obtained from images smoothed to have a circular beam in the deprojected plane.
In Band 9, the background emission outside the crescent is stronger, and the contrast of the crescent is smaller than at other wavelengths.
It can be because the surrounding emission becomes more optically thick and thus reduces the contrast of the crescent.

\section{Dust characterization} \label{sec:SED_fitting}

\subsection{Modeling and fitting method} \label{sec:SED_fitting_method}

We estimate the dust properties, such as temperature, surface density, and size distribution, from the observed intensity at multiple wavelengths.
If we ignore the vertical variations of dust properties such as vertical temperature gradients and differences in dust size distribution, the observed intensity at a given wavelength is a function of the dust temperature, surface density, and dust size distribution.
Therefore, by comparing the observed intensity with the model intensity, we estimate the dust properties.

The intensity from a dust slab with scattering can be expressed as \citep{Miyake1993,Carrasco-Gonzalez2019,Sierra2021}
\begin{equation}\label{eq:radiative_transfer}
    I_\nu(\tau_\nu,\mu)=B_\nu(T)\left[1-\exp\left(-\tau_\nu^{\mathrm{ext,eff}}/\mu\right)+\omega_\nu F(\tau_\nu^{\mathrm{ext,eff}},\omega_\nu^{\mathrm{eff}},\mu)\right].
\end{equation}
Here, $B_\nu(T)$ is the Planck function at temperature $T$ and frequency $\nu$, $\tau_\nu^{\mathrm{ext,eff}}=\Sigma_{\mathrm{dust}}\kappa_\nu^{\mathrm{ext,eff}}$ is the effective extinction optical depth, where $\kappa_\nu^{\mathrm{ext,eff}}=\kappa_\nu^{\mathrm{abs}}+\kappa_\nu^{\mathrm{sca,eff}}$ is the effective extinction opacity.
The label eff indicates quantities corrected for forward scattering, as described below.
$\kappa_\nu^{\mathrm{abs}}$ and $\kappa_\nu^{\mathrm{sca,eff}}$ are the absorption and effective scattering opacities, respectively.
$\mu=\cos i$, where $i$ is the disk inclination.
$\omega_\nu^{\mathrm{eff}}=\kappa_\nu^{\mathrm{sca,eff}}/\kappa_\nu^{\mathrm{ext,eff}}$ is the single scattering albedo.
$F(\tau_\nu^{\mathrm{ext,eff}},\omega_\nu,\mu)$ is given by
\begin{equation}\label{eq:F_factor}
\begin{gathered}
F(\tau_\nu^{\mathrm{ext,eff}},\omega_\nu,\mu)
= \dfrac{1}{\exp\left(-\sqrt{3}\epsilon_\nu\tau_\nu^{\mathrm{ext,eff}}\right)\left(\epsilon_\nu-1\right)-\left(\epsilon_\nu+1\right)}\times \\
\times\left[\dfrac{1-\exp\left(-\left(\sqrt{3}\epsilon_\nu+1/\mu\right)\tau_\nu^{\mathrm{ext,eff}}\right)}{\sqrt{3}\epsilon_\nu\mu+1} + \right.\\
\left. + \dfrac{\exp\left(-\tau_\nu^{\mathrm{ext,eff}}/\mu\right)-\exp\left(-\sqrt{3}\epsilon_\nu\tau_\nu^{\mathrm{ext,eff}}\right)}{\sqrt{3}\epsilon_\nu\mu-1}\right],
\end{gathered}
\end{equation}
where $\epsilon_\nu=\sqrt{1-\omega_\nu^{\mathrm{eff}}}$.

The dust opacities $\kappa_\nu^{\mathrm{ext,eff}}$, $\kappa_\nu^{\mathrm{abs}}$, and $\kappa_\nu^{\mathrm{sca,eff}}$ depend on the optical constants, which are determined by the dust composition, and the dust size distribution.
For the dust composition, we consider two models: one is the dust model used in DSHARP (hereafter, DSHARP default) \citep{Birnstiel2018}, and the other is a dust model that uses the same mass fraction as the DSHARP default model but replaces the carbonaceous material of the refractory organics \citep{Henning1996} with amorphous carbon (sample BE) from \citet{Zubko1996} (hereafter, DSHARP Zubko).
The BE sample from \citet{Zubko1996} has a larger imaginary part of the complex refractive index than the refractory organics from \citet{Henning1996}, resulting in a larger absorption coefficient and a lower albedo.
We also consider both compact dust $(p=0)$ and porous dust $(p=0.9)$, where $p$ is the porosity.
The optical properties of these dust models are summarized in Appendix \ref{app:optical_properties}.

We assume a power-law dust size distribution and consider the power-law index $q$ and the maximum dust size $a_{\mathrm{max}}$ as parameters.
Here, the power-law index $q$ is defined such that the dust size distribution follows $n(a) \propto a^{-q}$.
The minimum dust size is fixed at $1\ \mathrm{\mu m}$.
For porous dust, we express the dust size as an effective size $a_{\mathrm{eff}} = a\times (1-p)$, which allows comparison between compact and porous dust \citep{Kataoka2014}.

We computed the size-dependent dust opacities using Mie theory \citep{Mie1908} with the \texttt{dsharp\_opac} code \citep{Birnstiel2018}.
When the dust size becomes comparable to or larger than the wavelength, forward scattering dominates the scattering.
Therefore, we calculated the effective scattering opacity $\kappa_\nu^{\mathrm{sca,eff}} = (1-g_\nu)\kappa_\nu^{\mathrm{sca}}$ using the asymmetry parameter $g_\nu$ \citep{Henyey1941,Tazaki2019}.
Then, we integrated over the size distribution to calculate the dust absorption opacity $\kappa_\nu^{\mathrm{abs}}$, the effective scattering opacity $\kappa_\nu^{\mathrm{sca,eff}}$, and the extinction opacity $\kappa_\nu^{\mathrm{ext,eff}} = \kappa_\nu^{\mathrm{abs}} + \kappa_\nu^{\mathrm{sca,eff}}$.
By calculating the model intensity using these opacities and comparing it with the observed intensity, we can estimate the dust properties.

From Eq. \eqref{eq:radiative_transfer}, we calculate the model intensity for a given set of the dust temperature, surface density, maximum dust size, and power-law index.
The probability of a given set of parameters is then calculated from the difference between the observed and model intensities at each radius as
\begin{equation}
    P(T, \Sigma_{\mathrm{dust}}, a_{\mathrm{max}}, q) \propto \exp\left(-\chi^2 / 2\right),
\end{equation}
where
\begin{equation}\label{eq:chi2}
    \chi^2 = \sum_i \frac{\left(I_{\mathrm{obs},i} - I_{\mathrm{model},i}\right)^2}{\sigma^2_{\mathrm{tot},i}}.
\end{equation}
Here, $I_{\mathrm{obs},i}$ is the observed intensity, $I_{\mathrm{model},i}$ is the model intensity calculated from Eq. \eqref{eq:radiative_transfer}, and $\sigma_{\mathrm{tot},i}$ is the total uncertainty of the observed intensity.
The total uncertainty $\sigma_{\mathrm{tot}}$ is given by
\begin{equation}
    \sigma_{\mathrm{tot},i}^2 = \sigma_{\mathrm{thermal},i}^2 + (\delta_i I_{\mathrm{obs},i})^2,
\end{equation}
where $\delta_i$ is the absolute flux calibration uncertainty of the observed intensity.
Here, we adopt the nominal ALMA flux calibration uncertainty of 20\% for Band 9, 10\% for Band 7 and Band 6, and 5\% for Band 4 and Band 3 \citep{Cortes2025}.

We created a grid of the four parameters: temperature, surface density, maximum dust size, and power-law index, and calculated the model intensity at each grid point to compare with the observed intensity and evaluate the relative probability distribution.
The temperature grid ranges from $10^{0.5}$ to $10^{2.5}$ K, the surface density grid from $10^{-4}$ to $10^{2}$ g cm$^{-2}$, and the maximum dust size grid from $10^{-3}$ to $10^{2}$ cm. All grids are logarithmically spaced with 100 points per decade, resulting in a grid of $201 \times 601 \times 501$ points.
We performed the calculations for two cases where the power-law index is fixed at specific values or varied from 2.0 to 4.0 in steps of 0.1, creating a grid of 21 points for the power-law index.
By comparing the model intensity at these grid points with the observed intensity at each radius in the radial profile of Fig. \ref{fig:radial_obs}, we calculated the relative probability distribution of these parameters.
We note that the parameter grids define only the ranges to be explored.
As described below, we evaluate the parameter constraints using the profile likelihood, which does not involve any weighting within the grid.

\begin{figure*}[phtb]
  \centering
  \setlength{\tabcolsep}{0pt}
  \renewcommand{\arraystretch}{0}
  \begin{tabular}{cc}
    \includegraphics[width=8.8cm]{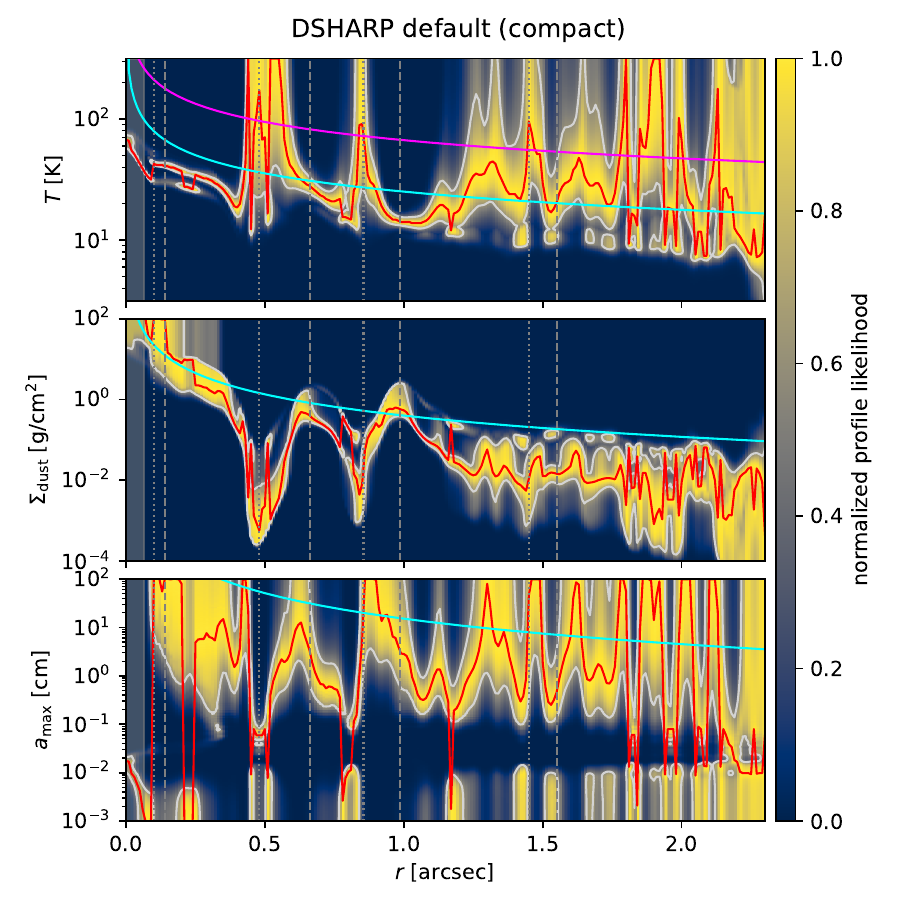} &
    \includegraphics[width=8.8cm]{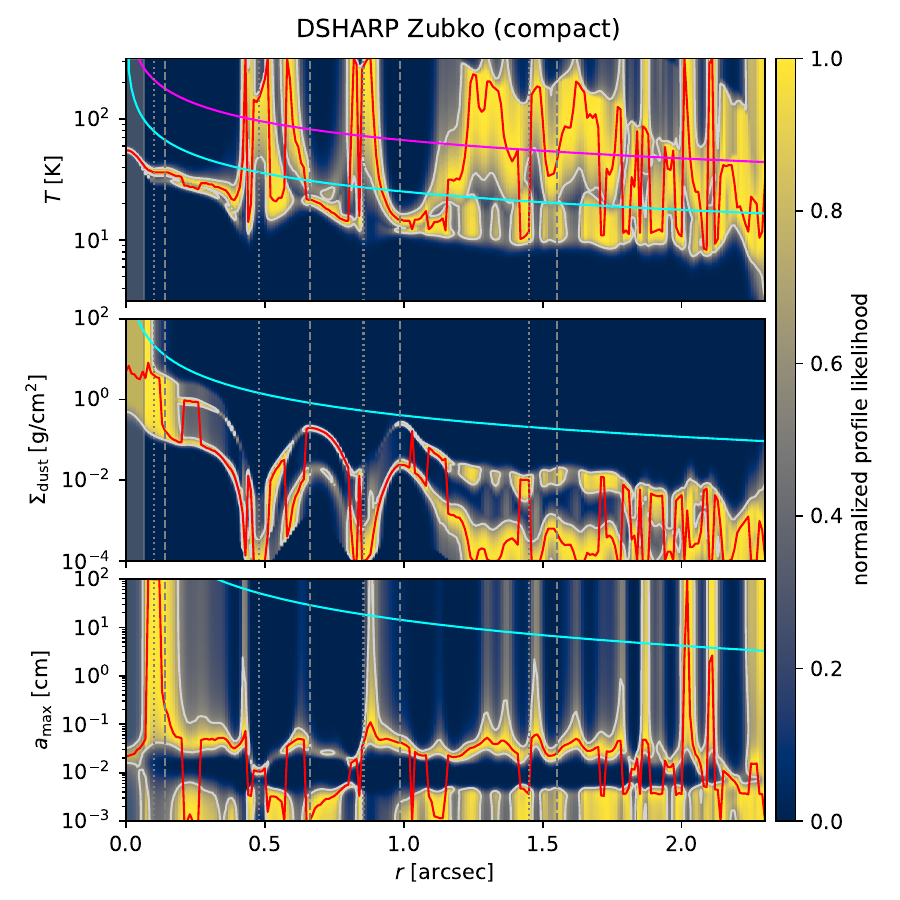} \\
    \includegraphics[width=8.8cm]{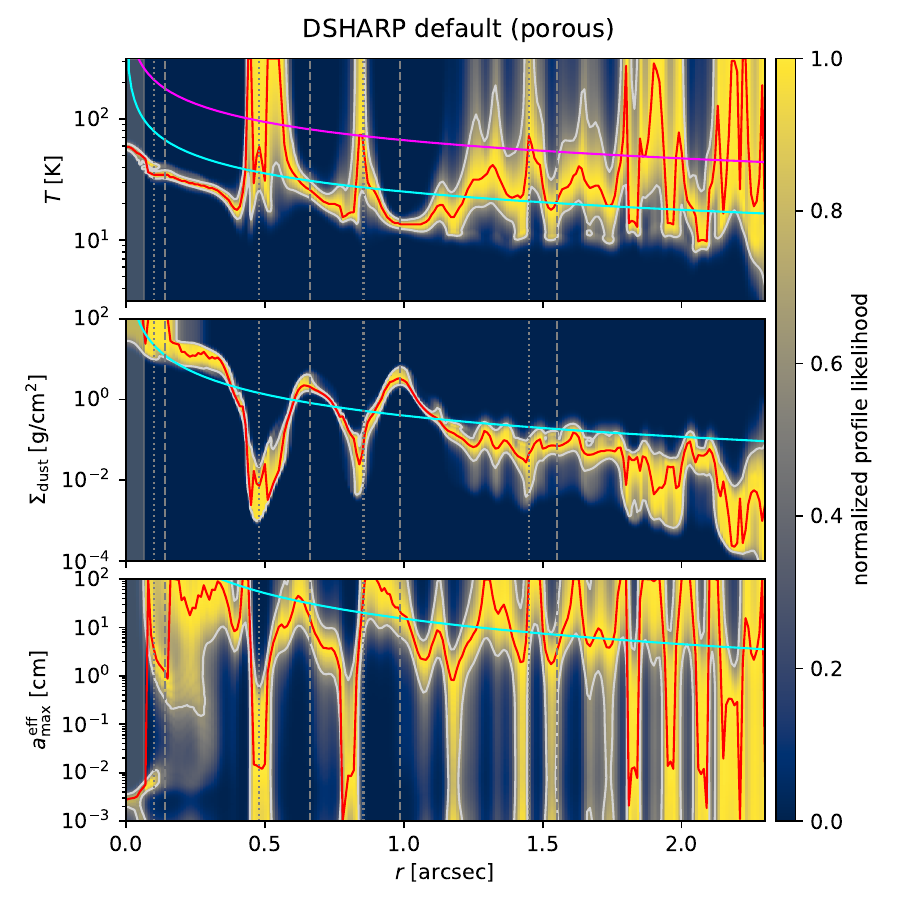} &
    \includegraphics[width=8.8cm]{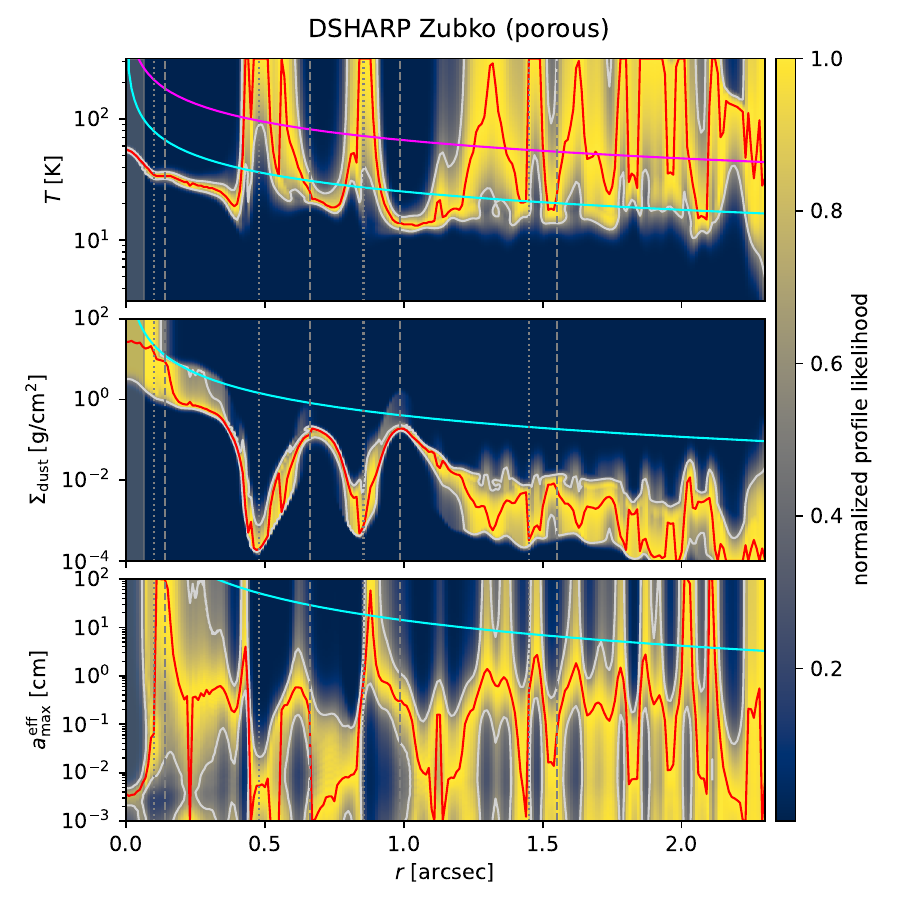} \\
  \end{tabular}
  \caption{
Normalized profile likelihoods from the SED fitting for each dust model.
In each panel, the top, middle, and bottom subplots show the likelihood profiles for temperature $T$, dust surface density $\Sigma_{\mathrm{dust}}$, and maximum dust size $a_{\mathrm{max}}$, respectively.
The red lines indicate the maximum likelihood estimates (MLEs), and the gray lines indicate the ranges derived from $\Delta\chi^2 < 1$.
In the $T$ subplots, the pink and cyan curves show the temperature profiles at the disk surface ($\varphi=1$) and midplane ($\varphi=0.02$), respectively, from a passive disk model \citep{Huang2018_ring}.
In the $\Sigma_{\mathrm{dust}}$ subplots, the cyan curve shows the upper limit of the dust surface density corresponding to a Toomre $Q=1$ disk with a dust-to-gas ratio of 0.01.
In the $a_{\mathrm{max}}$ subplots, the cyan curve shows the particle size for $\mathrm{St}=1$ for the same $Q=1$ disk.
For the porous-grain panels (bottom row), the maximum dust size is reported as an effective size, $a_{\mathrm{eff}} = a \times (1 - p)$, where $p$ is the porosity, for comparison with the results for compact grains.
The vertical gray dashed and dotted lines indicate the ring and gap radii, respectively, reported by \citet{Huang2018_ring}.
  }
  \label{fig:dust_characterization}
\end{figure*}

\begin{figure}[thbp]
    \begin{center}
        \includegraphics[width=9cm]{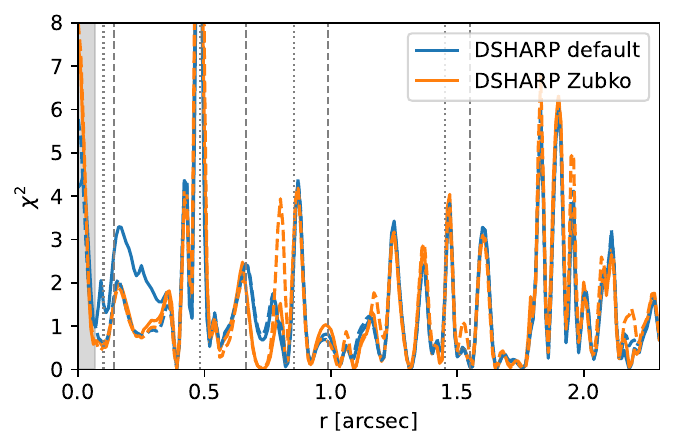}
    \end{center}
    \caption{
        Radial profiles of the minimum $\chi^2$ for each model.
        The blue lines are for the dust model with DSHARP default composition, and the orange lines are for the DSHARP Zubko composition. 
        The solid lines are for the compact grain models, and the dashed lines are for the porous grain models.
        The vertical gray dashed lines indicate the ring radii, and the gray dotted lines indicate the gap radii reported in \citep{Huang2018_ring}.
    }
    \label{fig:chi2}
\end{figure}

We used the relative profile likelihood normalized by the maximum likelihood for each parameter as a frequentist approach. The profile likelihood is obtained by maximizing the likelihood over the other parameters while fixing the parameter of interest.
The error range is defined as the range where $\Delta \chi^2 < 1$, with $\Delta \chi^2 = \chi^2 - \chi^2_{\mathrm{min}}$.
Here, $\chi^2$ is the value of Eq. \eqref{eq:chi2} calculated jointly for the parameter set $(T, \Sigma_{\mathrm{dust}}, a_{\mathrm{max}}, q)$, and $\chi^2_{\mathrm{min}}$ is its minimum at each radius.
When the $\Delta \chi^2 < 1$ range reaches the boundary of the parameter grid, we regard the parameter as unconstrained there.
In previous studies, parameter constraints are presented using marginalized distributions obtained by integrating the joint distribution over the other parameters. 
However, when multiple solutions exist, marginalization can make narrow but high-likelihood solutions appear less prominent, and the resulting intervals can depend on the adopted priors and parameterization. 
In this study, we use the relative profile likelihood to highlight the range of parameter values allowed by the data without integrating over the volume of the parameter space. 
A comparison between the profile likelihood and the marginalized distribution is shown in Appendix \ref{app:profile_marginal_comparison}.

\subsection{Fitted dust properties}

Here, we show the fitting results for the temperature $T$, surface density $\Sigma_{\mathrm{dust}}$, and maximum dust size $a_{\mathrm{max}}$ with the power-law index fixed at $q=3.5$.
The power-law index $q=3.5$ is the theoretically expected value when dust growth is limited by fragmentation \citep{Dohnanyi1969,Tanaka1996,Birnstiel2010_fragmentation}, and it is also consistent with the observed size distribution of interstellar dust \citep{Mathis1977} and previous estimate for the HD 163296 disk \citep{Doi2023}.
The fitting was performed using the radial profiles from the high-resolution images.
For the outer regions, where the high-resolution profiles are noisy, we also performed the SED fitting using the more sensitive low-resolution profiles, as shown in Appendix \ref{app:SED_low_res}.
In Appendix \ref{app:4d_fitting}, we show the fitting results for the temperature, surface density, dust size, and power-law index $q$ without assuming a specific value for the power-law index.
Also, to show the necessity of Band 9 observations, the fitting results without Band 9 profiles are shown in Appendix \ref{app:with_without_band9}.

\begin{figure*}[htbp]
  \centering
  \setlength{\tabcolsep}{0pt}
  \renewcommand{\arraystretch}{0}
  \begin{tabular}{cc}
    \includegraphics[width=9cm]{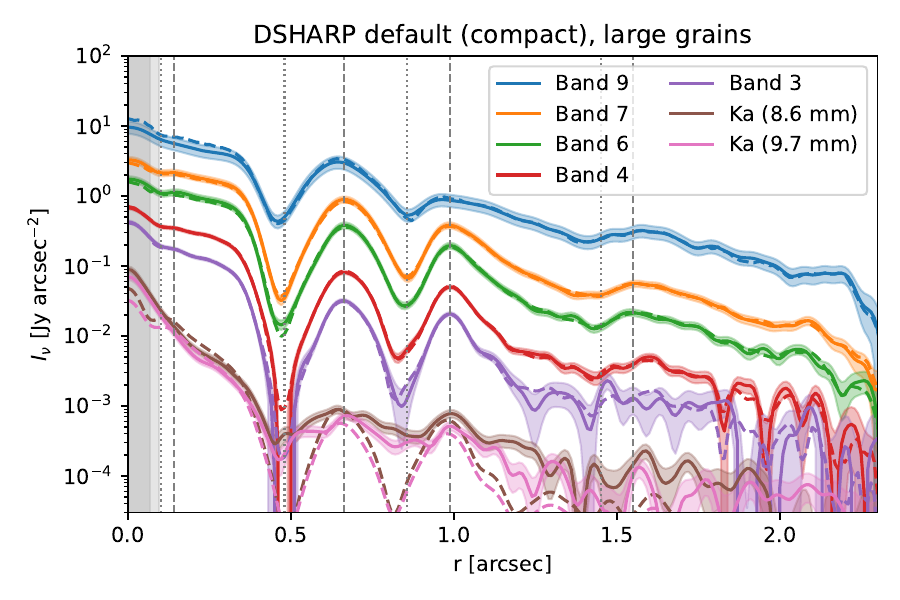} &
    \includegraphics[width=9cm]{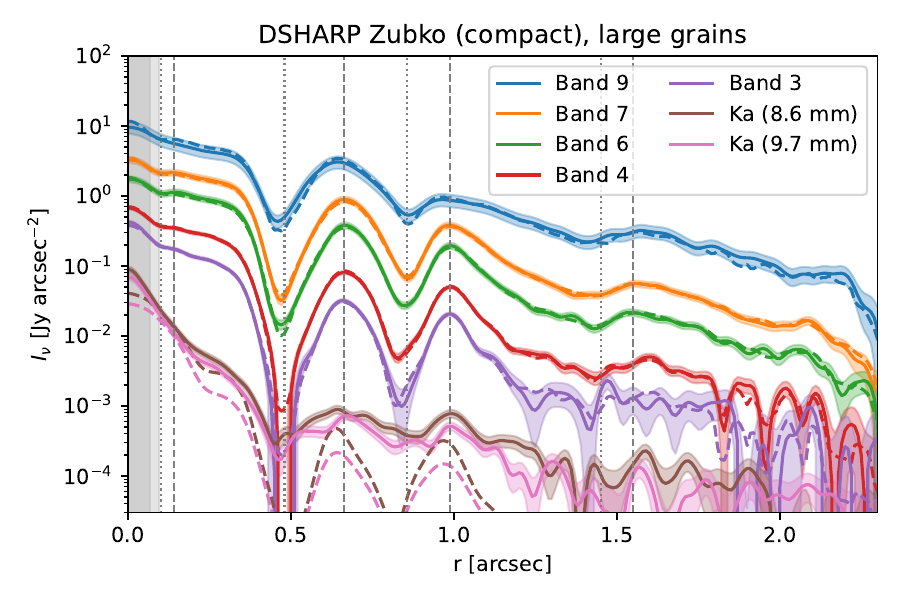} \\
    \makebox[9cm]{} &
    \includegraphics[width=9cm]{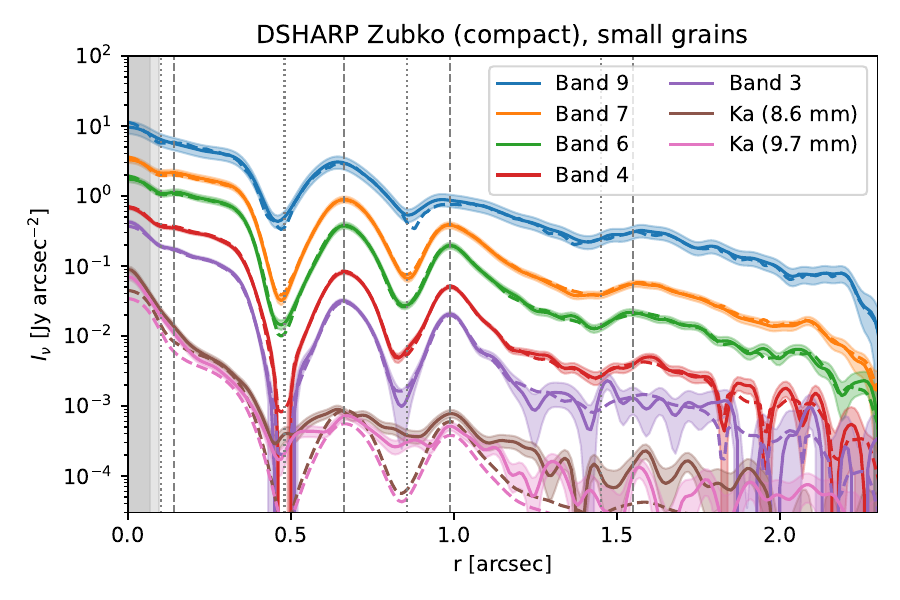} \\
    \includegraphics[width=9cm]{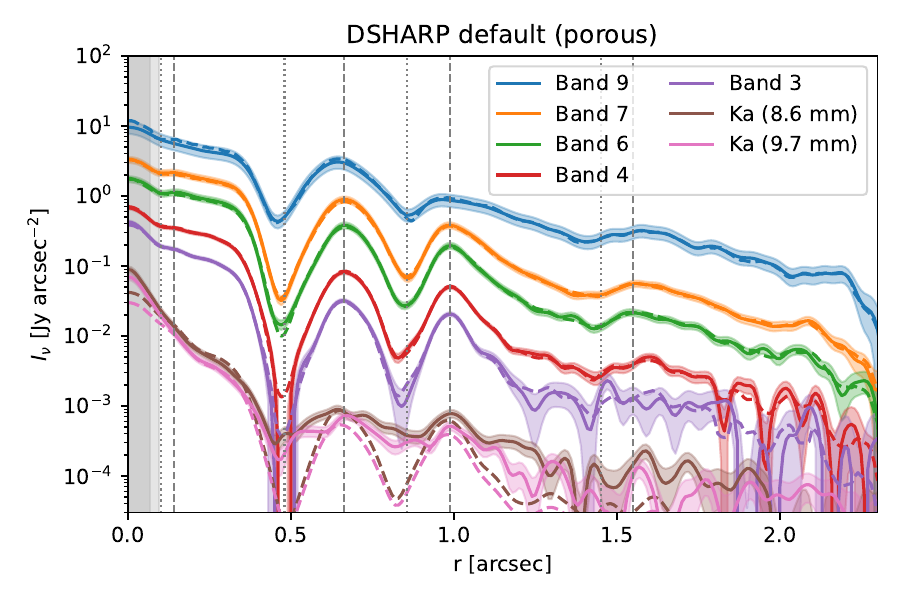} &
    \includegraphics[width=9cm]{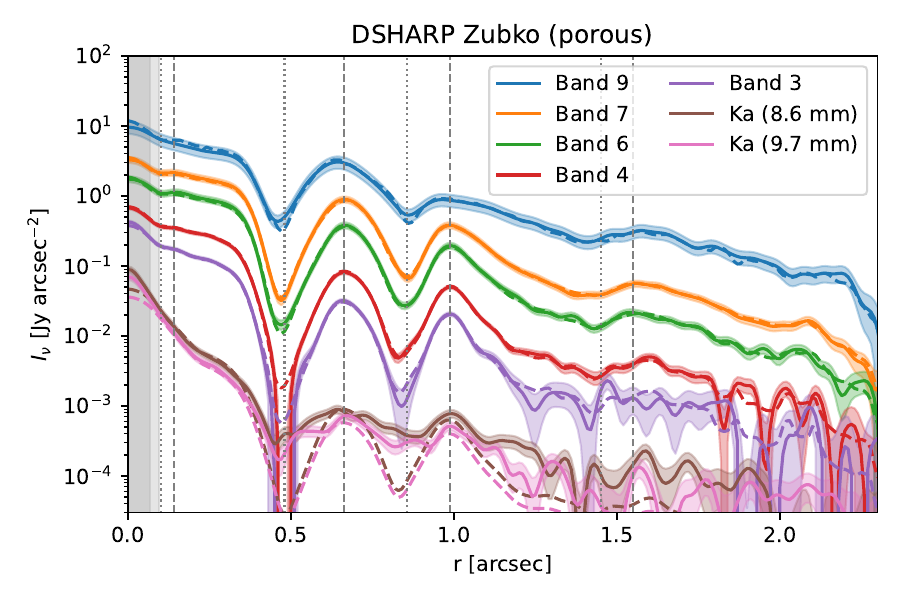}
  \end{tabular}
  \caption{
Comparison between the observed radial profiles (solid) and the model radial profiles (dashed) of the best-fit parameters from the SED fitting for each dust model.
For all panels, the shaded regions for the observed profiles show the total uncertainties, including both the thermal noise and the absolute flux calibration error.
The SED fitting was performed using the radial profiles at the ALMA wavelengths, and the VLA Ka-band radial profile is shown for comparison.
The dark-gray region on the left indicates the ALMA resolution, and the light-gray region indicates the VLA resolution.
The model radial profiles at the VLA wavelengths are smoothed to match the VLA angular resolution.
  }
  \label{fig:obs-model_comparison}
\end{figure*}

We show the relative profile likelihood for three parameters: temperature $T$, dust surface density $\Sigma_{\mathrm{dust}}$, and maximum dust size $a_{\mathrm{max}}$, obtained from the SED fitting with the power-law index fixed at $q=3.5$.
Here, we show the results of SED fitting assuming four dust models: DSHARP default (compact), DSHARP Zubko (compact), DSHARP default (porous), and DSHARP Zubko (porous).
At this stage, we do not assume any of these to be the fiducial dust model, and we show the results for each model.
Also, the fitting here is performed over the full parameter grids without imposing any physical constraints on the parameters.
We will discuss in Section \ref{subsec:dust_composition} which dust model is preferred, based on the ALMA and VLA observations and physical constraints.

Fig. \ref{fig:dust_characterization} shows the relative profile likelihood for three parameters: temperature $T$, dust surface density $\Sigma_{\mathrm{dust}}$, and maximum dust size $a_{\mathrm{max}}$, obtained from the SED fitting for the four dust models.
The left panels show the results for the DSHARP default composition, and the right panels show the results for the DSHARP Zubko composition.
The upper panels show the results for compact grains, and the lower panels show the results for porous grains.
We note that the MLEs (red lines) show radial oscillations, but this does not mean that the parameters are uniquely determined at these values, nor that the physical quantities truly oscillate.
The parameters are constrained within ranges of finite width, as indicated by the gray lines, and the estimates should be interpreted as these ranges rather than as single values.
As a reference for the temperature, we show the temperature profiles of the disk surface and midplane predicted by a passive disk model as follows
\begin{equation} \label{eq:temperature_profile}
    T(r) = \left( \frac{\varphi L_*}{8\pi r^2 \sigma_{\mathrm{SB}}} \right)^{1/4},
\end{equation}
where $\varphi$ is the flaring angle, $L_*$ is the stellar luminosity, and $\sigma_{\mathrm{SB}}$ is the Stefan-Boltzmann constant.
Here, we adopt $L_* = 17.0 L_\odot$ \citep{Fairlamb2015,Andrews2018} and $\varphi = 1$ for the disk surface and $\varphi = 0.02$ for the disk midplane \citep{Huang2018_ring}.
Furthermore, as an upper limit for the dust surface density, we show 1/100 of the gas surface density of a Toomre $Q=1$ disk, calculated assuming the midplane temperature profile described above, where Toomre $Q$ is
\begin{equation}
Q = \frac{c_s \Omega_{\mathrm{Kep}}}{\pi G \Sigma_{\mathrm{gas}}},
\end{equation}
where $c_s$ is the sound speed, $\Omega_{\mathrm{Kep}}$ is the Keplerian angular velocity, and $G$ is the gravitational constant \citep{Toomre1964}.
In addition, we show the particle size for $\mathrm{St}=1$ for the Toomre $Q=1$ disk as a reference for the dust size, where the Stokes number $\mathrm{St}$ is defined as $\mathrm{St} = \pi a \rho_{\mathrm{int}} / (2 \Sigma_{\mathrm{gas}})$, where $a$ is the dust size and $\rho_{\mathrm{int}}$ is the dust internal density.

The dust properties were constrained for each model, although the inferred values depend on the assumed dust model.
For the DSHARP default (compact) model, the fitted parameters at the inner ring $r=0\farcs67$ are 
$\Sigma_{\mathrm{dust}} = 0.35^{+0.30}_{-0.08}\ \mathrm{g/cm^2}$,
$a_{\mathrm{max}} = 1.91^{+11.58}_{-1.43}\ \mathrm{cm}$, and
$T = 26.92^{+3.99}_{-2.93}\ \mathrm{K}$, 
and those at the outer ring $r=1\farcs00$ are
$\Sigma_{\mathrm{dust}} = 0.54^{+1.86}_{-0.17}\ \mathrm{g/cm^2}$,
$a_{\mathrm{max}} > 0.51\ \mathrm{cm}$, and 
$T = 14.13^{+1.01}_{-0.94}\ \mathrm{K}$.
For the DSHARP Zubko (porous) model, the parameters at the inner ring $r=0\farcs67$ are
$\Sigma_{\mathrm{dust}} = 0.19^{+0.02}_{-0.08}\ \mathrm{g/cm^2}$,
$a_{\mathrm{max}}^{\mathrm{eff}}  < 0.42\ \mathrm{cm}$, and 
$T = 21.88^{+8.32}_{-1.46}\ \mathrm{K}$,
and those at the outer ring $r=1\farcs00$ are
$\Sigma_{\mathrm{dust}} = 0.19^{+0.03}_{-0.02}\ \mathrm{g/cm^2}$,
$a_{\mathrm{max}}^{\mathrm{eff}} = 0.26^{+0.72}_{-0.22}\ \mathrm{cm}$, and
$T = 13.49^{+2.00}_{-1.19}\ \mathrm{K}$.
The corner plot for $r=1\farcs00$ is shown in Fig. \ref{fig:corner_DSHARP_default_100} in Appendix \ref{app:corner_plots}.
On the other hand, at the gap locations and the outer extended regions, the parameters are not well constrained, especially the temperature, which reaches the upper limit of the grid.
This is because the observed intensity at these locations is weak and poorly detected, especially at longer wavelengths, and there is not enough information to constrain the parameters.

Fig. \ref{fig:chi2} shows the radial profiles of the minimum $\chi^2$ for each model.
As can be seen from this figure, all models fit well in most radii and show similar values of $\chi^2$, and it is not possible to determine which model is more appropriate based on the ALMA observational data alone.
We can see that the $\chi^2$ is larger at the gap locations.
This is because the SED fitting is performed on the smoothed intensity, and at the gap locations, the contamination by emission from the ring edges is large in the smoothed intensity, which leads to a larger deviation from the model.
Therefore, it should be noted that the estimates at these gap locations may not be reliable due to this contamination.

Regarding the temperature, at the outer ring location of $r=1\farcs00$, the estimated temperature is lower than the midplane temperature of a passive disk for all dust models.
This may be due to the effect of self-shadowing by the disk itself, and we will discuss this in Section \ref{subsec:temperature}.

The surface density differs by more than an order of magnitude depending on the dust model.
We will discuss the validity of these estimates in Section \ref{subsec:dust_composition} and the total dust mass in Section \ref{sec:dust_mass}.

Regarding the dust size, the estimates also differ significantly depending on the dust model.
From previous studies, it is known that when assuming compact dust, the results of SED fitting can yield both a large dust size solution and a small dust size solution, separated by $a_{\mathrm{threshold}} \sim \lambda / 2\pi$ \citep{Sierra2021,Macias2021}.
In our fitting, for the DSHARP Zubko (compact) model, we obtained both a large dust size solution and a small dust size solution, while for the DSHARP default (compact) model, the large dust size solution was preferred except for the gap regions where the estimates are less reliable.
For the porous dust models, the opacity and spectral index do not have a local peak at $a_{\mathrm{threshold}} \sim \lambda / 2\pi$ and have a monotonic dependence on dust size, as shown in Fig. \ref{fig:opacity}, so the fitting did not yield split solutions, although the constraints are loose.

To compare each of the solutions described above with the observations separately, we performed the SED fitting again within the parameter ranges restricted for each solution.
For the dust size, we evaluate the large- and small-grain solutions of the DSHARP Zubko (compact) model separately by imposing a lower or upper bound at $100\ \mathrm{\mu m}$, because both solutions are obtained with comparable likelihoods.
For the DSHARP default (compact) model, we impose a lower bound of $200\ \mathrm{\mu m}$ because the small-grain solution is not preferred in the fitting.
In addition, we limit the upper bound of the temperature to the surface temperature of a passive disk based on physical considerations.
Fig. \ref{fig:obs-model_comparison} shows a comparison of the radial profiles between observations and models based on these best-fit parameters.
The absorption and effective scattering optical depth profiles are shown in Fig. \ref{fig:tau_abs} and \ref{fig:tau_sca} in Appendix \ref{app:tau_profiles}.
The shaded regions for the observed profiles show the total uncertainties, including both the thermal noise and the absolute flux calibration error.
The SED fitting was performed using the radial profiles at the ALMA wavelengths, but we also plot the observed and modeled VLA Ka band radial profile for comparison.
The dark gray region on the left indicates the ALMA resolution, and the light gray region indicates the VLA resolution.
The model radial profiles at the VLA wavelengths are smoothed to match the VLA angular resolution.

For the ALMA wavelengths, we can see that the observed radial profiles and the model radial profiles based on the SED fitting results agree well throughout the disk for all dust models.
For the VLA wavelengths, we focus on the intensity at the ring locations because the VLA has characteristic sidelobes and lower sensitivity, which can result in leakage of intensity from bright regions into the gaps.
The intensity at the ring locations for the large-grain solution of the DSHARP Zubko (compact) model underestimates the observed intensity, whereas the other models are consistent with the observed ring intensities.
This suggests that the large-grain solution of the DSHARP Zubko (compact) model is not favored by the VLA observations.
The remaining four solutions are all broadly consistent with the available observational constraints and are therefore further assessed in Section \ref{subsec:dust_composition} based on physical considerations.

\section{Discussion} \label{sec:discussion}

\subsection{Preferred dust composition} \label{subsec:dust_composition}

In this study, we performed fitting assuming four dust models: the DSHARP default and DSHARP Zubko compositions, each with compact and porous grains.
The SED fitting using the five ALMA wavelengths shows that all dust models yield similar $\chi^2$ values for the best-fit models, and we cannot discriminate among any of the dust models.
Among these, the DSHARP Zubko (compact) model yields two solutions corresponding to large and small dust sizes, while the other three dust models yield one solution each, resulting in a total of five solutions.

From the comparison of the intensities at the VLA wavelengths, we found that the large grain solution of the DSHARP Zubko (compact) model underestimates the observed intensities.
Therefore, from the observations including both ALMA and VLA, the large grain solution of the DSHARP Zubko (compact) model is not favored.

We discuss which of these solutions is preferred based on other observational and physical constraints.
First, we assess the dust models from the perspective of the dust scale height.
The dust scale height of this disk has been shown to be puffed up to the gas scale height in the inner ring, while in the outer ring it is settled to less than 1/10 of the gas scale height \citep{Doi2021,Doi2023}.
If the small-grain solution of the DSHARP Zubko (compact) model is correct, the turbulence strength required for dust with $a \simeq 50 \mathrm{\mu m}$ to settle to about $1/10$ of the gas scale height in the outer ring would need to be unrealistically small \citep{Dubrulle1995,Youdin2007}.
Thus, we consider that the solution with smaller dust sizes for the DSHARP Zubko (compact) model is not favored.

We then examine the dust models from the viewpoint of gravitational stability.
For the DSHARP default (porous) model, the inferred disk would be gravitationally unstable under the canonical dust-to-gas ratio of 0.01, as shown in Fig. \ref{fig:dust_characterization}.
Especially for the outer ring, even if we assume a disk with $Q \sim 1$, the dust-to-gas ratio would be around 10 and the Stokes number would be around 1, which would make it difficult to maintain this state for a long time without triggering streaming instability \citep{Youdin2005}.
From these considerations, we regard the solution for the DSHARP default (porous) model as physically not favored.

We further assess the dust models from the perspective of the dust-to-gas ratio, using the gas disk mass derived from molecular line observations and the dust mass derived from our analysis.
The disk gas mass of HD 163296 has been estimated from observations of CO isotopologues \citep{Booth2019}, as well as from dynamical mass measurements based on the rotation curves of molecular lines \citep{Martire2024,Pezzotta2025}. 
We calculate the total dust-to-gas mass ratio of the entire disk by dividing the total dust mass derived in Section \ref{sec:dust_mass} and listed in Table \ref{tab:dust_mass}, including its $1\sigma$ uncertainty, by the gas mass of $0.12 M_\odot$ estimated by \citet{Pezzotta2025}.
For the DSHARP default model, the dust-to-gas ratio is $0.014 - 0.19$ for the compact-grain large-grain solution and $0.059 - 0.24$ for the porous model, both of which exceed the canonical value of $0.01$. 
In contrast, for the DSHARP Zubko model, the dust-to-gas ratio ranges from $0.00055 - 0.032$ for the compact-grain large-grain solution, $0.0066 - 0.037$ for the small-grain solution, and $0.0046 - 0.068$ for the porous model, all of which include the canonical value of $0.01$ within their ranges. 
Therefore, from the viewpoint of the disk-integrated dust-to-gas ratio, the DSHARP Zubko models are preferred to the DSHARP default models.
In summary, when combined with the constraints from the VLA intensities, dust scale height, and gravitational stability discussed above, the DSHARP Zubko (porous) model appears to be the most plausible solution for this system.

Comparing with other studies on dust composition in disks, \citet{Shi2026} performed SED fitting for MWC 480 using both ALMA and VLA data, and also favored the DSHARP default (compact) model and the DSHARP Zubko (porous) model, which is consistent with our results.
\citet{Zagaria2025} reported that dust models with amorphous carbon and relatively low porosity are preferred for the CI Tau disk, based on ALMA and VLA observations.
CI Tau shows a large spectral index of $\alpha > 4$, which may have led to the preference for compact models, while our observations did not show such high $\alpha$.
The population synthesis model of \citet{Delussu2024} shows a preference for compact dust models that include the carbonaceous material of \citet{Zubko1996} with a high absorption coefficient.
On the other hand, \citet{Yoshida2025albedo} suggests a high albedo for the TW Hya disk based on the estimation of gas temperature using molecular lines for the inner region of TW Hya, which does not favor the carbonaceous material of \citet{Zubko1996}.
Many millimeter polarization observations show signatures of self-scattering \citep[e.g.,][]{Kataoka2015,Stephens2017}, which also favor the refractory organics for carbonaceous material of \citet{Henning1996} with a high albedo.
\citet{Ueda2025} analyzed multi-wavelength continuum observations as well as polarization observations of HL Tau and the favored dust model is porous dust with refractory organics for carbonaceous material.
For HD 163296, polarization observations show polarization consistent with self-scattering \citep{Dent2019,Ohashi2019,Lin2020}, which might favor refractory organics for the carbonaceous material.
Thus, no consensus has yet been reached on the appropriate dust model for disks, and further research is needed.

\begin{table*}[htbp]
\caption{Dust mass within each radial region from the SED fitting.}
\centering
\renewcommand{\arraystretch}{1.2}
\begin{tabular}{lcccccccc}
\hline\hline
 & \multicolumn{4}{c}{radial region} & total\\
\hline
 & central disk & inner ring & outer ring & extended disk\\
$r$ [arcsec] & $<0.48$ & $0.48-0.85$ & $0.85-1.45$ & $1.45-2.3$ & $<2.3$   \\
\hline\hline
dust model & \multicolumn{5}{c}{dust mass $[M_{\mathrm{earth}}]$} & \\
\hline
DSHARP default (compact) (large) & $2551.8^{+3960.0}_{-2220.0}$ & $84.7^{+106.9}_{-23.0}$ & $205.3^{+365.9}_{-79.0}$ & $40.0^{+93.8}_{-27.3}$ & $2881.9^{+4526.6}_{-2349.2}$  \\
DSHARP Zubko (compact) (large) & $92.0^{+1118.9}_{-79.5}$ & $3.7^{+1.8}_{-1.2}$ & $7.4^{+3.1}_{-1.8}$ & $0.9^{+5.4}_{-0.2}$ & $104.0^{+1129.2}_{-82.6}$  \\
DSHARP Zubko (compact) (small) & $311.7^{+923.8}_{-183.0}$ & $47.9^{+7.9}_{-7.5}$ & $94.2^{+15.5}_{-17.4}$ & $24.1^{+12.9}_{-13.8}$ & $477.9^{+960.0}_{-221.8}$  \\
DSHARP default (porous) & $4543.3^{+1908.6}_{-3338.7}$ & $441.4^{+226.3}_{-102.3}$ & $980.2^{+675.5}_{-287.2}$ & $146.9^{+203.6}_{-85.4}$ & $6111.9^{+3014.0}_{-3813.5}$  \\
DSHARP Zubko (porous) & $499.0^{+1967.7}_{-393.0}$ & $35.0^{+11.2}_{-12.5}$ & $61.6^{+30.7}_{-14.4}$ & $8.2^{+23.4}_{-4.3}$ & $603.7^{+2033.0}_{-424.2}$  \\
\hline
\end{tabular}
\tablefoot{
The SED fitting can sometimes yield two different solutions for large and small dust grains.
For the DSHARP default model, the small grain solution is not favored, so we show only the large-grain solution. In contrast, for the DSHARP Zubko model, we show both the large and small grain solutions separately.
Here, we separate large and small dust grains with $a_{\mathrm{threshold}}=200\ \mathrm{\mu m}$ for the DSHARP default model and $a_{\mathrm{threshold}}=100\ \mathrm{\mu m}$ for the DSHARP Zubko model.
}
\label{tab:dust_mass}
\end{table*}

\subsection{Thermal structure} \label{subsec:temperature}

Our SED fitting constrained the temperature together with the other dust properties, without assuming a temperature profile or temperature priors.
The temperature is $T = 26.92^{+3.99}_{-2.93}\ \mathrm{K}$ at $0\farcs67$ and $T = 14.13^{+1.01}_{-0.94}\ \mathrm{K}$ at $1\farcs00$, and the ratio is $1.91^{+0.31}_{-0.24}$ for DSHARP default (compact) model.
For DSHARP Zubko (porous) model, the temperature is $T = 21.88^{+8.32}_{-1.46}\ \mathrm{K}$ at $0\farcs67$ and $T = 13.49^{+2.00}_{-1.19}\ \mathrm{K}$ at $1\farcs00$, and the ratio is $1.62^{+0.66}_{-0.18}$.
If we assume a passive disk, the temperature follows $T \propto r^{-1/2}$ for a constant grazing angle and $T \propto r^{-3/7}$ for a self-consistent temperature structure \citep{Chiang1997}, which would yield temperature ratios of $1.22$ and $1.18$ between $0\farcs67$ and $1\farcs00$, respectively.
This indicates that the temperature structure is steeper than the passive disk.

\begin{figure}[bhtp]
    \begin{center}
        \includegraphics[width=7cm]{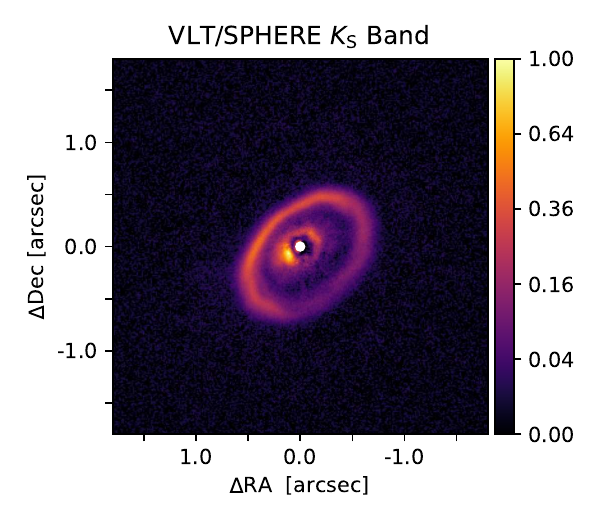}
        \includegraphics[width=9cm]{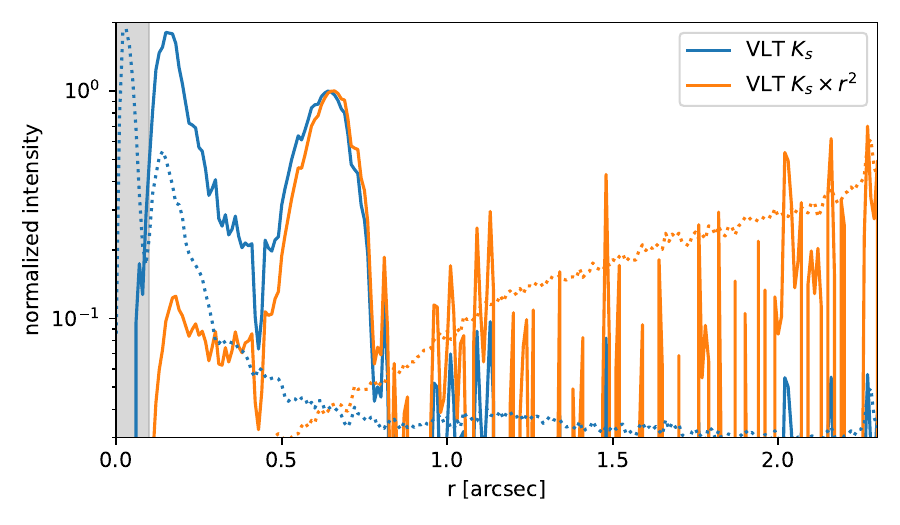}
    \end{center}
    \caption{
Observed Stokes $Q_{\phi}$ image in the $K_s$ band from VLT/SPHERE (top) and the intensity profile along the major axis (bottom).
The VLT image is taken from Fig. 1 of \citet{Ren2023} for the observation on 2021 June 3 and is normalized by its peak intensity.
The radial profile is a cross-section along the major axis toward the northwest, corrected for the scattering height.
The blue line shows the profile extracted from the VLT image, and the orange line shows the same profile multiplied by $r^2$.
The dotted line indicates the noise level, estimated at each radius as the azimuthal rms of the Stokes $U_{\phi}$ image, which is ideally expected to be zero for single scattering of the stellar light.
    }
    \label{fig:VLT}
\end{figure}

This result is consistent with previous optical and infrared observations. 
\citet{Muro-Arena2018, Ren2023} performed infrared observations of HD 163296 with VLT/SPHERE. 
As shown in Fig. \ref{fig:VLT}, they clearly detected the inner ring of this object, whereas the outer ring was not detected. 
This indicates that the outer ring is hidden in the shadow of the inner ring, preventing it from receiving direct light from the central star and resulting in locally lower temperatures.

The shadowing of the outer ring by the inner ring is also consistent with the vertical distribution of dust inferred from millimeter observations.
\citet{Doi2021,Doi2023} showed from high-resolution ALMA observations of HD 163296 that the dust is puffed up in the inner ring, while it is settled in the outer ring. 
Although the dust population traced by infrared observations is different, the dust in the outer ring is settled, which leads to shadowing.

The temperature of the central disk is also lower than the midplane temperature expected from Eq. (\ref{eq:temperature_profile}).
The bottom panel of Fig. \ref{fig:VLT} shows the radial profile along the major axis of the VLT observations and the profile rescaled by $r^2$.
If the grazing angle and the dust optical properties are constant, the scattered light intensity rescaled by $r^2$ should be constant.
However, the profile rescaled by $r^2$ (orange line) is about 10 times lower in intensity in the central disk compared to the peak of the inner ring.
This suggests that the radiation from the central star reaching the central disk is weaker than expected from a smooth disk, and as a result, the temperature of the central disk is lower than expected from a passive disk model.
This can be explained if the innermost structure casts a shadow on the central disk. 
Such shadowing by a puffed-up inner structure has been discussed theoretically \citep{Turner2014,Ueda2019}. 
This shadowing may be caused by the innermost structure observed with VLTI \citep{Varga2021}.

\subsection{Dust size and structure formation}

The inferred dust size depends on the adopted dust model. 
For the preferred DSHARP Zubko (porous) model, the dust size is only weakly constrained, with an upper limit of less than a centimeter, and no clear correlation is found between the dust size and the ring-gap structure. 
Theoretically, if the timescales of growth and fragmentation are shorter than those of accumulation and diffusion at a pressure maximum, the grain size may be determined primarily by local growth and fragmentation, resulting in little or no size segregation \citep{Jiang2024,Yang2025}. 
The result for the DSHARP Zubko (porous) model may indicate such a situation.

For comparison, if we adopt the DSHARP default (compact) model, which is not preferred, the dust size is estimated to be larger than $1\ \mathrm{mm}$ throughout the disk, suggesting that dust growth has already progressed even in the outer regions of the HD 163296 disk.
In particular, in the $0\farcs67$ ring, the dust size shows a local maximum near the ring center, which is consistent with dust accumulation at a gas pressure maximum \citep{Rosotti2020,Izquierdo2023}. 
Theoretically, larger and more weakly coupled grains are expected to be trapped more efficiently at gas pressure maxima, leading to larger characteristic grain sizes near the ring center \citep{Weidenschilling1977mnras,Pinilla2012a}. 

\subsection{Dust mass and planet formation} \label{sec:dust_mass}

We estimate the dust mass for each disk structure by integrating the surface density over the radial range of each structure as
\begin{equation}
M_{\mathrm{dust}} = 2\pi \int_{r_{\mathrm{in}}}^{r_{\mathrm{out}}} \Sigma_{\mathrm{dust}}(r) r dr,
\end{equation}
where $r_{\mathrm{in}}$ and $r_{\mathrm{out}}$ are the inner and outer radii of each structure, respectively.
As the uncertainty range, we take the difference between the mass obtained by integrating the upper limit of the $1\sigma$ error of the surface density and the mass obtained by integrating the lower limit of the $1\sigma$ error of the surface density.

Table \ref{tab:dust_mass} shows the estimated dust mass for each disk structure and each dust model.
The estimated dust mass varies significantly depending on the dust model.
The DSHARP default model shows a dust mass that is about an order of magnitude larger than that of the DSHARP Zubko model, which is due to the larger absorption opacity of the carbonaceous material of \citet{Zubko1996}.
Also, the porous models show a dust mass that is several times larger than that of the compact models, because the porous models do not have the peak of absorption opacity that the compact models have, as shown in Fig. \ref{fig:opacity}.
In all dust models, there is enough dust mass in the central disk and the two rings to form Earth-like planets.
In addition, the outer ring is more massive than the inner ring, and the central disk is about an order of magnitude more massive than the two rings.

\subsection{Comparison with previous studies} \label{subsec:previous_studies}

\citet{Sierra2021} estimated the dust size from ALMA Band 3 and 6 using the DSHARP default (compact) model, assuming a temperature profile constrained by molecular line observations \citep{Zhang2021}.
They estimated millimeter-sized dust grains in both rings, and also showed that the dust size is larger in the outer ring.
Our fitting also shows that the dust size is larger in the outer ring, although the estimated dust size is larger than that of \citet{Sierra2021}.
\citet{Sierra2021} estimated the total dust mass to be $2.8^{+4.5}_{-0.6} \times 10^2$ Earth masses.
This is comparable to the lower limit of the large dust solution of the DSHARP default (compact) model.

\citet{Guidi2022} reported that a dust size of $a \sim 200\ \mathrm{\mu m}$ is favored throughout the disk.
They used the radial profiles from ALMA Band 3, 4, 6, and 7, as well as the radial profiles from VLA Ka-band, and performed SED fitting to the radial profiles estimated by \texttt{frank}, a non-parametric visibility modeling for the radial profile assuming symmetric structure \citep{Jennings2020}.
However, in the \texttt{frank} profile of \citet{Guidi2022}, the inner ring is not detected in VLA Ka Band, and the difference in the radial profiles may have influenced the dust size estimation.
Furthermore, they adopted the DIANA dust model \citep{Min2016}.
The differences in assumptions can also systematically affect the estimated dust sizes.

\citet{Dent2019,Ohashi2019} constrained the dust size based on ALMA Band 7 polarization observations.
They showed that the polarization pattern is consistent with self-scattering, and the polarization fraction is maximum at the disk gap locations, while weak polarization is also detected in the central disk. 
Based on this, they proposed that the dust size is $\sim 140\ \mathrm{\mu m}$ in the gaps, while slightly different sizes of dust dominate the emission in the inner disk.
However, recent studies have shown that larger dust grains can also produce self-scattering polarization in the case of porous dust grains \citep{Tazaki2019} or irregularly shaped dust grains \citep{Lin2023pol,Lin2025}, so the dust-size constraints inferred from these polarization observations may depend on the assumed dust model.
The weak constraint on dust size obtained for the DSHARP Zubko (porous) model in this work does not contradict the constraints from the polarization observations.
  
\subsection{Caveats and future prospects} \label{sec:caveats}

The temperature, surface density, and dust size distribution inferred from the SED fitting are parameters that reproduce the observed beam-smoothed radial intensity profiles, rather than the intrinsic intensity distribution.
In particular, if unresolved sub-beam structures are present, fitting the smoothed intensity profiles can lead to systematic deviations from the true dust properties.
In this study, we performed the fitting using high-angular-resolution data with an angular resolution of $0\farcs066$, but the inferred parameters may still be biased if finer-scale structures remain unresolved.

Another major source of uncertainty arises from uncertainties in the dust composition and optical constants used in the SED fitting.
The dust composition is still poorly constrained, and the optical properties of each composition model also remain highly uncertain.
The optical constants for refractory organics adopted by \citet{Pollack1996,Henning1996} were constructed by combining measurements of different materials over different wavelength ranges.
In addition, the optical properties of amorphous carbon (BE) from \citet{Zubko1996} have been measured up to $2$ mm in wavelength, and beyond that, the optical constants rely on extrapolation. Therefore, the interpretation of the Band 3 and VLA data can be affected by this extrapolation.
Moreover, the optical properties of amorphous silicates may also depend on temperature \citep{Demyk2022}.
Our SED fitting results show that the inferred dust properties depend strongly on the adopted dust model.

In addition to the short wavelength observations, longer-wavelength observations may be effective for discriminating dust compositions or more strongly constraining dust properties.
The predicted intensity at 9 mm can differ depending on the assumed composition, and thus continuum observations at longer wavelengths can be effective for composition discrimination.
In this study, we did not include the VLA data in the fitting due to limitations in angular resolution, sensitivity, and PSF characteristics.
Since ALMA Cycle 11, Band 1 observations at 7-9 mm have become available and offer high sensitivity and image fidelity.
On the other hand, the best angular resolution of ALMA Band 1 is typically $\sim 0\farcs1$ \citep{Privon2025}, which is coarser than that achieved by the VLA in Ka and Q bands.
Therefore, combining ALMA Band 1 with VLA long-wavelength data can provide complementary improvements in sensitivity, angular resolution, and image fidelity, potentially enabling more robust SED fitting and stronger constraints on dust composition.
Ongoing observations with ALMA Band 1 and the VLA Q band for this target will allow stronger constraints on the dust properties.

In the longer term, the ngVLA is expected to deliver long-wavelength observations with substantially improved sensitivity and angular resolution, which will further refine constraints on dust properties including dust composition \citep{Murphy2018,Kadler2023}.
However, longer-wavelength observations become increasingly sensitive to uncertainties in the long-wavelength extrapolation of the optical constants.
Therefore, in parallel with advances in observational capabilities, improved laboratory measurements that better constrain the relevant optical properties will be essential.

\section{Conclusion} \label{sec:conclusion}

We performed high angular resolution observations of HD 163296 in ALMA Band 9 (0.45 mm).
Combined with existing high angular resolution observations in ALMA Band 3, 4, 6, and 7, we performed SED fitting using continuum observations over a wide wavelength range from 0.45 mm to 3 mm at a resolution of $0\farcs066$ to infer the dust properties of the HD 163296 disk.

Our main conclusions are as follows.
\begin{itemize}
    \item The high angular resolution observations in ALMA Band 9 have revealed the structure of the disk around HD 163296, including the central disk, two rings, and an extended outer disk. The rings appear broader and the gaps appear shallower at shorter wavelengths. The extended emission is brighter at higher frequencies, and emission is detected out to 2.3 arcsec. The crescent structure is observed with lower contrast at higher frequencies. These can be explained by optical-depth effects and/or size segregation.
    \item SED fitting using the ALMA Band 3, 4, 6, 7, and 9 images broke the degeneracy and constrained the temperature, surface density, and dust size for each dust model. However, the inferred dust properties depend strongly on the choice of dust model, and no model can be rejected based on the ALMA observations alone. Considering the comparison of the observed intensity with VLA Ka-band and physical requirements such as scale height, gravitational stability and the dust-to-gas ratio, we identified the DSHARP Zubko (porous) model as the preferred model.
    \item The estimated temperature is lower than that of a smooth passive disk model in the outer ring for all the models considered. This suggests that shadowing by the inner ring is lowering the temperature in the outer ring.
    \item The estimated dust size depends on the adopted dust model. In the DSHARP Zubko (porous) model, the constraint on the dust size is weak and there is no apparent size segregation. In the DSHARP default (compact) model, which is not preferred, grown dust is inferred throughout the disk, and there is a tendency for the size to be locally maximized at the ring centers, suggesting that size segregation due to accumulation at the gas pressure maximum may be occurring.
    \item The estimated dust mass also depends on the dust model. In the preferred DSHARP Zubko (porous) models, the central disk, the two rings, and the extended disk each contain more than a few Earth masses of dust.
\end{itemize}

These results demonstrate that high-frequency, high-resolution observations with ALMA Band 9 provide a powerful way to break the temperature degeneracy and constrain the dust properties of protoplanetary disks. 
At the same time, the inferred dust properties remain highly dependent on the adopted dust model, and future high-resolution observations at longer wavelengths, together with improved laboratory constraints on dust optical constants, will be essential for further constraining the dust properties.

\begin{acknowledgements}

We thank the anonymous referee for constructive comments that improved the manuscript.
We acknowledge Andrew Winter for the helpful discussions on statistics.
We also thank Bin Ren for helpful advice on the noise estimation of the infrared image.
M.B. has received funding from the European Research Council (ERC) under the European Union’s Horizon 2020 research and innovation program (PROTOPLANETS, grant agreement No. 101002188). 
C.D. acknowledges support from the ERC AdG project grant 101199769 GT4Pebbles.
Views and opinions expressed are, however, those of the author(s) only and do not necessarily reflect those of the European Union or the European Research Council. 
Neither the European Union nor the granting authority can be held responsible for them. 
This work was supported by JSPS KAKENHI Grant Nos. 22K03680, 23KJ1008, 24K07097, and 25K07351.

This paper makes use of the following ALMA data: 
ADS/JAO.ALMA\#2023.1.00578.S
ADS/JAO.ALMA\#2015.1.00616.S
ADS/JAO.ALMA\#2016.1.01086.S
ADS/JAO.ALMA\#2013.1.00366.S
ADS/JAO.ALMA\#2013.1.00601.S
ADS/JAO.ALMA\#2016.1.00484.L
ADS/JAO.ALMA\#2017.1.01682.S
ADS/JAO.ALMA\#2015.1.00725.S
ADS/JAO.ALMA\#2016.1.01086.S
ALMA is a partnership of ESO (representing its member states), NSF (USA), and NINS (Japan), together with NRC (Canada), MOST and ASIAA (Taiwan), and KASI (Republic of Korea), in cooperation with the Republic of Chile. 
The Joint ALMA Observatory is operated by ESO, AUI/NRAO, and NAOJ. 
The National Radio Astronomy Observatory and Green Bank Observatory are facilities of the U.S. National Science Foundation operated under cooperative agreement by Associated Universities, Inc.
The authors thank the DSHARP team and \citet{Guidi2022} for making their data publicly available.

\end{acknowledgements}

\bibliographystyle{aa}
\bibliography{doi_citation}

\begin{thebibliography}{119}
\expandafter\ifx\csname natexlab\endcsname\relax\def\natexlab#1{#1}\fi

\bibitem[{{ALMA Partnership} {et~al.}(2015){ALMA Partnership}, {Brogan},
  {P{\'e}rez}, {Hunter}, {Dent}, {Hales}, {Hills}, {Corder}, {Fomalont},
  {Vlahakis}, {Asaki}, {Barkats}, {Hirota}, {Hodge}, {Impellizzeri}, {Kneissl},
  {Liuzzo}, {Lucas}, {Marcelino}, {Matsushita}, {Nakanishi}, {Phillips},
  {Richards}, {Toledo}, {Aladro}, {Broguiere}, {Cortes}, {Cortes}, {Espada},
  {Galarza}, {Garcia-Appadoo}, {Guzman-Ramirez}, {Humphreys}, {Jung}, {Kameno},
  {Laing}, {Leon}, {Marconi}, {Mignano}, {Nikolic}, {Nyman}, {Radiszcz},
  {Remijan}, {Rod{\'o}n}, {Sawada}, {Takahashi}, {Tilanus}, {Vila Vilaro},
  {Watson}, {Wiklind}, {Akiyama}, {Chapillon}, {de Gregorio-Monsalvo}, {Di
  Francesco}, {Gueth}, {Kawamura}, {Lee}, {Nguyen Luong}, {Mangum}, {Pietu},
  {Sanhueza}, {Saigo}, {Takakuwa}, {Ubach}, {van Kempen}, {Wootten},
  {Castro-Carrizo}, {Francke}, {Gallardo}, {Garcia}, {Gonzalez}, {Hill},
  {Kaminski}, {Kurono}, {Liu}, {Lopez}, {Morales}, {Plarre}, {Schieven},
  {Testi}, {Videla}, {Villard}, {Andreani}, {Hibbard}, \&
  {Tatematsu}}]{alma2015}
{ALMA Partnership}, {Brogan}, C.~L., {P{\'e}rez}, L.~M., {et~al.} 2015, \apjl,
  808, L3

\bibitem[{{Andrews} {et~al.}(2018){Andrews}, {Huang}, {P{\'e}rez}, {Isella},
  {Dullemond}, {Kurtovic}, {Guzm{\'a}n}, {Carpenter}, {Wilner}, {Zhang}, {Zhu},
  {Birnstiel}, {Bai}, {Benisty}, {Hughes}, {{\"O}berg}, \&
  {Ricci}}]{Andrews2018}
{Andrews}, S.~M., {Huang}, J., {P{\'e}rez}, L.~M., {et~al.} 2018, \apjl, 869,
  L41

\bibitem[{{Benisty} {et~al.}(2023){Benisty}, {Dominik}, {Follette}, {Garufi},
  {Ginski}, {Hashimoto}, {Keppler}, {Kley}, \& {Monnier}}]{Benisty2023}
{Benisty}, M., {Dominik}, C., {Follette}, K., {et~al.} 2023, in Astronomical
  Society of the Pacific Conference Series, Vol. 534, Protostars and Planets
  VII, ed. S.~{Inutsuka}, Y.~{Aikawa}, T.~{Muto}, K.~{Tomida}, \& M.~{Tamura},
  605

\bibitem[{{Benisty} {et~al.}(2026){Benisty}, {Izquierdo}, {Stadler},
  {Galloway-Sprietsma}, {Facchini}, {Winter}, {Bae}, {Fukagawa}, {Teague},
  {Pinte}, {Andrews}, {Barraza-Alfaro}, {Cataldi}, {Curone}, {Czekala},
  {Fasano}, {Flock}, {Garg}, {Huang}, {Ilee}, {Kanagawa}, {Lawrence}, {Lesur},
  {Lodato}, {Longarini}, {Loomis}, {M{\'e}nard}, {Orihara}, {Price}, {Rosotti},
  {Wafflard-Fernandez}, {Wilner}, {W{\"o}lfer}, {Yen}, {Yoshida}, \&
  {Zawadzki}}]{Benisty2026}
{Benisty}, M., {Izquierdo}, A.~F., {Stadler}, J., {et~al.} 2026, \apjl, 1000,
  L14

\bibitem[{{Birnstiel}(2024)}]{Birnstiel2024ARAA}
{Birnstiel}, T. 2024, \araa, 62, 157

\bibitem[{{Birnstiel} {et~al.}(2018){Birnstiel}, {Dullemond}, {Zhu}, {Andrews},
  {Bai}, {Wilner}, {Carpenter}, {Huang}, {Isella}, {Benisty}, {P{\'e}rez}, \&
  {Zhang}}]{Birnstiel2018}
{Birnstiel}, T., {Dullemond}, C.~P., {Zhu}, Z., {et~al.} 2018, \apjl, 869, L45

\bibitem[{{Birnstiel} {et~al.}(2010){Birnstiel}, {Ricci}, {Trotta},
  {Dullemond}, {Natta}, {Testi}, {Dominik}, {Henning}, {Ormel}, \&
  {Zsom}}]{Birnstiel2010_fragmentation}
{Birnstiel}, T., {Ricci}, L., {Trotta}, F., {et~al.} 2010, \aap, 516, L14

\bibitem[{{Blum} \& {Wurm}(2008)}]{Blum2008ARAA}
{Blum}, J. \& {Wurm}, G. 2008, \araa, 46, 21

\bibitem[{{Booth} {et~al.}(2019){Booth}, {Walsh}, {Ilee}, {Notsu}, {Qi},
  {Nomura}, \& {Akiyama}}]{Booth2019}
{Booth}, A.~S., {Walsh}, C., {Ilee}, J.~D., {et~al.} 2019, \apjl, 882, L31

\bibitem[{{Brauer} {et~al.}(2008){Brauer}, {Dullemond}, \&
  {Henning}}]{Brauer2008}
{Brauer}, F., {Dullemond}, C.~P., \& {Henning}, T. 2008, \aap, 480, 859

\bibitem[{{Carrasco-Gonz{\'a}lez} {et~al.}(2019){Carrasco-Gonz{\'a}lez},
  {Sierra}, {Flock}, {Zhu}, {Henning}, {Chandler}, {Galv{\'a}n-Madrid},
  {Mac{\'\i}as}, {Anglada}, {Linz}, {Osorio}, {Rodr{\'\i}guez}, {Testi},
  {Torrelles}, {P{\'e}rez}, \& {Liu}}]{Carrasco-Gonzalez2019}
{Carrasco-Gonz{\'a}lez}, C., {Sierra}, A., {Flock}, M., {et~al.} 2019, \apj,
  883, 71

\bibitem[{{CASA Team} {et~al.}(2022){CASA Team}, {Bean}, {Bhatnagar}, {Castro},
  {Donovan Meyer}, {Emonts}, {Garcia}, {Garwood}, {Golap}, {Gonzalez Villalba},
  {Harris}, {Hayashi}, {Hoskins}, {Hsieh}, {Jagannathan}, {Kawasaki},
  {Keimpema}, {Kettenis}, {Lopez}, {Marvil}, {Masters}, {McNichols},
  {Mehringer}, {Miel}, {Moellenbrock}, {Montesino}, {Nakazato}, {Ott}, {Petry},
  {Pokorny}, {Raba}, {Rau}, {Schiebel}, {Schweighart}, {Sekhar}, {Shimada},
  {Small}, {Steeb}, {Sugimoto}, {Suoranta}, {Tsutsumi}, {van Bemmel},
  {Verkouter}, {Wells}, {Xiong}, {Szomoru}, {Griffith}, {Glendenning}, \&
  {Kern}}]{CASA2022}
{CASA Team}, {Bean}, B., {Bhatnagar}, S., {et~al.} 2022, \pasp, 134, 114501

\bibitem[{{Chiang} \& {Goldreich}(1997)}]{Chiang1997}
{Chiang}, E.~I. \& {Goldreich}, P. 1997, \apj, 490, 368

\bibitem[{Cortes {et~al.}(2025)Cortes, Carpenter, Kameno, Loomis, Vila~Vilaro,
  Immer, Vlahakis, Law, Stoehr, Saini, Hales, \& Kneissl}]{Cortes2025}
Cortes, P., Carpenter, J., Kameno, S., {et~al.} 2025, ALMA Cycle 12 Technical
  Handbook

\bibitem[{{Delussu} {et~al.}(2024){Delussu}, {Birnstiel}, {Miotello},
  {Pinilla}, {Rosotti}, \& {Andrews}}]{Delussu2024}
{Delussu}, L., {Birnstiel}, T., {Miotello}, A., {et~al.} 2024, \aap, 688, A81

\bibitem[{{Demyk} {et~al.}(2022){Demyk}, {Gromov}, {Meny}, {Ysard}, {Paradis},
  {Jones}, {Petitprez}, {Hubert}, {Leroux}, {Nayral}, \& {Delpech}}]{Demyk2022}
{Demyk}, K., {Gromov}, V., {Meny}, C., {et~al.} 2022, \aap, 666, A192

\bibitem[{{Dent} {et~al.}(2019){Dent}, {Pinte}, {Cortes}, {M{\'e}nard},
  {Hales}, {Fomalont}, \& {de Gregorio-Monsalvo}}]{Dent2019}
{Dent}, W.~R.~F., {Pinte}, C., {Cortes}, P.~C., {et~al.} 2019, \mnras, 482, L29

\bibitem[{{Dohnanyi}(1969)}]{Dohnanyi1969}
{Dohnanyi}, J.~S. 1969, \jgr, 74, 2531

\bibitem[{{Doi} \& {Kataoka}(2021)}]{Doi2021}
{Doi}, K. \& {Kataoka}, A. 2021, \apj, 912, 164

\bibitem[{{Doi} \& {Kataoka}(2023)}]{Doi2023}
{Doi}, K. \& {Kataoka}, A. 2023, \apj, 957, 11

\bibitem[{{Dominik} \& {Dullemond}(2024)}]{Dominik2024}
{Dominik}, C. \& {Dullemond}, C.~P. 2024, \aap, 682, A144

\bibitem[{{Dominik} {et~al.}(2021){Dominik}, {Min}, \& {Tazaki}}]{Dominik2021}
{Dominik}, C., {Min}, M., \& {Tazaki}, R. 2021, {OpTool: Command-line driven
  tool for creating complex dust opacities}, Astrophysics Source Code Library,
  record ascl:2104.010

\bibitem[{{Draine}(2003)}]{Draine2003ARAA}
{Draine}, B.~T. 2003, \araa, 41, 241

\bibitem[{{Dubrulle} {et~al.}(1995){Dubrulle}, {Morfill}, \&
  {Sterzik}}]{Dubrulle1995}
{Dubrulle}, B., {Morfill}, G., \& {Sterzik}, M. 1995, \icarus, 114, 237

\bibitem[{{Fairlamb} {et~al.}(2015){Fairlamb}, {Oudmaijer}, {Mendigut{\'\i}a},
  {Ilee}, \& {van den Ancker}}]{Fairlamb2015}
{Fairlamb}, J.~R., {Oudmaijer}, R.~D., {Mendigut{\'\i}a}, I., {Ilee}, J.~D., \&
  {van den Ancker}, M.~E. 2015, \mnras, 453, 976

\bibitem[{{Gaia Collaboration} {et~al.}(2023){Gaia Collaboration}, {Vallenari},
  {Brown}, {Prusti}, {de Bruijne}, {Arenou}, {Babusiaux}, {Biermann},
  {Creevey}, {Ducourant}, {Evans}, {Eyer}, {Guerra}, {Hutton}, {Jordi},
  {Klioner}, {Lammers}, {Lindegren}, {Luri}, {Mignard}, {Panem}, {Pourbaix},
  {Randich}, {Sartoretti}, {Soubiran}, {Tanga}, {Walton}, {Bailer-Jones},
  {Bastian}, {Drimmel}, {Jansen}, {Katz}, {Lattanzi}, {van Leeuwen}, {Bakker},
  {Cacciari}, {Casta{\~n}eda}, {De Angeli}, {Fabricius}, {Fouesneau},
  {Fr{\'e}mat}, {Galluccio}, {Guerrier}, {Heiter}, {Masana}, {Messineo},
  {Mowlavi}, {Nicolas}, {Nienartowicz}, {Pailler}, {Panuzzo}, {Riclet}, {Roux},
  {Seabroke}, {Sordo}, {Th{\'e}venin}, {Gracia-Abril}, {Portell}, {Teyssier},
  {Altmann}, {Andrae}, {Audard}, {Bellas-Velidis}, {Benson}, {Berthier},
  {Blomme}, {Burgess}, {Busonero}, {Busso}, {C{\'a}novas}, {Carry}, {Cellino},
  {Cheek}, {Clementini}, {Damerdji}, {Davidson}, {de Teodoro}, {Nu{\~n}ez
  Campos}, {Delchambre}, {Dell'Oro}, {Esquej}, {Fern{\'a}ndez-Hern{\'a}ndez},
  {Fraile}, {Garabato}, {Garc{\'\i}a-Lario}, {Gosset}, {Haigron}, {Halbwachs},
  {Hambly}, {Harrison}, {Hern{\'a}ndez}, {Hestroffer}, {Hodgkin}, {Holl},
  {Jan{\ss}en}, {Jevardat de Fombelle}, {Jordan}, {Krone-Martins}, {Lanzafame},
  {L{\"o}ffler}, {Marchal}, {Marrese}, {Moitinho}, {Muinonen}, {Osborne},
  {Pancino}, {Pauwels}, {Recio-Blanco}, {Reyl{\'e}}, {Riello}, {Rimoldini},
  {Roegiers}, {Rybizki}, {Sarro}, {Siopis}, {Smith}, {Sozzetti}, {Utrilla},
  {van Leeuwen}, {Abbas}, {{\'A}brah{\'a}m}, {Abreu Aramburu}, {Aerts},
  {Aguado}, {Ajaj}, {Aldea-Montero}, {Altavilla}, {{\'A}lvarez}, {Alves},
  {Anders}, {Anderson}, {Anglada Varela}, {Antoja}, {Baines}, {Baker},
  {Balaguer-N{\'u}{\~n}ez}, {Balbinot}, {Balog}, {Barache}, {Barbato},
  {Barros}, {Barstow}, {Bartolom{\'e}}, {Bassilana}, {Bauchet}, {Becciani},
  {Bellazzini}, {Berihuete}, {Bernet}, {Bertone}, {Bianchi}, {Binnenfeld},
  {Blanco-Cuaresma}, {Blazere}, {Boch}, {Bombrun}, {Bossini}, {Bouquillon},
  {Bragaglia}, {Bramante}, {Breedt}, {Bressan}, {Brouillet}, {Brugaletta},
  {Bucciarelli}, {Burlacu}, {Butkevich}, {Buzzi}, {Caffau}, {Cancelliere},
  {Cantat-Gaudin}, {Carballo}, {Carlucci}, {Carnerero}, {Carrasco},
  {Casamiquela}, {Castellani}, {Castro-Ginard}, {Chaoul}, {Charlot}, {Chemin},
  {Chiaramida}, {Chiavassa}, {Chornay}, {Comoretto}, {Contursi}, {Cooper},
  {Cornez}, {Cowell}, {Crifo}, {Cropper}, {Crosta}, {Crowley}, {Dafonte},
  {Dapergolas}, {David}, {David}, {de Laverny}, {De Luise}, {De March}, {De
  Ridder}, {de Souza}, {de Torres}, {del Peloso}, {del Pozo}, {Delbo},
  {Delgado}, {Delisle}, {Demouchy}, {Dharmawardena}, {Di Matteo}, {Diakite},
  {Diener}, {Distefano}, {Dolding}, {Edvardsson}, {Enke}, {Fabre}, {Fabrizio},
  {Faigler}, {Fedorets}, {Fernique}, {Fienga}, {Figueras}, {Fournier},
  {Fouron}, {Fragkoudi}, {Gai}, {Garcia-Gutierrez}, {Garcia-Reinaldos},
  {Garc{\'\i}a-Torres}, {Garofalo}, {Gavel}, {Gavras}, {Gerlach}, {Geyer},
  {Giacobbe}, {Gilmore}, {Girona}, {Giuffrida}, {Gomel}, {Gomez},
  {Gonz{\'a}lez-N{\'u}{\~n}ez}, {Gonz{\'a}lez-Santamar{\'\i}a},
  {Gonz{\'a}lez-Vidal}, {Granvik}, {Guillout}, {Guiraud},
  {Guti{\'e}rrez-S{\'a}nchez}, {Guy}, {Hatzidimitriou}, {Hauser}, {Haywood},
  {Helmer}, {Helmi}, {Sarmiento}, {Hidalgo}, {Hilger}, {H{\l}adczuk}, {Hobbs},
  {Holland}, {Huckle}, {Jardine}, {Jasniewicz}, {Jean-Antoine Piccolo},
  {Jim{\'e}nez-Arranz}, {Jorissen}, {Juaristi Campillo}, {Julbe}, {Karbevska},
  {Kervella}, {Khanna}, {Kontizas}, {Kordopatis}, {Korn}, {K{\'o}sp{\'a}l},
  {Kostrzewa-Rutkowska}, {Kruszy{\'n}ska}, {Kun}, {Laizeau}, {Lambert},
  {Lanza}, {Lasne}, {Le Campion}, {Lebreton}, {Lebzelter}, {Leccia}, {Leclerc},
  {Lecoeur-Taibi}, {Liao}, {Licata}, {Lindstr{\o}m}, {Lister}, {Livanou},
  {Lobel}, {Lorca}, {Loup}, {Madrero Pardo}, {Magdaleno Romeo}, {Managau},
  {Mann}, {Manteiga}, {Marchant}, {Marconi}, {Marcos}, {Marcos Santos},
  {Mar{\'\i}n Pina}, {Marinoni}, {Marocco}, {Marshall}, {Martin Polo},
  {Mart{\'\i}n-Fleitas}, {Marton}, {Mary}, {Masip}, {Massari},
  {Mastrobuono-Battisti}, {Mazeh}, {McMillan}, {Messina}, {Michalik}, {Millar},
  {Mints}, {Molina}, {Molinaro}, {Moln{\'a}r}, {Monari}, {Mongui{\'o}},
  {Montegriffo}, {Montero}, {Mor}, {Mora}, {Morbidelli}, {Morel}, {Morris},
  {Muraveva}, {Murphy}, {Musella}, {Nagy}, {Noval}, {Oca{\~n}a}, {Ogden},
  {Ordenovic}, {Osinde}, {Pagani}, {Pagano}, {Palaversa}, {Palicio},
  {Pallas-Quintela}, {Panahi}, {Payne-Wardenaar}, {Pe{\~n}alosa Esteller},
  {Penttil{\"a}}, {Pichon}, {Piersimoni}, {Pineau}, {Plachy}, {Plum}, {Poggio},
  {Pr{\v{s}}a}, {Pulone}, {Racero}, {Ragaini}, {Rainer}, {Raiteri}, {Rambaux},
  {Ramos}, {Ramos-Lerate}, {Re Fiorentin}, {Regibo}, {Richards}, {Rios Diaz},
  {Ripepi}, {Riva}, {Rix}, {Rixon}, {Robichon}, {Robin}, {Robin}, {Roelens},
  {Rogues}, {Rohrbasser}, {Romero-G{\'o}mez}, {Rowell}, {Royer}, {Ruz Mieres},
  {Rybicki}, {Sadowski}, {S{\'a}ez N{\'u}{\~n}ez}, {Sagrist{\`a} Sell{\'e}s},
  {Sahlmann}, {Salguero}, {Samaras}, {Sanchez Gimenez}, {Sanna},
  {Santove{\~n}a}, {Sarasso}, {Schultheis}, {Sciacca}, {Segol}, {Segovia},
  {S{\'e}gransan}, {Semeux}, {Shahaf}, {Siddiqui}, {Siebert}, {Siltala},
  {Silvelo}, {Slezak}, {Slezak}, {Smart}, {Snaith}, {Solano}, {Solitro},
  {Souami}, {Souchay}, {Spagna}, {Spina}, {Spoto}, {Steele},
  {Steidelm{\"u}ller}, {Stephenson}, {S{\"u}veges}, {Surdej}, {Szabados},
  {Szegedi-Elek}, {Taris}, {Taylor}, {Teixeira}, {Tolomei}, {Tonello}, {Torra},
  {Torra}, {Torralba Elipe}, {Trabucchi}, {Tsounis}, {Turon}, {Ulla}, {Unger},
  {Vaillant}, {van Dillen}, {van Reeven}, {Vanel}, {Vecchiato}, {Viala},
  {Vicente}, {Voutsinas}, {Weiler}, {Wevers}, {Wyrzykowski}, {Yoldas}, {Yvard},
  {Zhao}, {Zorec}, {Zucker}, \& {Zwitter}}]{Gaia2023}
{Gaia Collaboration}, {Vallenari}, A., {Brown}, A.~G.~A., {et~al.} 2023, \aap,
  674, A1

\bibitem[{{Garufi} {et~al.}(2017){Garufi}, {Meeus}, {Benisty}, {Quanz},
  {Banzatti}, {Kama}, {Canovas}, {Eiroa}, {Schmid}, {Stolker}, {Pohl},
  {Rigliaco}, {M{\'e}nard}, {Meyer}, {van Boekel}, \& {Dominik}}]{Garufi2017}
{Garufi}, A., {Meeus}, G., {Benisty}, M., {et~al.} 2017, \aap, 603, A21

\bibitem[{{Garufi} {et~al.}(2014){Garufi}, {Quanz}, {Schmid}, {Avenhaus},
  {Buenzli}, \& {Wolf}}]{Garufi2014}
{Garufi}, A., {Quanz}, S.~P., {Schmid}, H.~M., {et~al.} 2014, \aap, 568, A40

\bibitem[{{Grady} {et~al.}(2000){Grady}, {Devine}, {Woodgate}, {Kimble},
  {Bruhweiler}, {Boggess}, {Linsky}, {Plait}, {Clampin}, \&
  {Kalas}}]{Grady2000}
{Grady}, C.~A., {Devine}, D., {Woodgate}, B., {et~al.} 2000, \apj, 544, 895

\bibitem[{{Guerra-Alvarado} {et~al.}(2024){Guerra-Alvarado},
  {Carrasco-Gonz{\'a}lez}, {Mac{\'\i}as}, {van der Marel}, {Houge}, {Maud},
  {Pinilla}, {Villenave}, {Asaki}, \& {Humphreys}}]{Guerra-Alvarado2024}
{Guerra-Alvarado}, O.~M., {Carrasco-Gonz{\'a}lez}, C., {Mac{\'\i}as}, E.,
  {et~al.} 2024, \aap, 686, A298

\bibitem[{{Guidi} {et~al.}(2022){Guidi}, {Isella}, {Testi}, {Chandler}, {Liu},
  {Schmid}, {Rosotti}, {Meng}, {Jennings}, {Williams}, {Carpenter}, {de
  Gregorio-Monsalvo}, {Li}, {Liu}, {Ortolani}, {Quanz}, {Ricci}, \&
  {Tazzari}}]{Guidi2022}
{Guidi}, G., {Isella}, A., {Testi}, L., {et~al.} 2022, \aap, 664, A137

\bibitem[{{Guidi} {et~al.}(2026){Guidi}, {Menard}, {Price}, {Villenave}, \&
  {Ma}}]{Guidi2026}
{Guidi}, G., {Menard}, F., {Price}, D.~J., {Villenave}, M., \& {Ma}, J. 2026,
  Frontiers in Astronomy and Space Sciences, 13, 1727532

\bibitem[{{Guidi} {et~al.}(2018){Guidi}, {Ruane}, {Williams}, {Mawet}, {Testi},
  {Zurlo}, {Absil}, {Bottom}, {Choquet}, {Christiaens}, {Femen{\'\i}a
  Castell{\'a}}, {Huby}, {Isella}, {Kastner}, {Meshkat}, {Reggiani}, {Riggs},
  {Serabyn}, \& {Wallack}}]{Guidi2018}
{Guidi}, G., {Ruane}, G., {Williams}, J.~P., {et~al.} 2018, \mnras, 479, 1505

\bibitem[{{Guidi} {et~al.}(2016){Guidi}, {Tazzari}, {Testi}, {de
  Gregorio-Monsalvo}, {Chandler}, {P{\'e}rez}, {Isella}, {Natta}, {Ortolani},
  {Henning}, {Corder}, {Linz}, {Andrews}, {Wilner}, {Ricci}, {Carpenter},
  {Sargent}, {Mundy}, {Storm}, {Calvet}, {Dullemond}, {Greaves}, {Lazio},
  {Deller}, \& {Kwon}}]{Guidi2016}
{Guidi}, G., {Tazzari}, M., {Testi}, L., {et~al.} 2016, \aap, 588, A112

\bibitem[{{Hayashi}(1981)}]{Hayashi1981}
{Hayashi}, C. 1981, Progress of Theoretical Physics Supplement, 70, 35

\bibitem[{{Henning} \& {Stognienko}(1996)}]{Henning1996}
{Henning}, T. \& {Stognienko}, R. 1996, \aap, 311, 291

\bibitem[{{Henyey} \& {Greenstein}(1941)}]{Henyey1941}
{Henyey}, L.~G. \& {Greenstein}, J.~L. 1941, \apj, 93, 70

\bibitem[{{Huang} {et~al.}(2018){Huang}, {Andrews}, {Dullemond}, {Isella},
  {P{\'e}rez}, {Guzm{\'a}n}, {{\"O}berg}, {Zhu}, {Zhang}, {Bai}, {Benisty},
  {Birnstiel}, {Carpenter}, {Hughes}, {Ricci}, {Weaver}, \&
  {Wilner}}]{Huang2018_ring}
{Huang}, J., {Andrews}, S.~M., {Dullemond}, C.~P., {et~al.} 2018, \apjl, 869,
  L42

\bibitem[{{Isella} {et~al.}(2016){Isella}, {Guidi}, {Testi}, {Liu}, {Li}, {Li},
  {Weaver}, {Boehler}, {Carperter}, {De Gregorio-Monsalvo}, {Manara}, {Natta},
  {P{\'e}rez}, {Ricci}, {Sargent}, {Tazzari}, \& {Turner}}]{Isella2016}
{Isella}, A., {Guidi}, G., {Testi}, L., {et~al.} 2016, \prl, 117, 251101

\bibitem[{{Isella} {et~al.}(2018){Isella}, {Huang}, {Andrews}, {Dullemond},
  {Birnstiel}, {Zhang}, {Zhu}, {Guzm{\'a}n}, {P{\'e}rez}, {Bai}, {Benisty},
  {Carpenter}, {Ricci}, \& {Wilner}}]{Isella2018}
{Isella}, A., {Huang}, J., {Andrews}, S.~M., {et~al.} 2018, \apjl, 869, L49

\bibitem[{{Izquierdo} {et~al.}(2026){Izquierdo}, {Bae}, {Galloway-Sprietsma},
  {van Dishoeck}, {Facchini}, {Rosotti}, {Stadler}, {Benisty}, \&
  {Testi}}]{Izquierdo2026}
{Izquierdo}, A.~F., {Bae}, J., {Galloway-Sprietsma}, M., {et~al.} 2026, \apjl,
  997, L2

\bibitem[{{Izquierdo} {et~al.}(2022){Izquierdo}, {Facchini}, {Rosotti}, {van
  Dishoeck}, \& {Testi}}]{Izquierdo2022}
{Izquierdo}, A.~F., {Facchini}, S., {Rosotti}, G.~P., {van Dishoeck}, E.~F., \&
  {Testi}, L. 2022, \apj, 928, 2

\bibitem[{{Izquierdo} {et~al.}(2023){Izquierdo}, {Testi}, {Facchini},
  {Rosotti}, {van Dishoeck}, {W{\"o}lfer}, \&
  {Paneque-Carre{\~n}o}}]{Izquierdo2023}
{Izquierdo}, A.~F., {Testi}, L., {Facchini}, S., {et~al.} 2023, \aap, 674, A113

\bibitem[{{Jennings} {et~al.}(2020){Jennings}, {Booth}, {Tazzari}, {Rosotti},
  \& {Clarke}}]{Jennings2020}
{Jennings}, J., {Booth}, R.~A., {Tazzari}, M., {Rosotti}, G.~P., \& {Clarke},
  C.~J. 2020, \mnras, 495, 3209

\bibitem[{{Jiang} {et~al.}(2024){Jiang}, {Mac{\'\i}as}, {Guerra-Alvarado}, \&
  {Carrasco-Gonz{\'a}lez}}]{Jiang2024}
{Jiang}, H., {Mac{\'\i}as}, E., {Guerra-Alvarado}, O.~M., \&
  {Carrasco-Gonz{\'a}lez}, C. 2024, \aap, 682, A32

\bibitem[{{Johansen} {et~al.}(2007){Johansen}, {Oishi}, {Mac Low}, {Klahr},
  {Henning}, \& {Youdin}}]{Johansen2007}
{Johansen}, A., {Oishi}, J.~S., {Mac Low}, M.-M., {et~al.} 2007, \nat, 448,
  1022

\bibitem[{{Kadler} {et~al.}(2023){Kadler}, {Riechers}, {Agarwal}, {Baczko},
  {Beuther}, {Bigiel}, {Birnstiel}, {Boccardi}, {Bomans}, {Boogaard}, {Braun},
  {Britzen}, {Br{\"u}ggen}, {Brunthaler}, {Caselli}, {Els{\"a}sser}, {von
  Fellenberg}, {Flock}, {Fromm}, {Fuhrmann}, {Hartogh}, {Hoeft}, {Keenan},
  {Kovalev}, {Kreckel}, {Livingston}, {Lobanov}, {M{\"u}ller}, {Ros},
  {Schilke}, {De Simone}, {Spitler}, {Ueda}, {Vardoulaki}, {Vegetti}, {Weis},
  {Wendel}, {Xu}, {Zhao}, {Albrecht}, {Basu}, {Becker Tjus}, {Bernhart},
  {Blum}, {Bonnassieux}, {Bredendiek}, {van Delden}, {Di Gennaro}, {Enders},
  {Eppel}, {Hase}, {Hoang}, {Hugentobler}, {Kaasinen}, {Krupp}, {Kun},
  {Laubach}, {Lin}, {Mannheim}, {Menten}, {Perkuhn}, {Pohl}, {Powell},
  {Rezzolla}, {Ricci}, {Schinnerer}, {Schmidt}, {Sch{\"o}pfel}, {Stanko},
  {Stein}, {Sulzenauer}, {Taziaux}, {Tursunov}, {Walter}, {Weiss}, {Witzel},
  {Wolf}, {Zensus}, {Mus}, {Toth}, {Alberdi}, {Benisty}, {Cox}, {Guirado},
  {Johnson}, {Juvela}, {Neeleman}, {Pashchenko}, {P{\'e}rez Torres}, {Perraut},
  \& {Zajacek}}]{Kadler2023}
{Kadler}, M., {Riechers}, D.~A., {Agarwal}, J., {et~al.} 2023, arXiv e-prints,
  arXiv:2311.10056

\bibitem[{{Kataoka} {et~al.}(2015){Kataoka}, {Muto}, {Momose}, {Tsukagoshi},
  {Fukagawa}, {Shibai}, {Hanawa}, {Murakawa}, \& {Dullemond}}]{Kataoka2015}
{Kataoka}, A., {Muto}, T., {Momose}, M., {et~al.} 2015, \apj, 809, 78

\bibitem[{{Kataoka} {et~al.}(2014){Kataoka}, {Okuzumi}, {Tanaka}, \&
  {Nomura}}]{Kataoka2014}
{Kataoka}, A., {Okuzumi}, S., {Tanaka}, H., \& {Nomura}, H. 2014, \aap, 568,
  A42

\bibitem[{{Kim} {et~al.}(2019){Kim}, {Nomura}, {Tsukagoshi}, {Kawabe}, \&
  {Muto}}]{Kim2019}
{Kim}, S., {Nomura}, H., {Tsukagoshi}, T., {Kawabe}, R., \& {Muto}, T. 2019,
  \apj, 872, 179

\bibitem[{{Li} \& {Youdin}(2021)}]{Li2021}
{Li}, R. \& {Youdin}, A.~N. 2021, \apj, 919, 107

\bibitem[{{Lim} {et~al.}(2024){Lim}, {Simon}, {Li}, {Armitage}, {Carrera},
  {Lyra}, {Rea}, {Yang}, \& {Youdin}}]{Lim2024}
{Lim}, J., {Simon}, J.~B., {Li}, R., {et~al.} 2024, \apj, 969, 130

\bibitem[{{Lin} {et~al.}(2020){Lin}, {Li}, {Yang}, {Looney}, {Stephens}, \&
  {Hull}}]{Lin2020}
{Lin}, Z.-Y.~D., {Li}, Z.-Y., {Yang}, H., {et~al.} 2020, \mnras, 496, 169

\bibitem[{{Lin} {et~al.}(2023){Lin}, {Li}, {Yang}, {Mu{\~n}oz}, {Looney},
  {Stephens}, {Hull}, {Fern{\'a}ndez-L{\'o}pez}, \& {Harrison}}]{Lin2023pol}
{Lin}, Z.-Y.~D., {Li}, Z.-Y., {Yang}, H., {et~al.} 2023, \mnras, 520, 1210

\bibitem[{{Lin} {et~al.}(2025){Lin}, {Weinberger}, {Zubko}, {Arnold}, \&
  {Videen}}]{Lin2025}
{Lin}, Z.-Y.~D., {Weinberger}, A.~J., {Zubko}, E., {Arnold}, J.~A., \&
  {Videen}, G. 2025, \pasp, 137, 124502

\bibitem[{{Liu} {et~al.}(2018){Liu}, {Jin}, {Li}, {Isella}, \& {Li}}]{Liu2018}
{Liu}, S.-F., {Jin}, S., {Li}, S., {Isella}, A., \& {Li}, H. 2018, \apj, 857,
  87

\bibitem[{{Liu} {et~al.}(2022){Liu}, {Bertrang}, {Flock}, {Rosotti}, {van
  Dishoeck}, {Boehler}, {Facchini}, {Cui}, {Wolf}, \& {Fang}}]{Liu_Yao2022}
{Liu}, Y., {Bertrang}, G. H.-M., {Flock}, M., {et~al.} 2022, Science China
  Physics, Mechanics, and Astronomy, 65, 129511

\bibitem[{{Long} {et~al.}(2018){Long}, {Pinilla}, {Herczeg}, {Harsono},
  {Dipierro}, {Pascucci}, {Hendler}, {Tazzari}, {Ragusa}, {Salyk}, {Edwards},
  {Lodato}, {van de Plas}, {Johnstone}, {Liu}, {Boehler}, {Cabrit}, {Manara},
  {Menard}, {Mulders}, {Nisini}, {Fischer}, {Rigliaco}, {Banzatti}, {Avenhaus},
  \& {Gully-Santiago}}]{Long2018}
{Long}, F., {Pinilla}, P., {Herczeg}, G.~J., {et~al.} 2018, \apj, 869, 17

\bibitem[{{Loomis} {et~al.}(2025){Loomis}, {Facchini}, {Benisty}, {Curone},
  {Ilee}, {Cataldi}, {Yen}, {Teague}, {Pinte}, {Huang}, \& et~al.}]{Loomis2025}
{Loomis}, R.~A., {Facchini}, S., {Benisty}, M., {et~al.} 2025, \apjl, 984, L7

\bibitem[{{Mac{\'\i}as} {et~al.}(2021){Mac{\'\i}as}, {Guerra-Alvarado},
  {Carrasco-Gonz{\'a}lez}, {Ribas}, {Espaillat}, {Huang}, \&
  {Andrews}}]{Macias2021}
{Mac{\'\i}as}, E., {Guerra-Alvarado}, O., {Carrasco-Gonz{\'a}lez}, C., {et~al.}
  2021, \aap, 648, A33

\bibitem[{{Martire} {et~al.}(2024){Martire}, {Longarini}, {Lodato}, {Rosotti},
  {Winter}, {Facchini}, {Hardiman}, {Benisty}, {Stadler}, {Izquierdo}, \&
  {Testi}}]{Martire2024}
{Martire}, P., {Longarini}, C., {Lodato}, G., {et~al.} 2024, \aap, 686, A9

\bibitem[{{Mathis} {et~al.}(1977){Mathis}, {Rumpl}, \&
  {Nordsieck}}]{Mathis1977}
{Mathis}, J.~S., {Rumpl}, W., \& {Nordsieck}, K.~H. 1977, \apj, 217, 425

\bibitem[{{Mesa} {et~al.}(2019){Mesa}, {Keppler}, {Cantalloube}, {Rodet},
  {Charnay}, {Gratton}, {Langlois}, {Boccaletti}, {Bonnefoy}, {Vigan},
  {Flasseur}, {Bae}, {Benisty}, {Chauvin}, {de Boer}, {Desidera}, {Henning},
  {Lagrange}, {Meyer}, {Milli}, {M{\"u}ller}, {Pairet}, {Zurlo}, {Antoniucci},
  {Baudino}, {Brown Sevilla}, {Cascone}, {Cheetham}, {Claudi}, {Delorme},
  {D'Orazi}, {Feldt}, {Hagelberg}, {Janson}, {Kral}, {Lagadec}, {Lazzoni},
  {Ligi}, {Maire}, {Martinez}, {Menard}, {Meunier}, {Perrot}, {Petrus},
  {Pinte}, {Rickman}, {Rochat}, {Rouan}, {Samland}, {Sauvage}, {Schmidt},
  {Udry}, {Weber}, \& {Wildi}}]{Mesa2019}
{Mesa}, D., {Keppler}, M., {Cantalloube}, F., {et~al.} 2019, \aap, 632, A25

\bibitem[{{Mie}(1908)}]{Mie1908}
{Mie}, G. 1908, Annalen der Physik, 330, 377

\bibitem[{{Min} {et~al.}(2016){Min}, {Rab}, {Woitke}, {Dominik}, \&
  {M{\'e}nard}}]{Min2016}
{Min}, M., {Rab}, C., {Woitke}, P., {Dominik}, C., \& {M{\'e}nard}, F. 2016,
  \aap, 585, A13

\bibitem[{{Miyake} \& {Nakagawa}(1993)}]{Miyake1993}
{Miyake}, K. \& {Nakagawa}, Y. 1993, \icarus, 106, 20

\bibitem[{{Monnier} {et~al.}(2017){Monnier}, {Harries}, {Aarnio}, {Adams},
  {Andrews}, {Calvet}, {Espaillat}, {Hartmann}, {Hinkley}, {Kraus}, {McClure},
  {Oppenheimer}, {Perrin}, \& {Wilner}}]{Monnier2017}
{Monnier}, J.~D., {Harries}, T.~J., {Aarnio}, A., {et~al.} 2017, \apj, 838, 20

\bibitem[{{Mullin} {et~al.}(2026){Mullin}, {Lucas}, {Dong}, {Hashimoto},
  {Jiang}, {Johnstone}, {Lawson}, {Brittain}, {Guyon}, {Kudo}, {Lozi},
  {Najita}, {Sun}, {Tamura}, \& {Wagner}}]{Mullin2026}
{Mullin}, C., {Lucas}, M., {Dong}, R., {et~al.} 2026, \aj, 171, 241

\bibitem[{{Muro-Arena} {et~al.}(2018){Muro-Arena}, {Dominik}, {Waters}, {Min},
  {Klarmann}, {Ginski}, {Isella}, {Benisty}, {Pohl}, {Garufi}, {Hagelberg},
  {Langlois}, {Menard}, {Pinte}, {Sezestre}, {van der Plas}, {Villenave},
  {Delboulb{\'e}}, {Magnard}, {M{\"o}ller-Nilsson}, {Pragt}, {Rabou}, \&
  {Roelfsema}}]{Muro-Arena2018}
{Muro-Arena}, G.~A., {Dominik}, C., {Waters}, L.~B.~F.~M., {et~al.} 2018, \aap,
  614, A24

\bibitem[{{Murphy} {et~al.}(2018){Murphy}, {Bolatto}, {Chatterjee}, {Casey},
  {Chomiuk}, {Dale}, {de Pater}, {Dickinson}, {Francesco}, {Hallinan},
  {Isella}, {Kohno}, {Kulkarni}, {Lang}, {Lazio}, {Leroy}, {Loinard},
  {Maccarone}, {Matthews}, {Osten}, {Reid}, {Riechers}, {Sakai}, {Walter}, \&
  {Wilner}}]{Murphy2018}
{Murphy}, E.~J., {Bolatto}, A., {Chatterjee}, S., {et~al.} 2018, in
  Astronomical Society of the Pacific Conference Series, Vol. 517, Science with
  a Next Generation Very Large Array, ed. E.~{Murphy}, 3

\bibitem[{{Ohashi} \& {Kataoka}(2019)}]{Ohashi2019}
{Ohashi}, S. \& {Kataoka}, A. 2019, \apj, 886, 103

\bibitem[{{Ormel} \& {Cuzzi}(2007)}]{Ormel2007}
{Ormel}, C.~W. \& {Cuzzi}, J.~N. 2007, \aap, 466, 413

\bibitem[{{Pezzotta} {et~al.}(2025){Pezzotta}, {Facchini}, {Longarini},
  {Lodato}, \& {Martire}}]{Pezzotta2025}
{Pezzotta}, V., {Facchini}, S., {Longarini}, C., {Lodato}, G., \& {Martire}, P.
  2025, \aap, 694, A108

\bibitem[{{Pinilla} {et~al.}(2012){Pinilla}, {Benisty}, \&
  {Birnstiel}}]{Pinilla2012a}
{Pinilla}, P., {Benisty}, M., \& {Birnstiel}, T. 2012, \aap, 545, A81

\bibitem[{{Pinte} {et~al.}(2016){Pinte}, {Dent}, {M{\'e}nard}, {Hales}, {Hill},
  {Cortes}, \& {de Gregorio-Monsalvo}}]{Pinte2016}
{Pinte}, C., {Dent}, W.~R.~F., {M{\'e}nard}, F., {et~al.} 2016, \apj, 816, 25

\bibitem[{{Pinte} {et~al.}(2018){Pinte}, {Price}, {M{\'e}nard}, {Duch{\^e}ne},
  {Dent}, {Hill}, {de Gregorio-Monsalvo}, {Hales}, \&
  {Mentiplay}}]{Pinte2018_planet}
{Pinte}, C., {Price}, D.~J., {M{\'e}nard}, F., {et~al.} 2018, \apjl, 860, L13

\bibitem[{{Planck}(1901)}]{Planck1901}
{Planck}, M. 1901, Annalen der Physik, 309, 553

\bibitem[{{Pollack} {et~al.}(1996){Pollack}, {Hubickyj}, {Bodenheimer},
  {Lissauer}, {Podolak}, \& {Greenzweig}}]{Pollack1996}
{Pollack}, J.~B., {Hubickyj}, O., {Bodenheimer}, P., {et~al.} 1996, \icarus,
  124, 62

\bibitem[{Privon {et~al.}(2025)Privon, Cerrigone, Corvillon, Kawamura, Popping,
  \& Rebolledo}]{Privon2025}
Privon, G., Cerrigone, L., Corvillon, A., {et~al.} 2025, ALMA Cycle 12
  Proposer's Guide

\bibitem[{{Rau} \& {Cornwell}(2011)}]{Rau2011}
{Rau}, U. \& {Cornwell}, T.~J. 2011, \aap, 532, A71

\bibitem[{{Ren} {et~al.}(2023){Ren}, {Benisty}, {Ginski}, {Tazaki}, {Wallack},
  {Milli}, {Garufi}, {Bae}, {Facchini}, {M{\'e}nard}, {Pinilla}, {Swastik},
  {Teague}, \& {Wahhaj}}]{Ren2023}
{Ren}, B.~B., {Benisty}, M., {Ginski}, C., {et~al.} 2023, \aap, 680, A114

\bibitem[{{Ribas} {et~al.}(2024){Ribas}, {Clarke}, \& {Zagaria}}]{Ribas2024}
{Ribas}, {\'A}., {Clarke}, C.~J., \& {Zagaria}, F. 2024, \mnras, 532, 1752

\bibitem[{{Rich} {et~al.}(2019){Rich}, {Wisniewski}, {Currie}, {Fukagawa},
  {Grady}, {Sitko}, {Pikhartova}, {Hashimoto}, {Abe}, {Brandner}, {Brandt},
  {Carson}, {Chilcote}, {Dong}, {Feldt}, {Goto}, {Groff}, {Guyon}, {Hayano},
  {Hayashi}, {Hayashi}, {Henning}, {Hodapp}, {Ishii}, {Iye}, {Janson},
  {Jovanovic}, {Kandori}, {Kasdin}, {Knapp}, {Kudo}, {Kusakabe}, {Kuzuhara},
  {Kwon}, {Lozi}, {Martinache}, {Matsuo}, {Mayama}, {McElwain}, {Miyama},
  {Morino}, {Moro-Martin}, {Nakagawa}, {Nishimura}, {Pyo}, {Serabyn}, {Suto},
  {Russel}, {Suzuki}, {Takami}, {Takato}, {Terada}, {Thalmann}, {Turner},
  {Uyama}, {Wagner}, {Watanabe}, {Yamada}, {Takami}, {Usuda}, \&
  {Tamura}}]{Rich2019}
{Rich}, E.~A., {Wisniewski}, J.~P., {Currie}, T., {et~al.} 2019, \apj, 875, 38

\bibitem[{{Rosotti} {et~al.}(2020){Rosotti}, {Teague}, {Dullemond}, {Booth}, \&
  {Clarke}}]{Rosotti2020}
{Rosotti}, G.~P., {Teague}, R., {Dullemond}, C., {Booth}, R.~A., \& {Clarke},
  C.~J. 2020, \mnras, 495, 173

\bibitem[{{Rybicki} \& {Lightman}(1979)}]{Rybicki1979}
{Rybicki}, G.~B. \& {Lightman}, A.~P. 1979, {Radiative processes in
  astrophysics}

\bibitem[{{Safronov}(1972)}]{Safronov1972}
{Safronov}, V.~S. 1972, {Evolution of the protoplanetary cloud and formation of
  the earth and planets.}

\bibitem[{{Sekiya}(1983)}]{Sekiya1983}
{Sekiya}, M. 1983, Progress of Theoretical Physics, 69, 1116

\bibitem[{{Shi} {et~al.}(2026){Shi}, {Long}, {Mac{\'\i}as}, {Herczeg},
  {Pinilla}, {Andrews}, {Wilner}, {Jiang}, {Dong}, {Teague}, {Pascucci},
  {Toci}, {Aikawa}, {Harsono}, \& {Liu}}]{Shi2026}
{Shi}, Y., {Long}, F., {Mac{\'\i}as}, E., {et~al.} 2026, \apj, 1001, 27

\bibitem[{{Sierra} {et~al.}(2021){Sierra}, {P{\'e}rez}, {Zhang}, {Law},
  {Guzm{\'a}n}, {Qi}, {Bosman}, {{\"O}berg}, {Andrews}, {Long}, {Teague},
  {Booth}, {Walsh}, {Wilner}, {M{\'e}nard}, {Cataldi}, {Czekala}, {Bae},
  {Huang}, {Bergner}, {Ilee}, {Benisty}, {Le Gal}, {Loomis}, {Tsukagoshi},
  {Liu}, {Yamato}, \& {Aikawa}}]{Sierra2021}
{Sierra}, A., {P{\'e}rez}, L.~M., {Zhang}, K., {et~al.} 2021, \apjs, 257, 14

\bibitem[{{Stephens} {et~al.}(2017){Stephens}, {Yang}, {Li}, {Looney},
  {Kataoka}, {Kwon}, {Fern{\'a}ndez-L{\'o}pez}, {Hull}, {Hughes}, {Segura-Cox},
  {Mundy}, {Crutcher}, \& {Rao}}]{Stephens2017}
{Stephens}, I.~W., {Yang}, H., {Li}, Z.-Y., {et~al.} 2017, \apj, 851, 55

\bibitem[{{Tanaka} {et~al.}(1996){Tanaka}, {Inaba}, \& {Nakazawa}}]{Tanaka1996}
{Tanaka}, H., {Inaba}, S., \& {Nakazawa}, K. 1996, \icarus, 123, 450

\bibitem[{{Tazaki} {et~al.}(2019){Tazaki}, {Tanaka}, {Kataoka}, {Okuzumi}, \&
  {Muto}}]{Tazaki2019}
{Tazaki}, R., {Tanaka}, H., {Kataoka}, A., {Okuzumi}, S., \& {Muto}, T. 2019,
  \apj, 885, 52

\bibitem[{{Teague} {et~al.}(2019){Teague}, {Bae}, \& {Bergin}}]{Teague2019_Nat}
{Teague}, R., {Bae}, J., \& {Bergin}, E.~A. 2019, \nat, 574, 378

\bibitem[{{Teague} {et~al.}(2018){Teague}, {Bae}, {Bergin}, {Birnstiel}, \&
  {Foreman-Mackey}}]{Teague2018_HD163296}
{Teague}, R., {Bae}, J., {Bergin}, E.~A., {Birnstiel}, T., \& {Foreman-Mackey},
  D. 2018, \apjl, 860, L12

\bibitem[{{Testi} {et~al.}(2014){Testi}, {Birnstiel}, {Ricci}, {Andrews},
  {Blum}, {Carpenter}, {Dominik}, {Isella}, {Natta}, {Williams}, \&
  {Wilner}}]{Testi2014review}
{Testi}, L., {Birnstiel}, T., {Ricci}, L., {et~al.} 2014, in Protostars and
  Planets VI, ed. H.~{Beuther}, R.~S. {Klessen}, C.~P. {Dullemond}, \&
  T.~{Henning}, 339

\bibitem[{{Toomre}(1964)}]{Toomre1964}
{Toomre}, A. 1964, \apj, 139, 1217

\bibitem[{{Turner} {et~al.}(2014){Turner}, {Benisty}, {Dullemond}, \&
  {Hirose}}]{Turner2014}
{Turner}, N.~J., {Benisty}, M., {Dullemond}, C.~P., \& {Hirose}, S. 2014, \apj,
  780, 42

\bibitem[{{Ueda} {et~al.}(2025){Ueda}, {Andrews}, {Carrasco-Gonz{\'a}lez},
  {Guerra-Alvarado}, {Okuzumi}, {Tazaki}, \& {Kataoka}}]{Ueda2025}
{Ueda}, T., {Andrews}, S.~M., {Carrasco-Gonz{\'a}lez}, C., {et~al.} 2025, \apj,
  990, 183

\bibitem[{{Ueda} {et~al.}(2019){Ueda}, {Flock}, \& {Okuzumi}}]{Ueda2019}
{Ueda}, T., {Flock}, M., \& {Okuzumi}, S. 2019, \apj, 871, 10

\bibitem[{{Ueda} {et~al.}(2022){Ueda}, {Kataoka}, \& {Tsukagoshi}}]{Ueda2022}
{Ueda}, T., {Kataoka}, A., \& {Tsukagoshi}, T. 2022, \apj, 930, 56

\bibitem[{{Uyama} {et~al.}(2025){Uyama}, {Ricci}, {Ygouf}, {Andrews},
  {Gallagher}, {Huang}, {Isella}, {Mawet}, {P{\'e}rez}, {Robberto}, {Ruane},
  {Zhang}, \& {Zhu}}]{Uyama2025}
{Uyama}, T., {Ricci}, L., {Ygouf}, M., {et~al.} 2025, \aj, 169, 287

\bibitem[{{Varga} {et~al.}(2021){Varga}, {Hogerheijde}, {van Boekel},
  {Klarmann}, {Petrov}, {Waters}, {Lagarde}, {Pantin}, {Berio}, {Weigelt},
  {Robbe-Dubois}, {Lopez}, {Millour}, {Augereau}, {Meheut}, {Meilland},
  {Henning}, {Jaffe}, {Bettonvil}, {Bristow}, {Hofmann}, {Matter}, {Zins},
  {Wolf}, {Allouche}, {Donnan}, {Schertl}, {Dominik}, {Heininger}, {Lehmitz},
  {Cruzal{\`e}bes}, {Glindemann}, {Meisenheimer}, {Paladini}, {Sch{\"o}ller},
  {Woillez}, {Venema}, {Kokoulina}, {Yoffe}, {{\'A}brah{\'a}m}, {Abadie},
  {Abuter}, {Accardo}, {Adler}, {Ag{\'o}cs}, {Antonelli}, {B{\"o}hm}, {Bailet},
  {Bazin}, {Beckmann}, {Beltran}, {Boland}, {Bourget}, {Brast}, {Bresson},
  {Burtscher}, {Castillo}, {Chelli}, {Cid}, {Clausse}, {Connot}, {Conzelmann},
  {Danchi}, {De Haan}, {Delbo}, {Ebert}, {Elswijk}, {Fantei}, {Frahm},
  {G{\'a}mez Rosas}, {Gabasch}, {Gallenne}, {Garces}, {Girard}, {Gont{\'e}},
  {Gonz{\'a}lez Herrera}, {Graser}, {Guajardo}, {Guitton}, {Haubois}, {Hron},
  {Hubin}, {Huerta}, {Isbell}, {Ives}, {Jakob}, {Jask{\'o}}, {Jochum}, {Klein},
  {Kragt}, {Kroes}, {Kuindersma}, {Labadie}, {Laun}, {Le Poole}, {Leinert},
  {Lizon}, {Lopez}, {M{\'e}rand}, {Marcotto}, {Mauclert}, {Maurer}, {Mehrgan},
  {Meisner}, {Meixner}, {Mellein}, {Mohr}, {Morel}, {Mosoni}, {Navarro},
  {Neumann}, {Nu{\ss}baum}, {Pallanca}, {Pasquini}, {Percheron}, {Pott},
  {Pozna}, {Ridinger}, {Rigal}, {Riquelme}, {Rivinius}, {Roelfsema}, {Rohloff},
  {Rousseau}, {Schuhler}, {Schuil}, {Soulain}, {Stee}, {Stephan}, {ter Horst},
  {Tromp}, {Vakili}, {van Duin}, {Vinther}, {Wittkowski}, \&
  {Wrhel}}]{Varga2021}
{Varga}, J., {Hogerheijde}, M., {van Boekel}, R., {et~al.} 2021, \aap, 647, A56

\bibitem[{{Villenave}(2025)}]{Villenave2025review}
{Villenave}, M. 2025, \pasp, 137, 103001

\bibitem[{{Wada} {et~al.}(2009){Wada}, {Tanaka}, {Suyama}, {Kimura}, \&
  {Yamamoto}}]{Wada2009}
{Wada}, K., {Tanaka}, H., {Suyama}, T., {Kimura}, H., \& {Yamamoto}, T. 2009,
  \apj, 702, 1490

\bibitem[{{Warren} \& {Brandt}(2008)}]{Warren2008}
{Warren}, S.~G. \& {Brandt}, R.~E. 2008, Journal of Geophysical Research
  (Atmospheres), 113, D14220

\bibitem[{{Weidenschilling}(1977)}]{Weidenschilling1977mnras}
{Weidenschilling}, S.~J. 1977, \mnras, 180, 57

\bibitem[{{Whipple}(1972)}]{Whipple1972}
{Whipple}, F.~L. 1972, in From Plasma to Planet, ed. A.~{Elvius}, 211

\bibitem[{{Wisniewski} {et~al.}(2008){Wisniewski}, {Clampin}, {Grady},
  {Ardila}, {Ford}, {Golimowski}, {Illingworth}, \& {Krist}}]{Wisniewski2008}
{Wisniewski}, J.~P., {Clampin}, M., {Grady}, C.~A., {et~al.} 2008, \apj, 682,
  548

\bibitem[{{Yang} {et~al.}(2025){Yang}, {Li}, {Dong}, {Doi}, {Liu}, \&
  {Huang}}]{Yang2025}
{Yang}, L., {Li}, Y.-P., {Dong}, R., {et~al.} 2025, \apj, 989, 176

\bibitem[{{Yoshida} {et~al.}(2025){Yoshida}, {Nomura}, {Tsukagoshi}, {Doi},
  {Furuya}, \& {Kataoka}}]{Yoshida2025albedo}
{Yoshida}, T.~C., {Nomura}, H., {Tsukagoshi}, T., {et~al.} 2025, \apj, 980, 50

\bibitem[{{Youdin} \& {Goodman}(2005)}]{Youdin2005}
{Youdin}, A.~N. \& {Goodman}, J. 2005, \apj, 620, 459

\bibitem[{{Youdin} \& {Lithwick}(2007)}]{Youdin2007}
{Youdin}, A.~N. \& {Lithwick}, Y. 2007, \icarus, 192, 588

\bibitem[{{Youdin} \& {Shu}(2002)}]{Youdin2002}
{Youdin}, A.~N. \& {Shu}, F.~H. 2002, \apj, 580, 494

\bibitem[{{Zagaria} {et~al.}(2025){Zagaria}, {Facchini}, {Curone}, {Williams},
  {Clarke}, {Ribas}, {Tazzari}, {Mac{\'\i}as}, {Booth}, {Rosotti}, \&
  {Testi}}]{Zagaria2025}
{Zagaria}, F., {Facchini}, S., {Curone}, P., {et~al.} 2025, \aap, 702, A56

\bibitem[{{Zhang} {et~al.}(2021){Zhang}, {Booth}, {Law}, {Bosman}, {Schwarz},
  {Bergin}, {{\"O}berg}, {Andrews}, {Guzm{\'a}n}, {Walsh}, {Qi}, {van't Hoff},
  {Long}, {Wilner}, {Huang}, {Czekala}, {Ilee}, {Cataldi}, {Bergner}, {Aikawa},
  {Teague}, {Bae}, {Loomis}, {Calahan}, {Alarc{\'o}n}, {M{\'e}nard}, {Le Gal},
  {Sierra}, {Yamato}, {Nomura}, {Tsukagoshi}, {P{\'e}rez}, {Trapman}, {Liu}, \&
  {Furuya}}]{Zhang2021}
{Zhang}, K., {Booth}, A.~S., {Law}, C.~J., {et~al.} 2021, \apjs, 257, 5

\bibitem[{{Zhang} {et~al.}(2023){Zhang}, {Zhu}, {Ueda}, {Kataoka}, {Sierra},
  {Carrasco-Gonz{\'a}lez}, \& {Mac{\'\i}as}}]{Zhang2023}
{Zhang}, S., {Zhu}, Z., {Ueda}, T., {et~al.} 2023, \apj, 953, 96

\bibitem[{{Zhu} {et~al.}(2012){Zhu}, {Nelson}, {Dong}, {Espaillat}, \&
  {Hartmann}}]{Zhu2012}
{Zhu}, Z., {Nelson}, R.~P., {Dong}, R., {Espaillat}, C., \& {Hartmann}, L.
  2012, \apj, 755, 6

\bibitem[{{Zsom} {et~al.}(2010){Zsom}, {Ormel}, {G{\"u}ttler}, {Blum}, \&
  {Dullemond}}]{Zsom2010}
{Zsom}, A., {Ormel}, C.~W., {G{\"u}ttler}, C., {Blum}, J., \& {Dullemond},
  C.~P. 2010, \aap, 513, A57

\bibitem[{{Zubko} {et~al.}(1996){Zubko}, {Mennella}, {Colangeli}, \&
  {Bussoletti}}]{Zubko1996}
{Zubko}, V.~G., {Mennella}, V., {Colangeli}, L., \& {Bussoletti}, E. 1996,
  \mnras, 282, 1321

\end{thebibliography}

\begin{appendix}

\section{Details of the observational data} \label{app:ALMA_obs}
Table \ref{tab:observations} summarizes the ALMA project codes, array configurations, on-source times, and spectral setups used in this work.

\begin{sidewaystable*}
\caption{Summary of the ALMA observations used in this work.}
\centering
\begin{tabular}{lcccccc}
\hline\hline
Project code & PI & Configuration & On-source time & Central frequency & Bandwidth\\
& & & [min] & [GHz] & [MHz]\\
\hline
Band 9       \\
\multirow{2}{*}{2023.1.00578.S} & \multirow{2}{*}{K. Doi}   & \multirow{2}{*}{ACA} & \multirow{2}{*}{39.9} & 660.021, 662.021, 663.979, 665.979,  & 2000, 2000, 2000, 2000, \\ 
& & & & 676.021, 678.021, 679.979, 681.979 & 2000, 2000, 2000, 2000\\
\multirow{2}{*}{2023.1.00578.S} & \multirow{2}{*}{K. Doi}   & \multirow{2}{*}{C-3} & \multirow{2}{*}{50.7} & 660.021, 662.021, 663.979, 665.979, & 1875, 1875, 1875, 1875, \\
& & & & 676.021, 678.021, 679.979, 681.979 & 1875, 1875, 1875, 1875 \\
\multirow{2}{*}{2023.1.00578.S} & \multirow{2}{*}{K. Doi}  & \multirow{2}{*}{C-6} & \multirow{2}{*}{33.3} & 660.021, 662.021, 663.979, 665.979, & 1875, 1875, 1875, 1875, \\
& & & & 676.021, 678.021, 679.979, 681.979 & 1875, 1875, 1875, 1875\\
\hline
Band 7       \\
2015.1.00616.S & C. Pinte   & C40-5 & 63.7\tablefootmark{a} & 336.495, 338.432, 348.495, 350.495 & 2000, 2000, 2000, 2000\\
2016.1.01086.S & A. Isella  & C40-7 & 28.4 & 329.305, 330.559, 340.976, 342.856\tablefootmark{b} & 1875, 1875, 1875, 1875 \\
\hline
Band 6       \\
2013.1.00366.S & M. Hughes   & C34-4 & 249.4 & 216.123, 219.571, 230.549, 232.711 & 58.594, 58.594, 58.594, 2000 \\
2013.1.00601.S & A. Isella   & C34-7/(6) &148.2 & 219.549, 220.387, 230.526, 231.588 & 58.594, 58.594, 58.594, 2000 \\
2016.1.00484.L & S. Andrews  & C40-8 & 53.8 & 230.521, 232.583, 244.971, 246.888 & 937.5, 2000, 2000, 2000 \\
\hline
Band 4       \\
2017.1.01682.S & G. Guidi   & C43-5 & 14.6 & 134.003, 135.940, 145.982, 146.979 & 1875, 1875, 1875, 58.594 \\
2017.1.01682.S & G. Guidi   & C43-8 & 90.7 & 134.003, 135.940, 145.982, 146.979 & 1875, 1875, 1875, 58.594 \\
\hline
Band 3       \\
2015.1.00725.S & G. Guidi   & C40-6    & 131.5 & 91.143, 93.166, 103.006, 104.694 & 1875, 58.5938, 1875, 1875\\
2016.1.01086.S & A. Isella   & C40-8/9 & 91.3  & 95.991, 97.946, 107.989, 109.771, 110.190 & 2000, 1875, 2000, 937.5, 937.5 \\
\hline
\end{tabular}
\tablefoot{
\tablefoottext{a}{While this MOUS has four EBs, we use only the first two because the other two are of lower quality and show spurious large-scale structures.}
\tablefoottext{b}{Two SPWs overlap in frequency. We manually flagged the overlapping channels.}
}
\label{tab:observations}
\end{sidewaystable*}

\section{Blackbody emission at (sub)millimeter wavelengths} \label{app:planck}

\begin{figure}[htbp]
    \begin{center}
        \includegraphics[width=8cm]{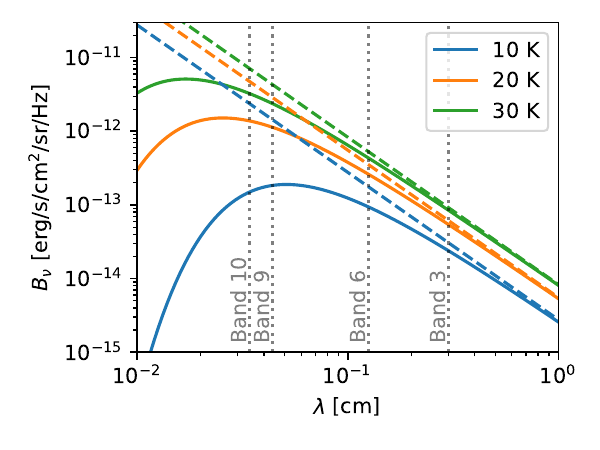}
        \vspace{-2em}
    \end{center}
    \caption{
        Spectrum of blackbody emission (solid lines) and the Rayleigh-Jeans approximation (dashed lines) at typical disk temperatures (10, 20, and 30 K).}
    \label{fig:planck_RJ}
\end{figure}

Figure \ref{fig:planck_RJ} compares the Planck function and the Rayleigh-Jeans approximation.
At typical disk temperatures of 10, 20, and 30 K, the Rayleigh-Jeans approximation closely approximates the blackbody emission at the longer wavelengths observed by ALMA, where the intensity is proportional to the temperature, while it significantly deviates from the Planck function at ALMA Band 9 (0.45 mm) and Band 10 (0.35 mm), where the intensity has a strong dependence on the temperature.
This is why observations at ALMA Band 9/10 are effective for breaking the degeneracy with temperature and constraining the dust properties.

\section{Fitting of geometric parameters} \label{app:geometry}

Geometric parameters of the disk, such as the center position, inclination, and position angle, are key parameters that affect the analysis based on the disk geometry, and accurate estimation is necessary.
Relative positional offsets between images at different wavelengths can affect the creation of spectral index maps.
Absolute offsets in the center position and the disk orientation (inclination and position angle) can affect the creation of radial profiles that require deprojection.

In this study, we estimated the absolute offset by minimizing the residuals between the original image and the 180°-rotated image within a mask region defined on the image plane.
If the image is point-symmetric, centering can be achieved by minimizing the imaginary part of the visibility.
However, while the continuum image of HD 163296 is broadly axisymmetric, it exhibits clear local asymmetries such as the crescent-like structure \citep{Isella2018}, and these structures may bias the centering based on the visibility analysis.
To mitigate this effect, we estimated the absolute center offset by minimizing the residuals between the original image and the $180^\circ$-rotated image within a mask excluding the asymmetries.

We defined an elliptical mask with an outer radius of $1\farcs1$.
We excluded the region at radii $0\farcs5 - 0\farcs6$ and $\pm 45^{\circ}$ along the southeast major axis to exclude the crescent, and also excluded the inner $0\farcs15$ region to exclude the central asymmetry.
The inclination and position angle used to define these ellipses were based on \citet{Huang2018_ring}.

\begin{figure*}[htbp]
    \begin{center}
        \includegraphics[width=16cm]{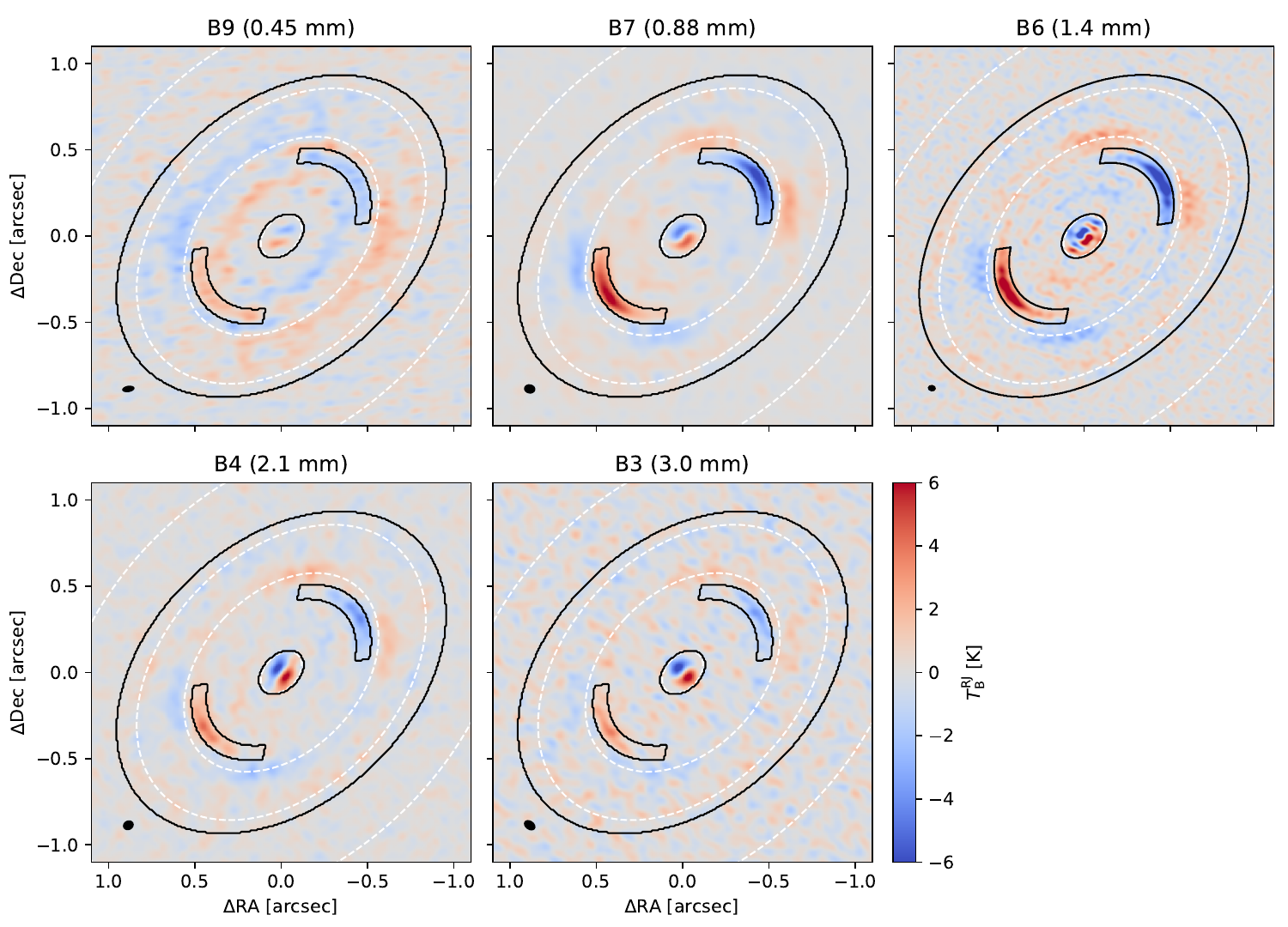}
    \end{center}
    \caption{
Residual images obtained by subtracting the $180^\circ$ rotated image from the original image at each wavelength.
The black line indicates the mask used to estimate the center position. The white dashed lines indicate the ring radii at $r=0\farcs140$, $0\farcs664$, $0\farcs987$, and $1\farcs550$, as estimated by \citet{Huang2018_ring}.
    }
    \label{fig:image_5panels_rotated_difference}
\end{figure*}

The inferred center offsets at each wavelength are summarized in Table \ref{tab:geometry}.
Fig. \ref{fig:image_5panels_rotated_difference} shows the residual map after the $180^\circ$ rotation.
At all wavelengths, the offset is $<0\farcs01$, indicating that, although asymmetric components are present, the images were already well aligned through centering by minimizing the visibility imaginary part.

We also estimated the inclination and position angle by minimizing the residual between rotated images.
Previously, these parameters have been estimated from ellipse fits to the overall emission, fits to the locations of ring peaks, or visibility modeling.
To utilize the full structural information of an approximately axisymmetric disk, we estimated the inclination and position angle by minimizing the residuals between the original image and the $90^\circ$-rotated image within the mask region excluding asymmetric components.
To reduce the impact of an elliptical beam on the evaluation of these rotational residuals, we performed the fit using images smoothed with \texttt{imsmooth} to a circular beam in the deprojected plane.
We performed this fitting for the Band 6 image due to its higher resolution. We used the same mask as for the centering.
These minimizations were performed with \texttt{scipy.optimize.least\_squares}.
The errors in Table \ref{tab:geometry} are derived from the covariance matrix computed from the Jacobian at the best-fit solution, taking into account the noise correlation within the beam.

\begin{figure*}[btp]
    \sidecaption
    \includegraphics[width=12.5cm]{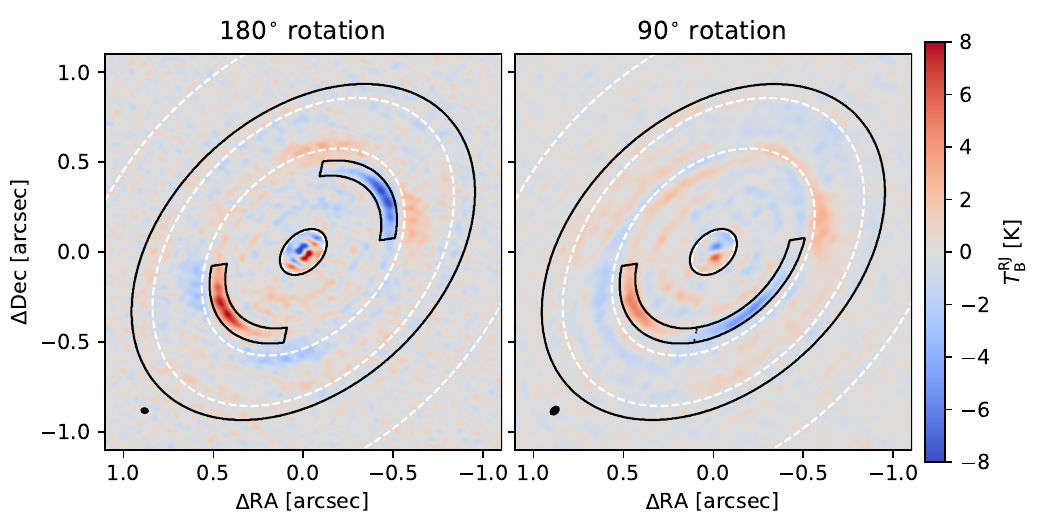}
    \caption{
Residual images after subtracting the $180^\circ$ rotated image (left) and the $90^\circ$ rotated image (right) from the Band 6 image.
The original image is used for the $180^\circ$ rotation, while the image smoothed to a circular beam in the deprojected plane is used for the $90^\circ$ rotation.
The black line indicates the mask used to estimate the center position. The white dashed lines indicate the ring radii at $r=0\farcs140$, $0\farcs664$, $0\farcs987$, and $1\farcs550$, as estimated by \citet{Huang2018_ring}.
    }
    \label{fig:rotated_difference}
\end{figure*}

Table \ref{tab:geometry} summarizes the inferred inclination and position angle, and Fig. \ref{fig:rotated_difference} shows the residual maps obtained by subtracting the $90^\circ$-rotated Band 6 image from the original image. 
The inferred inclination and position angle are close to those reported by \citet{Huang2018_ring} at all wavelengths, although the values from Band 9 are slightly lower. 
In the following analysis, we adopted the inclination and position angle derived from the Band 6 image, which has the highest angular resolution.

\begin{table*}[bth!]
\caption{Fitted center offsets and disk inclination and position angle.}
\centering
\begin{tabular}{lccccccc}
\hline\hline
 & $\Delta x$ [arcsec]  & $\Delta y$ [arcsec] & inclination [deg] & P.A. [deg] \\
\hline
Band 9 & $0.0034 \pm 0.0006$ & $0.0062 \pm 0.0005$ & $45.58 \pm 0.08$ & $132.77 \pm 0.11$\\
Band 7 & $0.0006 \pm 0.0002$ & $0.0071 \pm 0.0001$ & $46.27 \pm 0.02$ & $132.99 \pm 0.03$\\
Band 6 & $0.0062 \pm 0.0002$ & $0.0075 \pm 0.0002$ & $46.48 \pm 0.03$ & $133.27 \pm 0.04$\\
Band 4 & $0.0068 \pm 0.0003$ & $0.0020 \pm 0.0003$ & $46.53 \pm 0.05$ & $133.36 \pm 0.07$\\
Band 3 & $0.0039 \pm 0.0007$ & $0.0051 \pm 0.0007$ & $46.30 \pm 0.13$ & $133.33 \pm 0.19$\\
\hline
\label{tab:geometry}
\end{tabular}
\end{table*}

\section{Optical properties} \label{app:optical_properties}

The dust composition and complex refractive indices of the constituent materials are important factors that determine the optical properties of the dust, and here we describe the dust models used in this study.
In this study, we used two dust models: the DSHARP default model and the DSHARP Zubko model \citep{Birnstiel2018,Birnstiel2024ARAA}.
For each model, we considered both compact dust and porous dust with a porosity of 0.9, resulting in a total of four dust models.
All models assume the same mass fractions of the dust composition as in the DSHARP default model \citep{Birnstiel2018}, but the DSHARP Zubko model uses amorphous carbon (sample BE) from \citet{Zubko1996} instead of the refractory organics from \citet{Henning1996} for the carbonaceous material.
This difference in the carbonaceous material leads to significantly different optical properties between the two models, and the results of the SED fitting also differ significantly.

\begin{figure}[btp]
    \begin{center}
        \includegraphics[width=9cm]{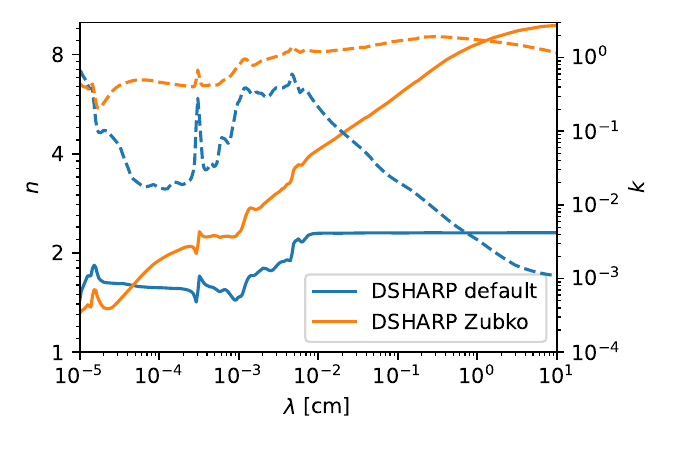}
    \end{center}
    \caption{
Complex refractive indices of the dust models used in this study. 
The real part is shown as a solid line and corresponds to the left axis, while the imaginary part is shown as a dashed line and corresponds to the right axis. 
Both the DSHARP default model and the DSHARP Zubko model are shown for the compact dust case.
    }
    \label{fig:optical_constants}
\end{figure}

Table \ref{tab:composition} shows the density and mass fraction of the dust components for the DSHARP default model and the DSHARP Zubko model.
Fig. \ref{fig:optical_constants} shows the complex refractive indices when these components are mixed together.
The complex refractive indices differ significantly between these two models, especially in the millimeter wavelength range, where the imaginary part differs by an order of magnitude. 

\begin{figure*}[btp]
    \begin{center}
        \includegraphics[width=17cm]{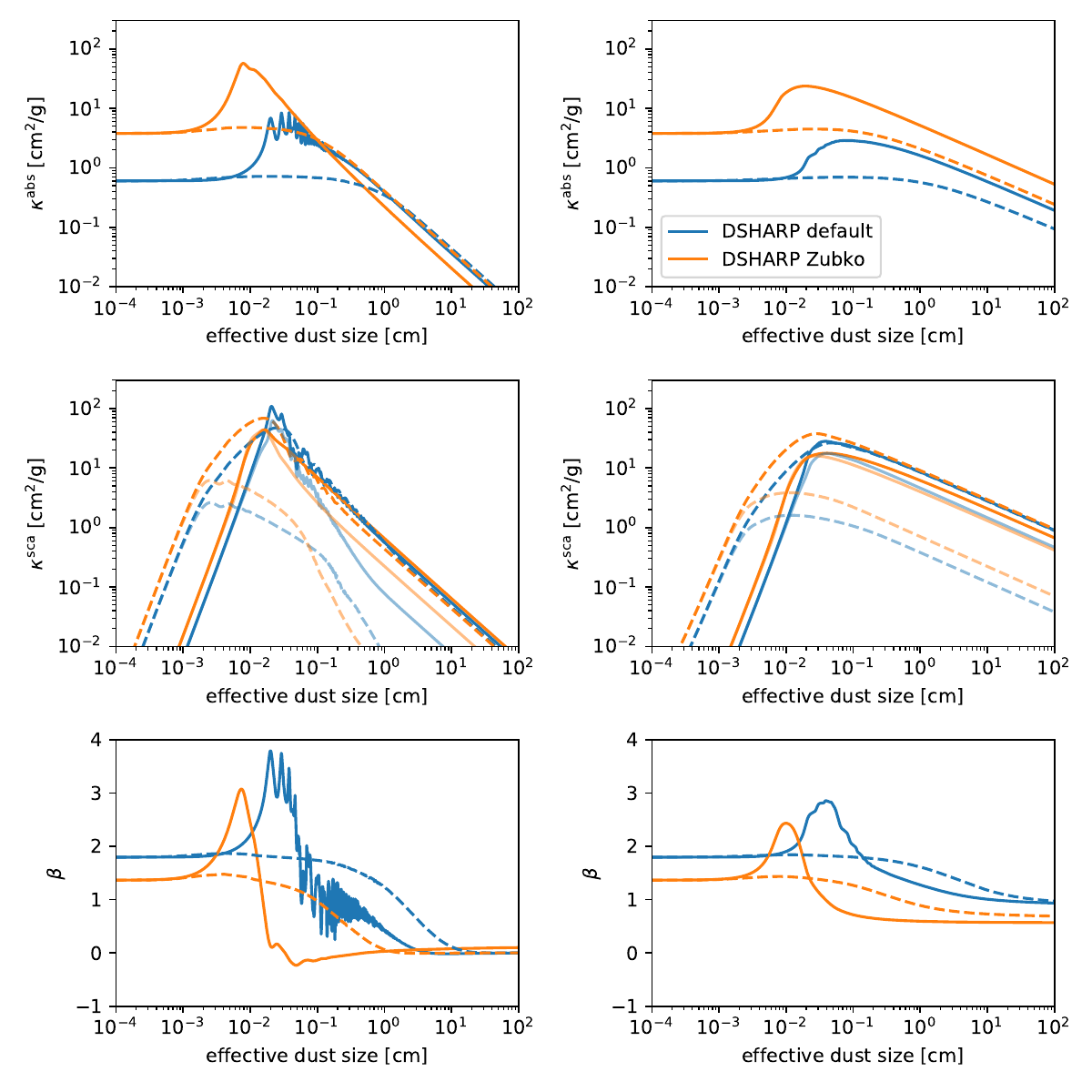}
    \end{center}
    \caption{
Absorption opacities (top), scattering opacities (middle), and the spectral index $\beta$ over the wavelength range of 1 - 3 mm (bottom) for the dust models used in this study. 
The left panels show the values without size averaging, while the right panels show the values averaged over a size distribution with $q=3.5$. 
The horizontal axis is the effective dust size, $a_\mathrm{eff} = a (1-p)$. 
The blue lines represent the DSHARP default model, while the orange lines represent the DSHARP Zubko model. Solid lines indicate the compact dust models, while dashed lines indicate the porous dust models. 
The faint lines in the scattering-opacity panels show the effective scattering opacities that take into account the asymmetry parameter $g$.
    }
    \label{fig:opacity}
\end{figure*}

Fig. \ref{fig:opacity} shows the absorption and scattering opacities at 1 mm and the spectral index $\beta$ over the wavelength range of 1–3 mm for each dust model.
The left side shows the values without size averaging, while the right side shows the values averaged over a size distribution with $q=3.5$.
These quantities also differ significantly between the dust models.
The dust size at which the absorption cross-section peaks differs among the compositions, and the absorption cross-section in the small-size limit can differ by an order of magnitude depending on the composition.
Also, the spectral index $\beta$ in the small-size limit differs depending on the composition.

\begin{table*}[bht!]
\caption{Density and mass fraction of dust components for the dust models used in this study.}
\centering
\begin{tabular}{lccccccc}
\hline\hline
Model    &  & DSHARP default & DSHARP Zubko\\
Component & Density $\mathrm{[g/cm^3]}$ &  \multicolumn{2}{c}{Mass Fraction} & Reference \\
\hline
Water ice               & 0.92 & 0.2000 & 0.2000 & \citet{Warren2008} \\
Astronomical sillicate  & 3.30 & 0.3291 & 0.3291 & \citet{Draine2003ARAA} \\
Troilite                & 4.83 & 0.0743 & 0.0743 & \citet{Henning1996} \\
Refractory organics     & 1.50 & 0.3966 & -      & \citet{Henning1996} \\
Amorphous carbon (BE)   & 1.80\tablefootmark{a} & - & 0.3966      & \citet{Zubko1996} \\
\hline
\end{tabular}
\tablefoot{\tablefoottext{a}{We adopt the density for amorphous carbon (BE) as used in \citet{Dominik2021}.}}
\label{tab:composition}
\end{table*}

\section{Low resolution} \label{sec:low_res_image}

We present lower-resolution but higher-sensitivity images obtained by using a higher robust parameter to focus on the extended emission. We used a robust parameter of 0.5 in this appendix.
Based on these images, we also created radial profiles and show the results of SED fitting.

\subsection{Images and their radial profile} \label{app:low_res_image}

\begin{figure*}[thbp]
    \begin{center}
        \includegraphics[width=18cm]{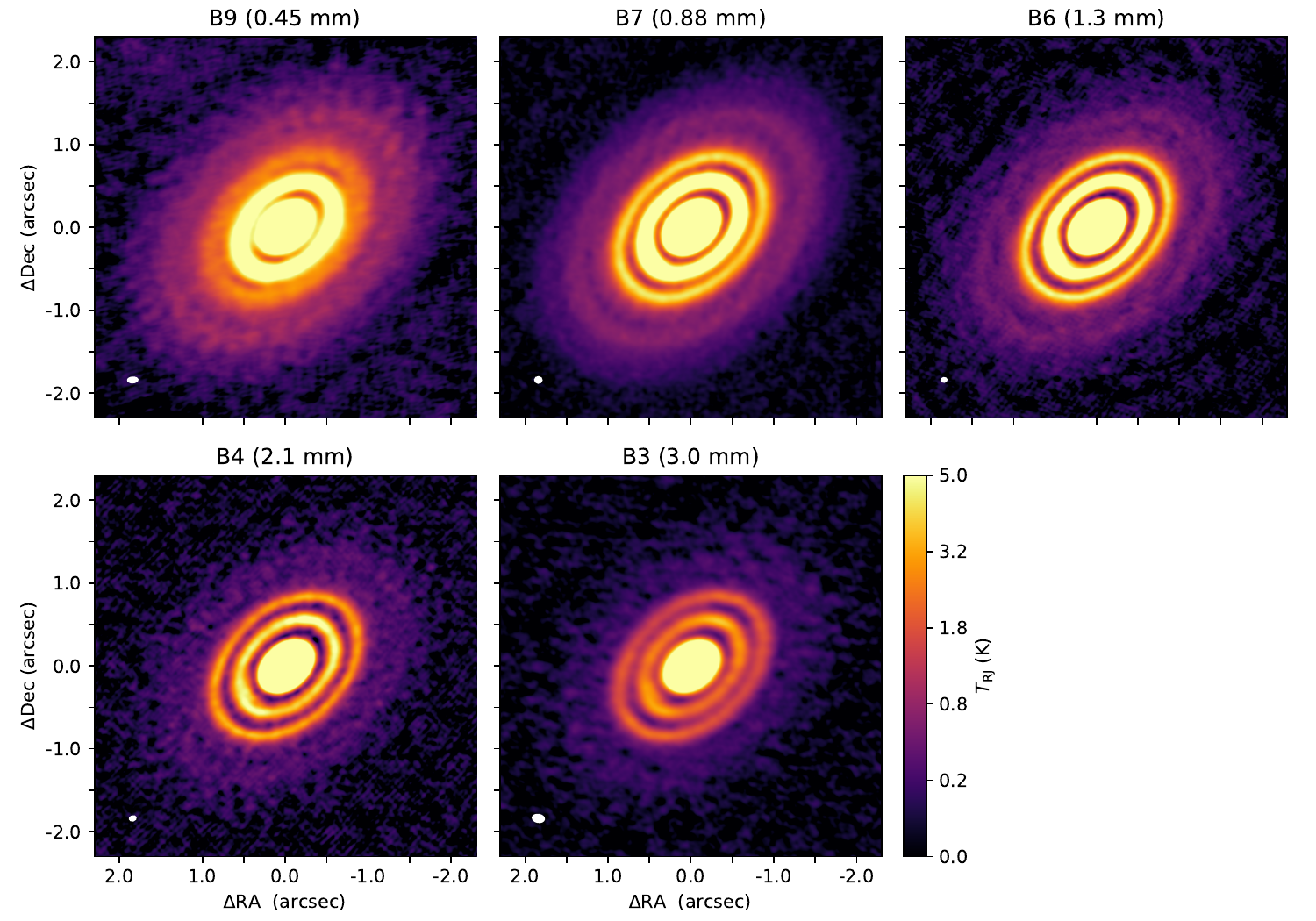}
    \end{center}
    \caption{
Same as Fig. \ref{fig:image_5panels}, but for the low-resolution images. The imaging settings, rms noise, and image intensities are summarized in Table \ref{tab:imaging_results_low}. The white ellipses at the bottom left of each panel indicate the beam size.
    }
    \label{fig:image_5panels_low}
\end{figure*}

Fig. \ref{fig:image_5panels_low} shows the lower-resolution images.
Since these images focus on the faint outer regions, the color range is set to match the intensity in the outer regions.
The imaging settings, rms noise, and intensity of these images are summarized in Table \ref{tab:imaging_results_low}.
As for the fiducial images, we applied the primary beam correction to these images.
The rms noise levels and SNRs in Table \ref{tab:imaging_results_low} were measured on the images before the primary beam correction, while the integrated fluxes were measured on the primary-beam-corrected images.
At all wavelengths, the total flux obtained is in good agreement with the fiducial high-resolution images.
From the images, extended emission can be seen outside the $1\arcsec$ ring at all wavelengths.

\begin{figure}[htbp]
    \begin{center}
        \includegraphics[width=8cm]{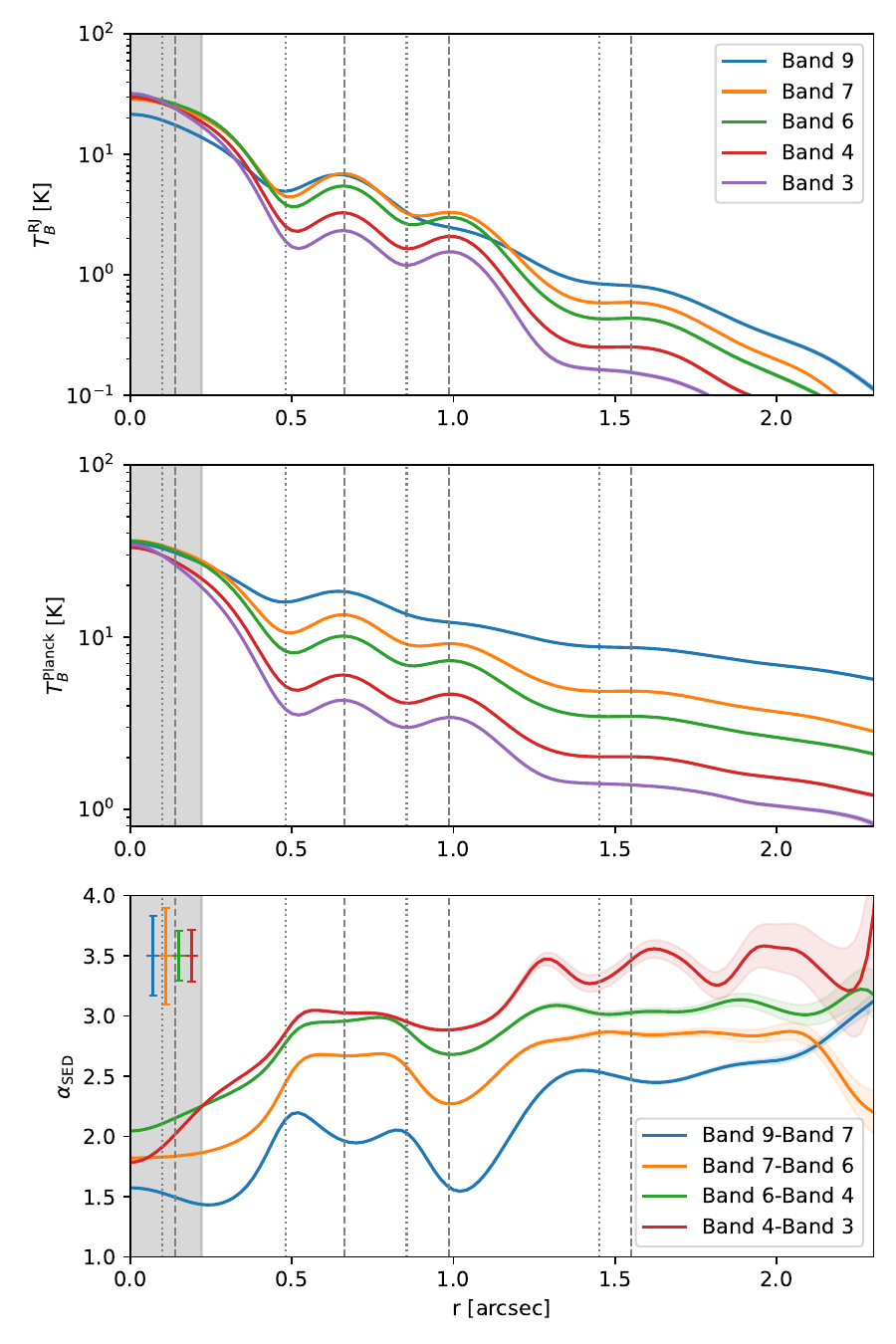}
    \end{center}
    \caption{
Azimuthally averaged radial profiles averaged over the full azimuthal angle and spectral index, as shown in Fig. \ref{fig:radial_obs}, but for the low-resolution images.
    }
    \label{fig:radial_obs_low}
\end{figure}

Fig. \ref{fig:radial_obs_low} shows the radial profiles and spectral index profiles based on the lower-resolution images.
Here, we smoothed the above images to a common beam that becomes circular with a diameter of $0\farcs22$ after deprojection, and created radial profiles by averaging over the full azimuthal angle, sampled at radial intervals of $0\farcs02$.
The noise is estimated by dividing the image rms noise by the square root of the number of beams within the azimuthal angle, to estimate the noise after azimuthal averaging.

\begin{table*}[htbp]
\caption{Imaging results.}
\centering
\begin{tabular}{lcccccccc}
\hline\hline
Band & Frequency & Robust & Beam size & Beam PA & rms noise & Peak intensity & SNR & Integrated flux \\
& [GHz] & & [arcsec] & [deg] & [$\mathrm{\mu Jy\ beam^{-1}}$] & [mJy beam$^{-1}$] & & [mJy] \\
\hline
Band 9 & $670.98$ & $0.5$ & $0.139 \times 0.083$ & $-86.4$ & $248.4$ & $111.99$ & $450.9$ & $6137.4$ \\
Band 7 & $339.94$ & $0.5$ & $0.105 \times 0.095$ & $85.8$ & $23.4$ & $33.17$ & $1417.7$ & $1650.5$ \\
Band 6 & $238.98$ & $0.5$ & $0.089 \times 0.071$ & $-82.2$ & $16.6$ & $12.17$ & $733.0$ & $716.7$  \\
Band 4 & $140.02$ & $0.5$ & $0.099 \times 0.077$ & $-75.7$ & $8.4$ & $5.30$ & $630.7$ & $174.0$ \\
Band 3 & $100.44$ & $0.5$ & $0.162 \times 0.111$ & $80.8$ & $6.5$ & $5.71$ & $879.2$ & $71.7$\\
\hline
\end{tabular}
\label{tab:imaging_results_low}
\end{table*}

\subsection{SED fitting for low resolution images} \label{app:SED_low_res}

\begin{figure*}[phtb]
  \centering
  \setlength{\tabcolsep}{0pt}
  \renewcommand{\arraystretch}{0}

  \begin{tabular}{cc}
    \includegraphics[width=8.8cm]{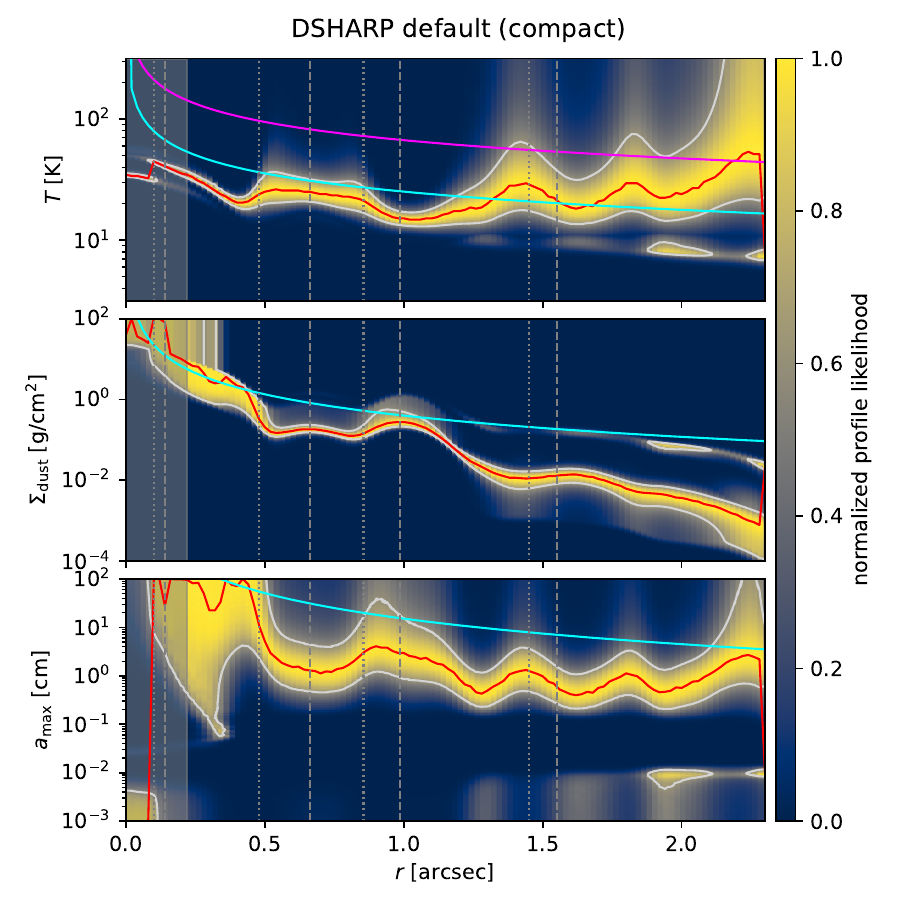} &
    \includegraphics[width=8.8cm]{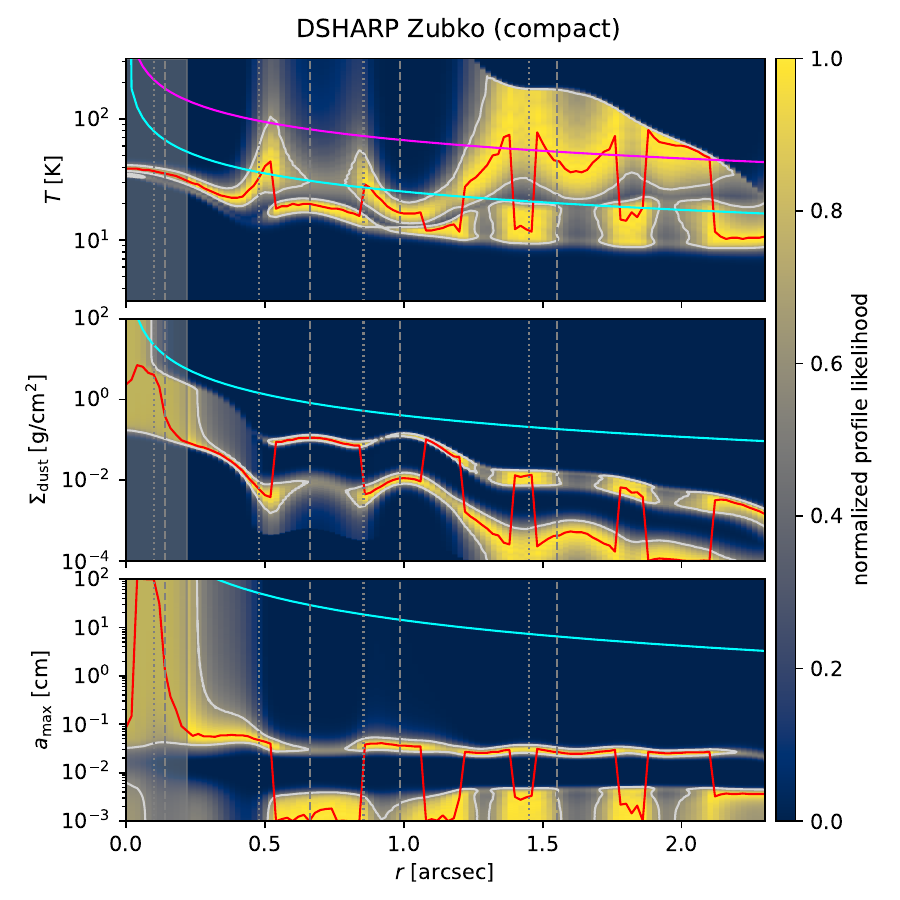} \\
    \includegraphics[width=8.8cm]{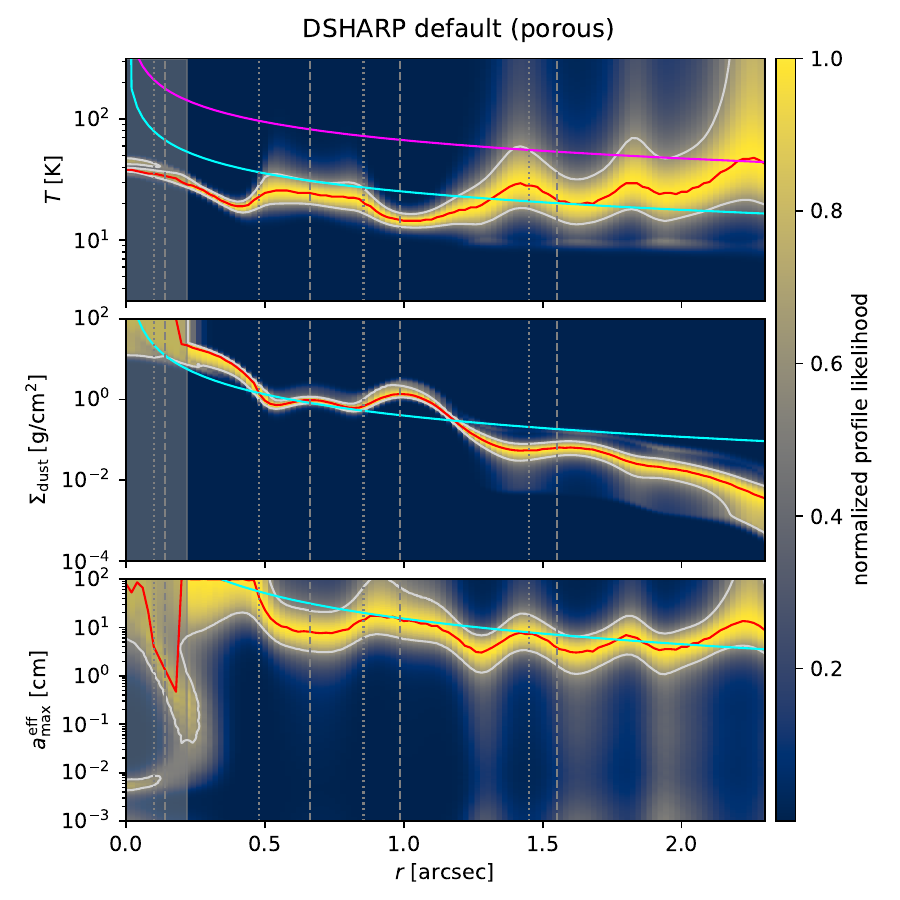} &
    \includegraphics[width=8.8cm]{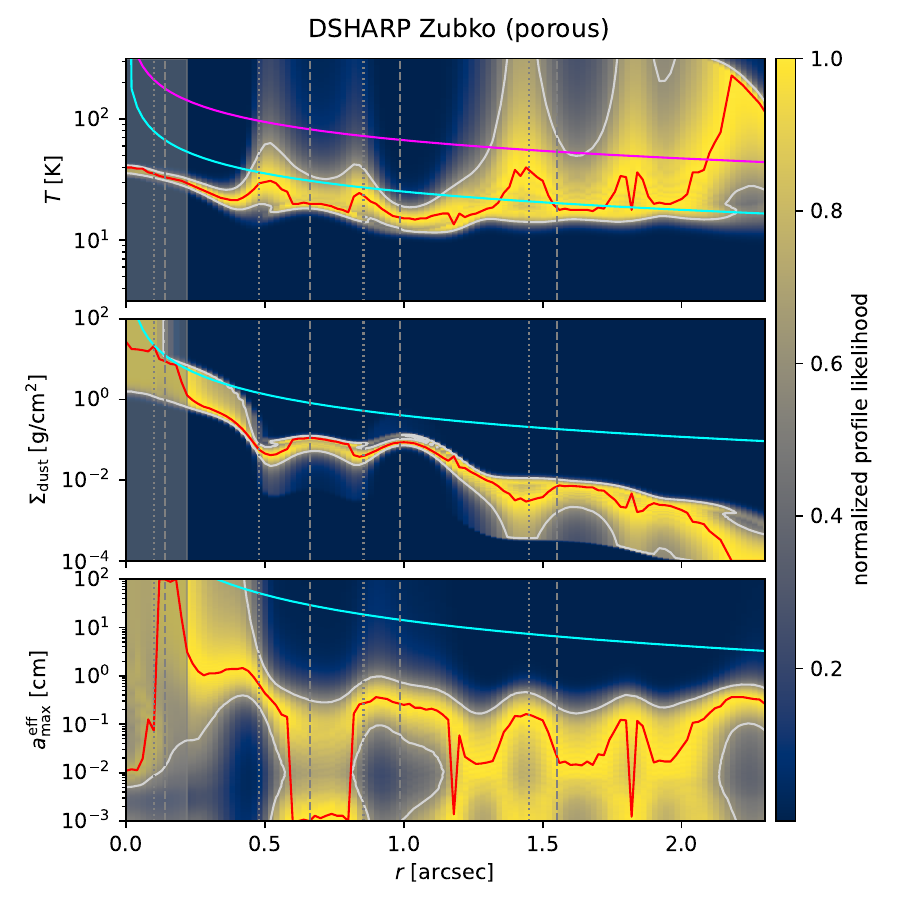} \\
  \end{tabular}
  \caption{
        Normalized profile likelihoods from the SED fitting based on the low-resolution images, shown in the same format as Fig. \ref{fig:dust_characterization}.
        }
  \label{fig:dust_characterization_low}
\end{figure*}

Fig. \ref{fig:dust_characterization_low} shows the results of SED fitting using the lower-resolution profiles.
These low-resolution profiles are more sensitive than the high-resolution profiles and thus provide better constraints on the dust properties in the faint outer regions.
On the other hand, the ring and gap structures are strongly smoothed at this resolution, and the intensities there are contaminated by emission from the surrounding regions.
Therefore, the estimates in the ring and gap regions can deviate from those obtained with the high-resolution profiles and are less reliable, and the focus here is on the outer extended emission regions.
From this analysis, even without using priors, the strong temperature dependence of Band 9 enables the temperature to be constrained even in these optically thin regions.

\section{Fitting for 4 parameters} \label{app:4d_fitting}

\begin{figure*}[phtb]
  \centering
  \setlength{\tabcolsep}{0pt}
  \renewcommand{\arraystretch}{0}
  \begin{tabular}{cc}
    \includegraphics[width=8.8cm]{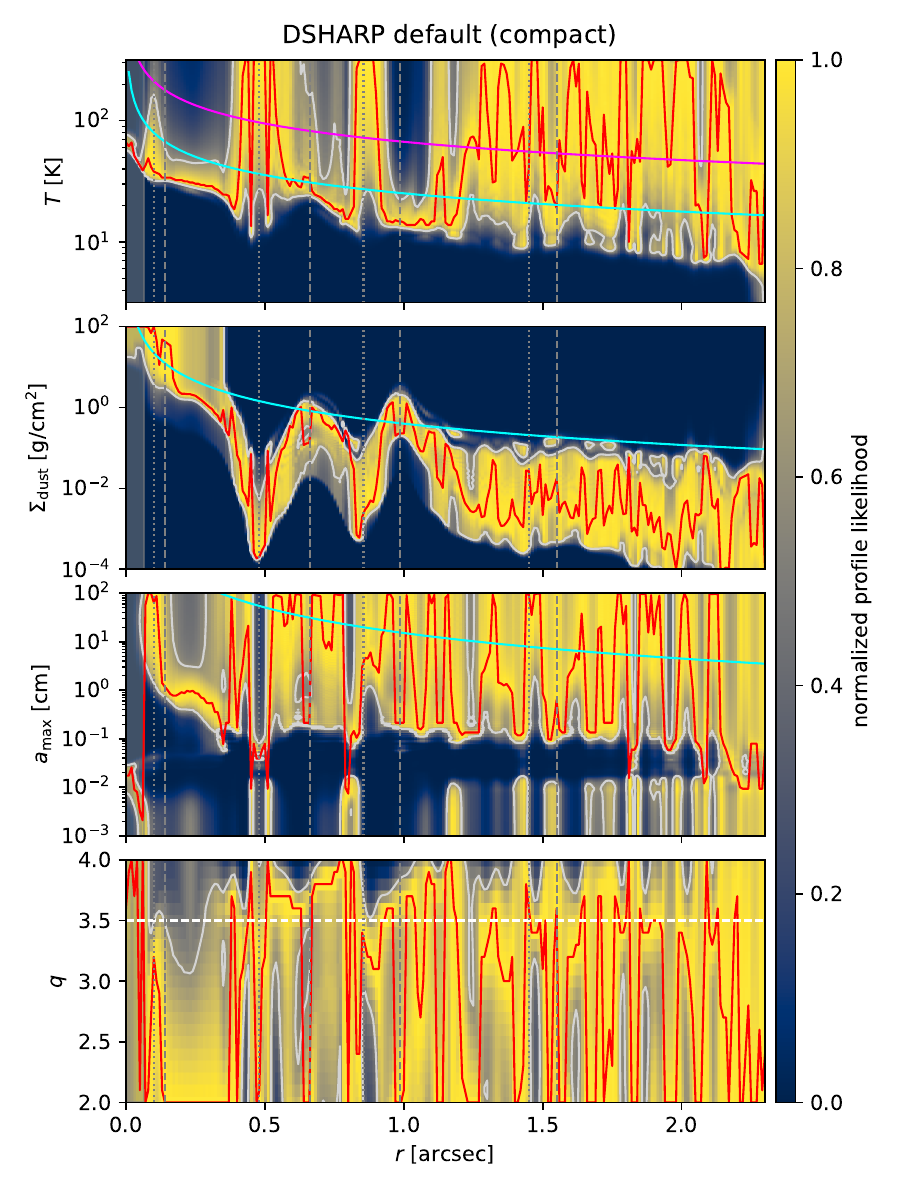} &
    \includegraphics[width=8.8cm]{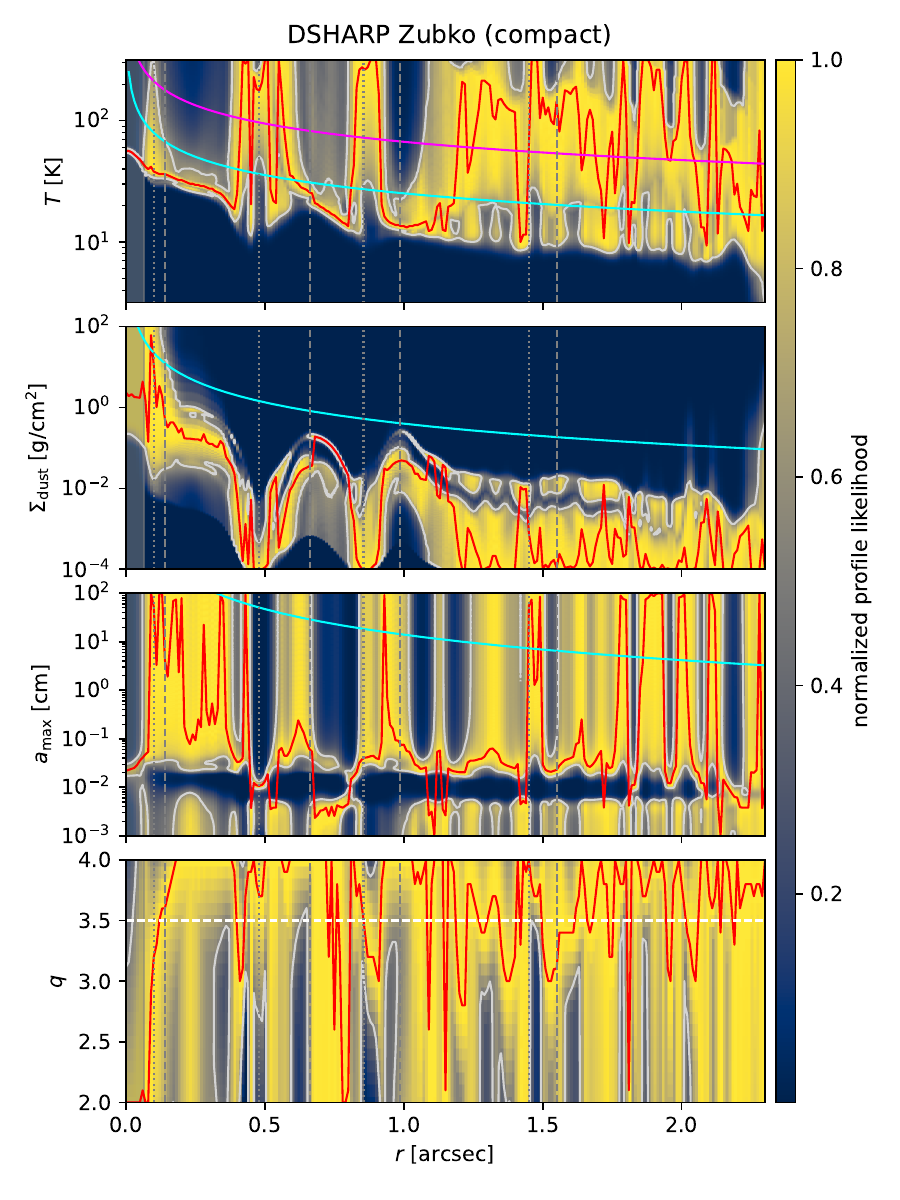} \\
    \includegraphics[width=8.8cm]{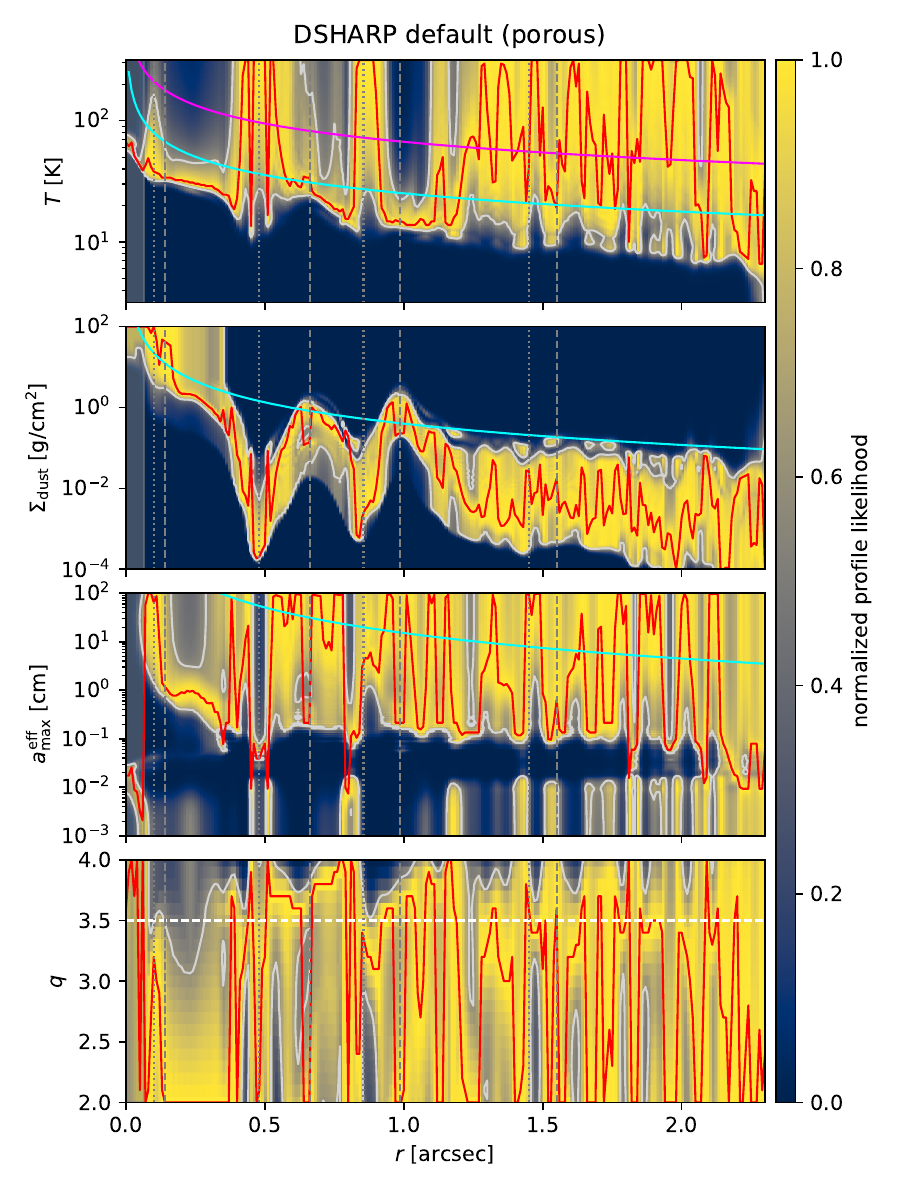} &
    \includegraphics[width=8.8cm]{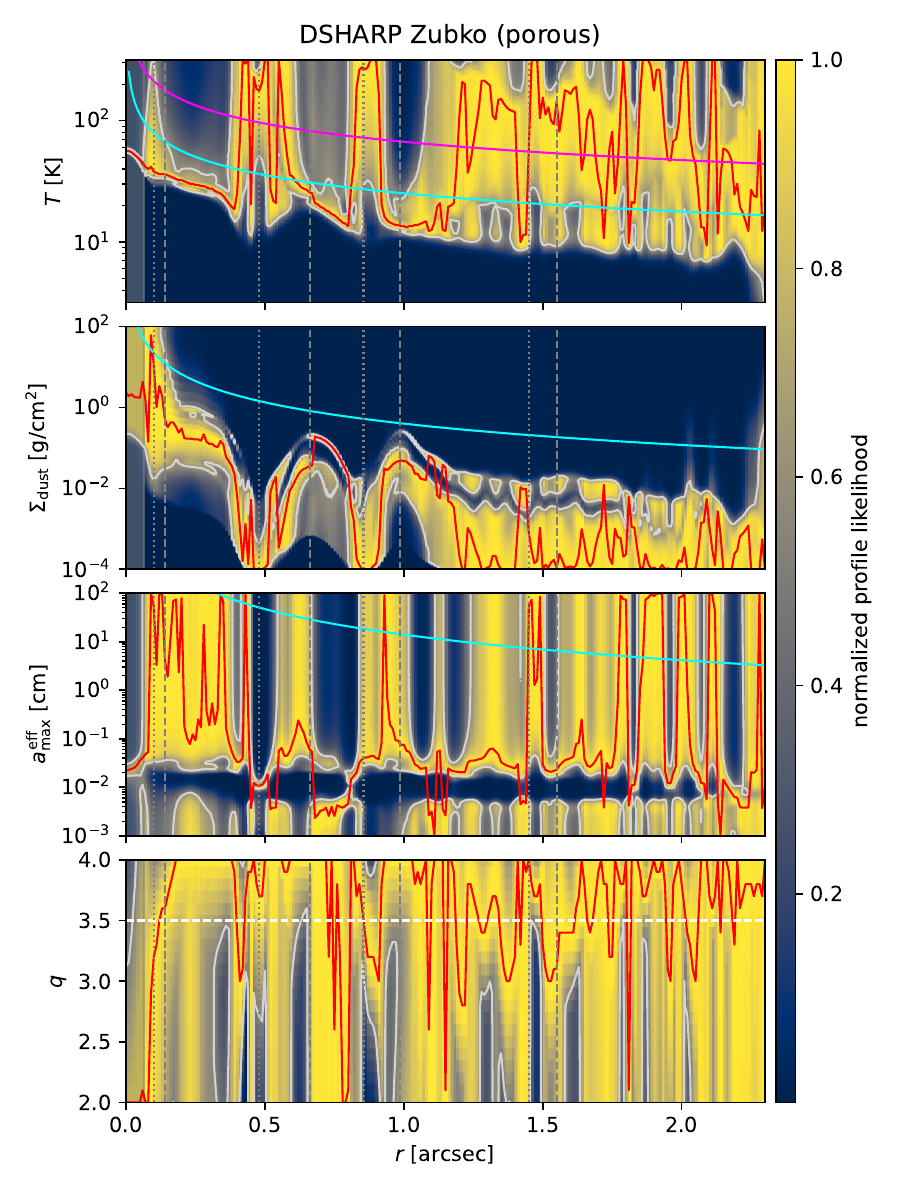} \\
  \end{tabular}
  \caption{
        Normalized profile likelihoods of temperature $T$, dust surface density $\Sigma_{\mathrm{dust}}$, maximum dust size $a_{\max}$, and power-law index $q$ from the SED fitting, shown in the same format as Fig. \ref{fig:dust_characterization}.}
  \label{fig:dust_characterization_4d}
\end{figure*}

Fig. \ref{fig:dust_characterization_4d} shows the fitting results for four parameters, including the power-law index of the dust size distribution $q$, in addition to the temperature $T$, surface density $\Sigma_{\mathrm{dust}}$, and maximum dust size $a_{\mathrm{max}}$ for each dust model.
For the temperature and surface density, we can see a similar trend to Fig. \ref{fig:dust_characterization}, but with larger uncertainties.
For the dust size and the power-law index, a wide range of values is allowed.
Fig. \ref{fig:corner_DSHARP_4param_100} in Appendix \ref{app:corner_plots} shows the corner plot for the four parameters at $r=1\farcs00$.
From this figure, we can see that there is a strong degeneracy between the maximum dust size and the power-law index, which results in similar opacity spectral indices.

\section{Fitting with/without Band 9} \label{app:with_without_band9}

\begin{figure*}[phtb]
  \centering
  \setlength{\tabcolsep}{0pt}
  \renewcommand{\arraystretch}{0}
  \begin{tabular}{cc}
    \includegraphics[width=8.8cm]{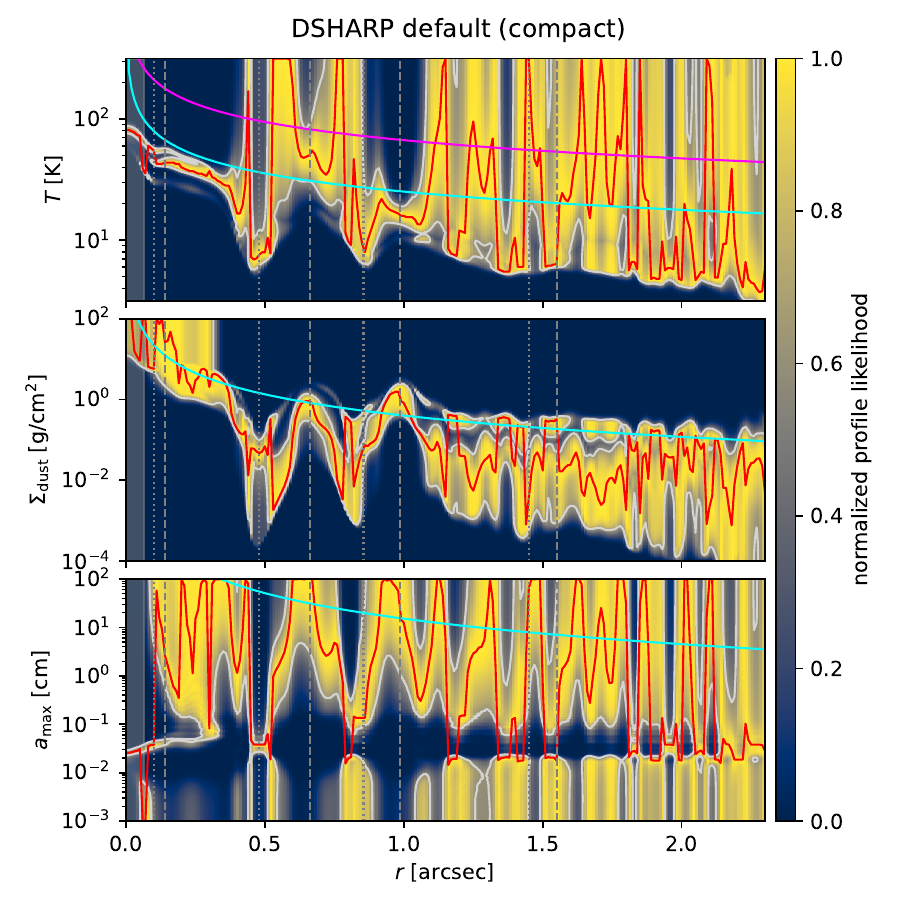} &
    \includegraphics[width=8.8cm]{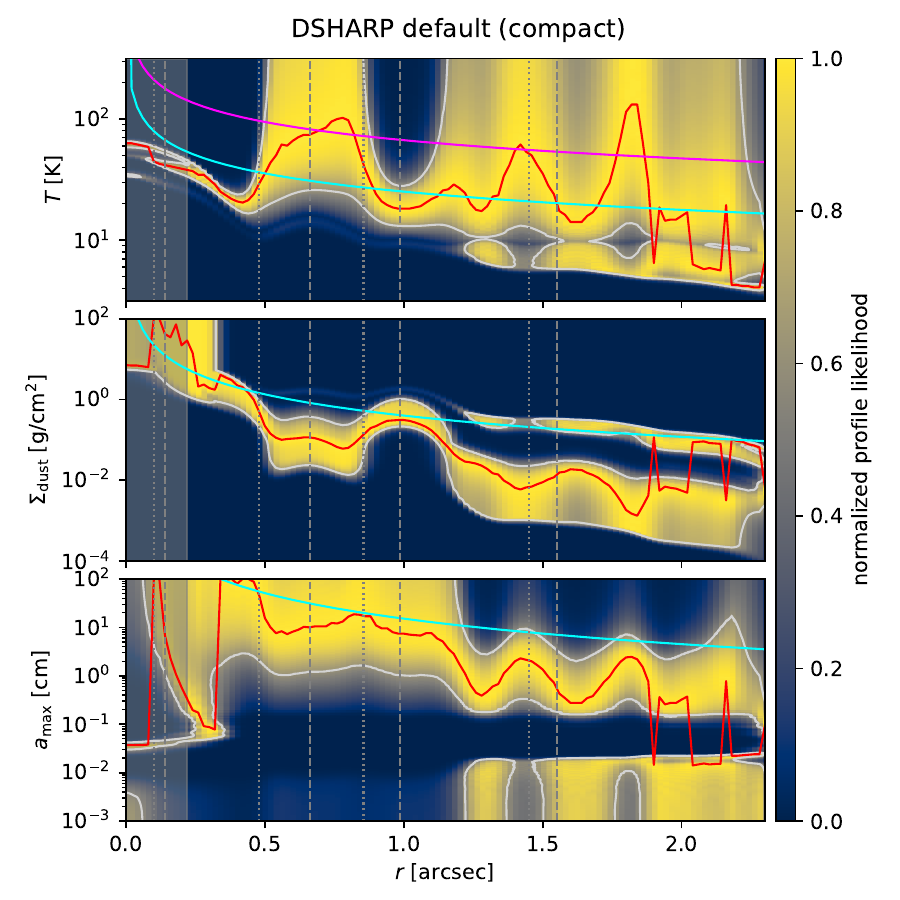} \\
  \end{tabular}
  \caption{
    Same as Fig. \ref{fig:dust_characterization}, but without Band 9. The left and right panels show the results for the high-resolution and low-resolution profiles, respectively, for the DSHARP default (compact) model.
        }
  \label{fig:dust_characterization_wo_B9}
\end{figure*}

To demonstrate the importance of the Band 9 observations, we also performed SED fitting using only the four longer-wavelength bands, Bands 3, 4, 6, and 7, excluding Band 9. 
Figure \ref{fig:dust_characterization_wo_B9} presents the SED fitting results for the DSHARP default compact dust model for both the high-resolution and low-resolution images for the case without Band 9.

For the high-resolution image at the inner ring at $r = 0\farcs67$, the temperature is constrained to $T = 26.92^{+3.99}_{-2.93}$ when Band 9 is included, whereas it is constrained to $ T = 54.95^{+49.76}_{-24.75} $ without Band 9.
Thus, the inclusion of Band 9 provides an order of magnitude tighter constraint on the temperature, which in turn leads to substantially tighter constraints on the dust surface density and grain size.
A similar trend is seen in the low-resolution SED fitting. 
In the extended outer region, only a lower limit on the temperature is obtained without Band 9, whereas both a lower limit and an upper limit are obtained when Band 9 is included.

\section{Profile likelihoods and marginal distributions} \label{app:profile_marginal_comparison}

\begin{figure*}[phtb]
  \centering
  \setlength{\tabcolsep}{0pt}
  \renewcommand{\arraystretch}{0}
  \begin{tabular}{cc}
    \includegraphics[width=8.8cm]{figure/dsharp_zubko_compact_full_range-highres_3panels_T_Sigma_a.pdf} &
    \includegraphics[width=8.8cm]{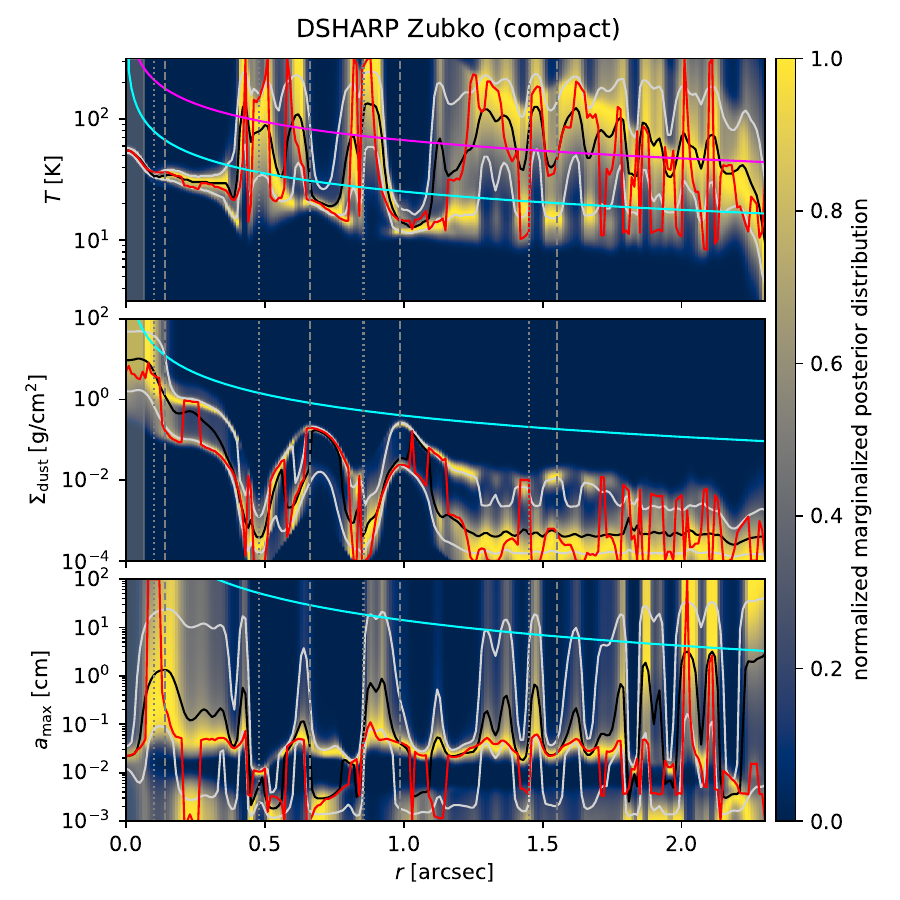} \\
  \end{tabular}
  \caption{
        Comparison of the profile likelihoods (left) and marginalized distributions (right), shown in the same format as Fig. \ref{fig:dust_characterization}, for the DSHARP Zubko (compact) dust model for the high-resolution images. 
        In the marginalized distributions, the gray contours indicate the 16th and 84th percentiles, and the black contour indicates the 50th percentile.
  }
  \label{fig:dust_characterization_profile_marginal}
\end{figure*}

Fig. \ref{fig:dust_characterization_profile_marginal} shows the profile likelihoods and marginalized distributions for the DSHARP Zubko (compact) model. 
In marginalization, flat priors are assumed for each parameter. 
Among the two solutions seen in the profile likelihood, with $a<100\ \mathrm{\mu m}$ and $a>100\ \mathrm{\mu m}$, the $a<100\ \mathrm{\mu m}$ solution becomes less pronounced in the marginalized distribution. 
This is because the small-grain solution occupies only a narrow region of parameter space and is therefore assigned less posterior weight by marginalization. 
Thus, marginalization tends to emphasize solutions that are supported over a broader volume of parameter space, but it can also reduce the weight of solutions confined to a narrow region of parameter space.

\section{Optical depth profiles} \label{app:tau_profiles}

\begin{figure*}[htbp]
  \centering
  \setlength{\tabcolsep}{0pt}
  \renewcommand{\arraystretch}{0}
  \begin{tabular}{cc}
    \includegraphics[width=9cm]{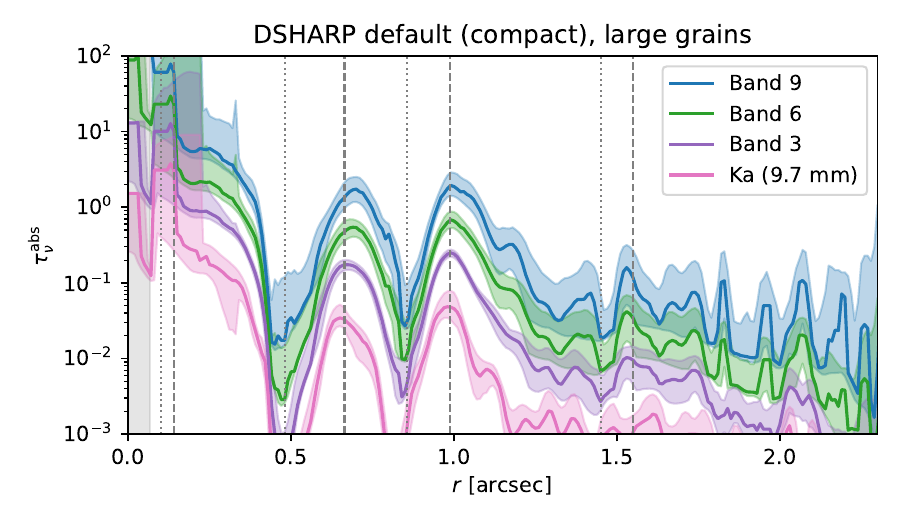} &
    \includegraphics[width=9cm]{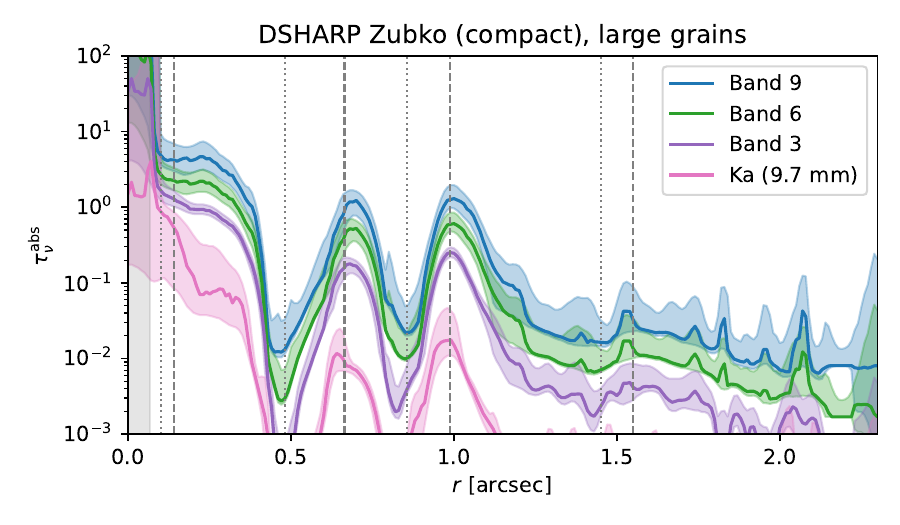} \\
    \makebox[9cm]{} &
    \includegraphics[width=9cm]{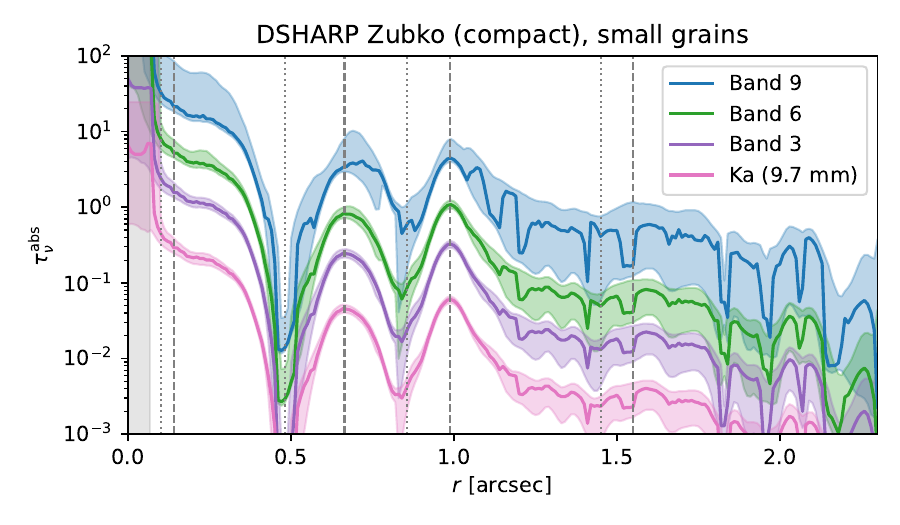} \\
    \includegraphics[width=9cm]{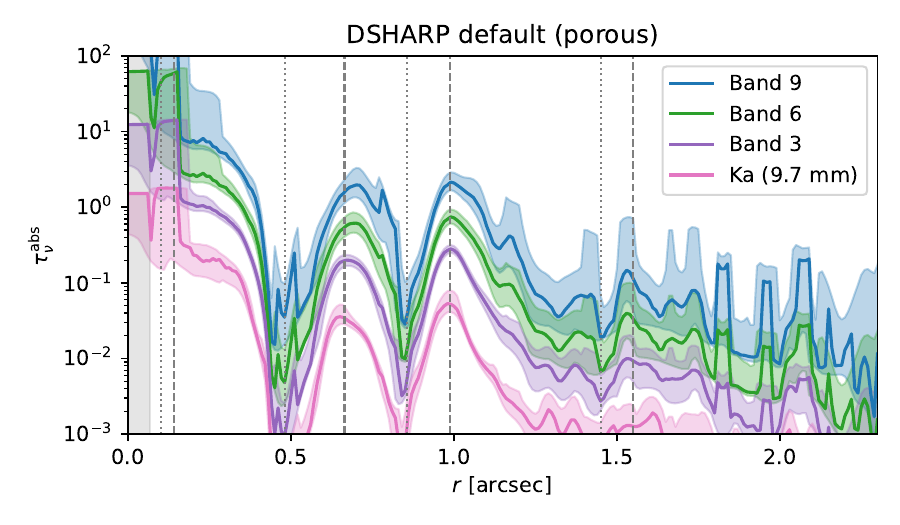} &
    \includegraphics[width=9cm]{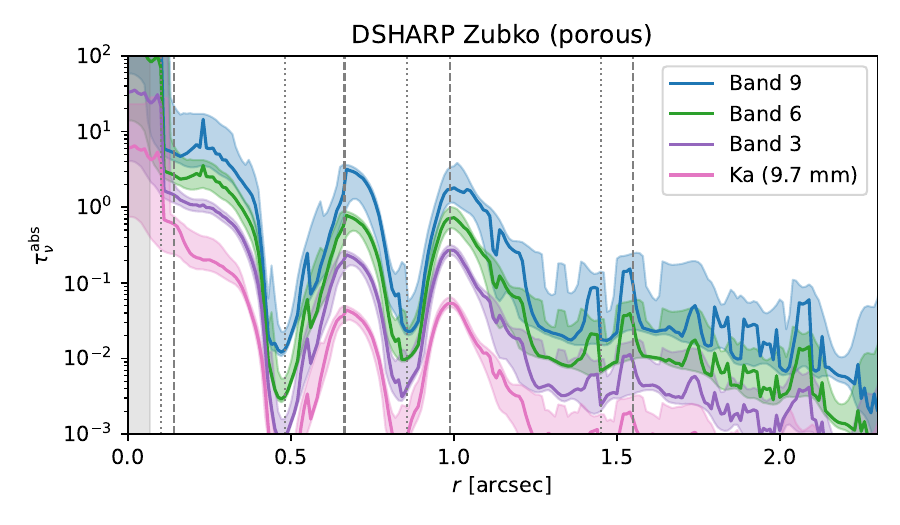}
  \end{tabular}
  \caption{
    Radial profiles of the absorption optical depth for each dust model.
    For the compact dust model, the results are shown separately for the large-grain and small-grain solutions, as in Fig. \ref{fig:obs-model_comparison}.
    The profiles are shown only for every other band to avoid overcrowding.
  }
  \label{fig:tau_abs}
\end{figure*}

\begin{figure*}[htbp]
  \centering
  \setlength{\tabcolsep}{0pt}
  \renewcommand{\arraystretch}{0}
  \begin{tabular}{cc}
    \includegraphics[width=9cm]{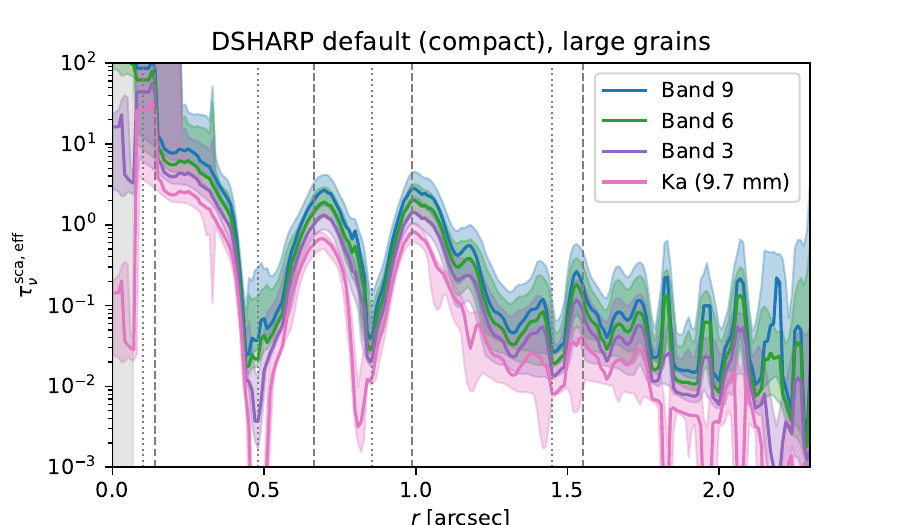} &
    \includegraphics[width=9cm]{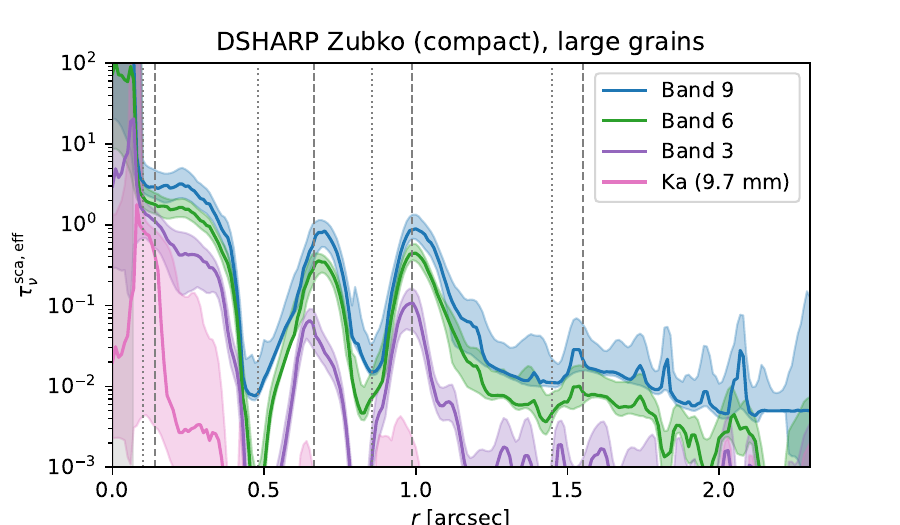} \\
    \makebox[9cm]{} &
    \includegraphics[width=9cm]{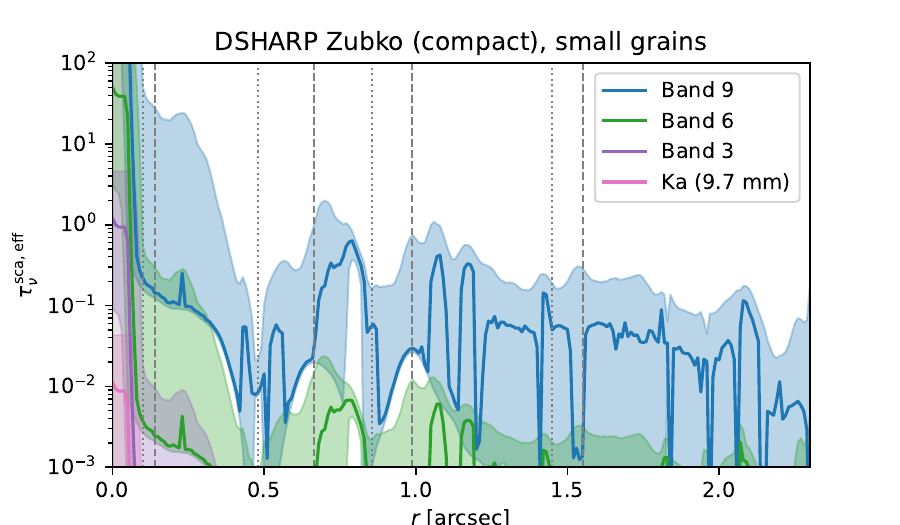} \\
    \includegraphics[width=9cm]{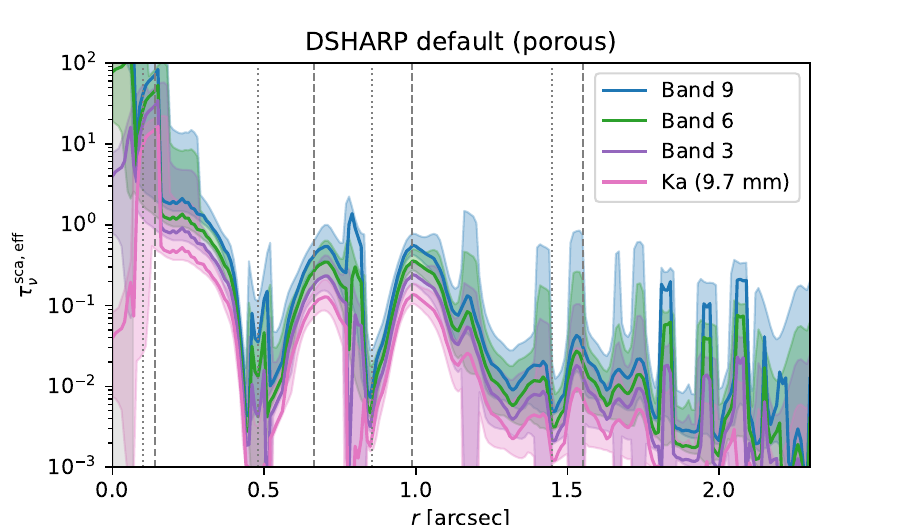} &
    \includegraphics[width=9cm]{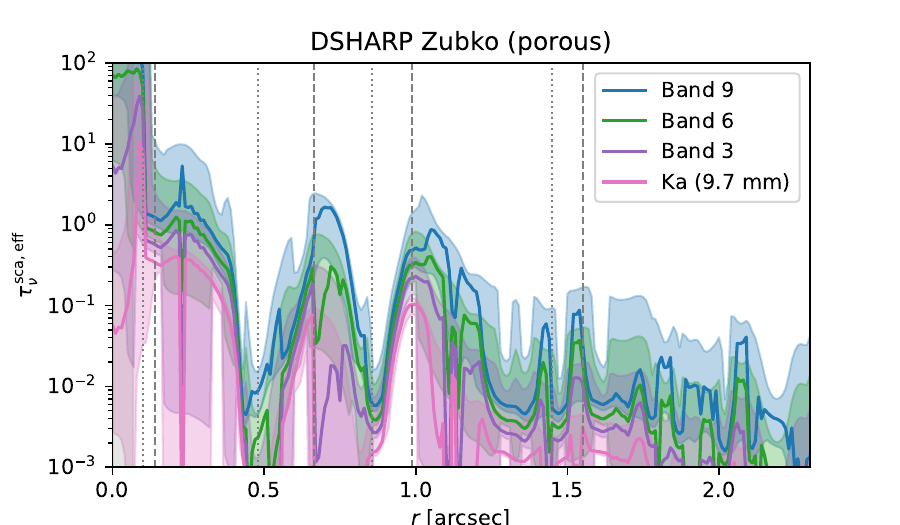}
  \end{tabular}
  \caption{
    Radial profiles of the effective scattering optical depth for each dust model.
    The results for the compact dust models are shown separately for the large-grain and small-grain solutions, as in Fig. \ref{fig:obs-model_comparison} and \ref{fig:tau_abs}.
    The profiles are shown only for every other band to avoid overcrowding.
  }
  \label{fig:tau_sca}
\end{figure*}

Figures \ref{fig:tau_abs} and \ref{fig:tau_sca} show the radial profiles of the absorption optical depth and the effective scattering optical depth for each dust model, respectively.
At the ring locations, the absorption optical depth at Band 9 is larger than 1 for all dust models, indicating that the Band 9 observations are suitable as a tracer of temperature.
On the other hand, at Band 3, the absorption optical depth is less than 1 for all dust models, which shows that the observations over this wide wavelength range enable the simultaneous estimation of temperature, surface density, and dust size.
On the other hand, at the central disk, the absorption optical depth is marginally optically thick even at Band 3, suggesting the need for observations at longer wavelengths.
The effective scattering optical depth varies significantly depending on the model, and especially for the compact small grain solutions, the scattering is very small.
Also, for the compact models, the DSHARP default model has larger scattering than the DSHARP Zubko model.

\section{Corner plots} \label{app:corner_plots}

\begin{figure}[htbp]
    \begin{center}
        \includegraphics[width=9cm]{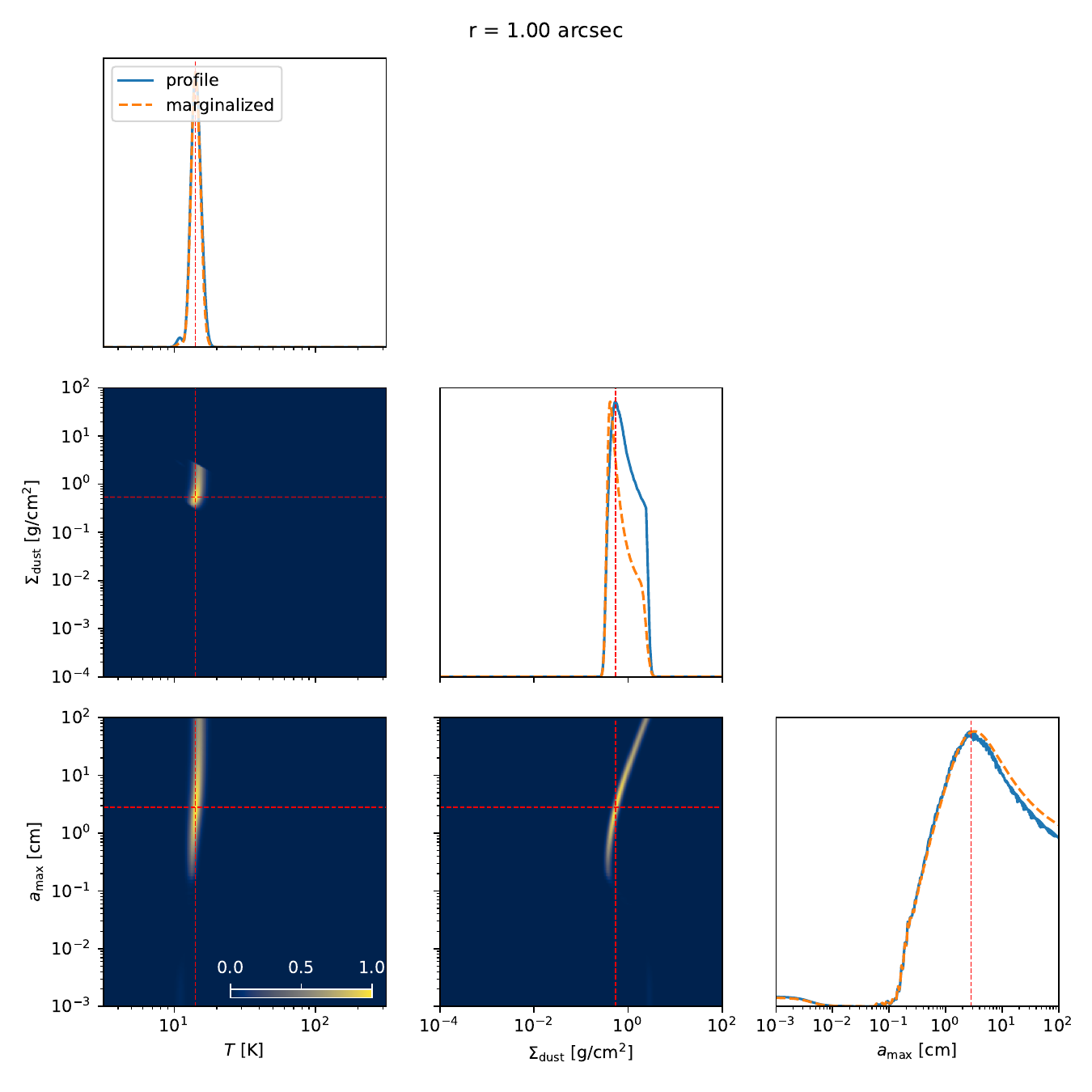}
        \includegraphics[width=9cm]{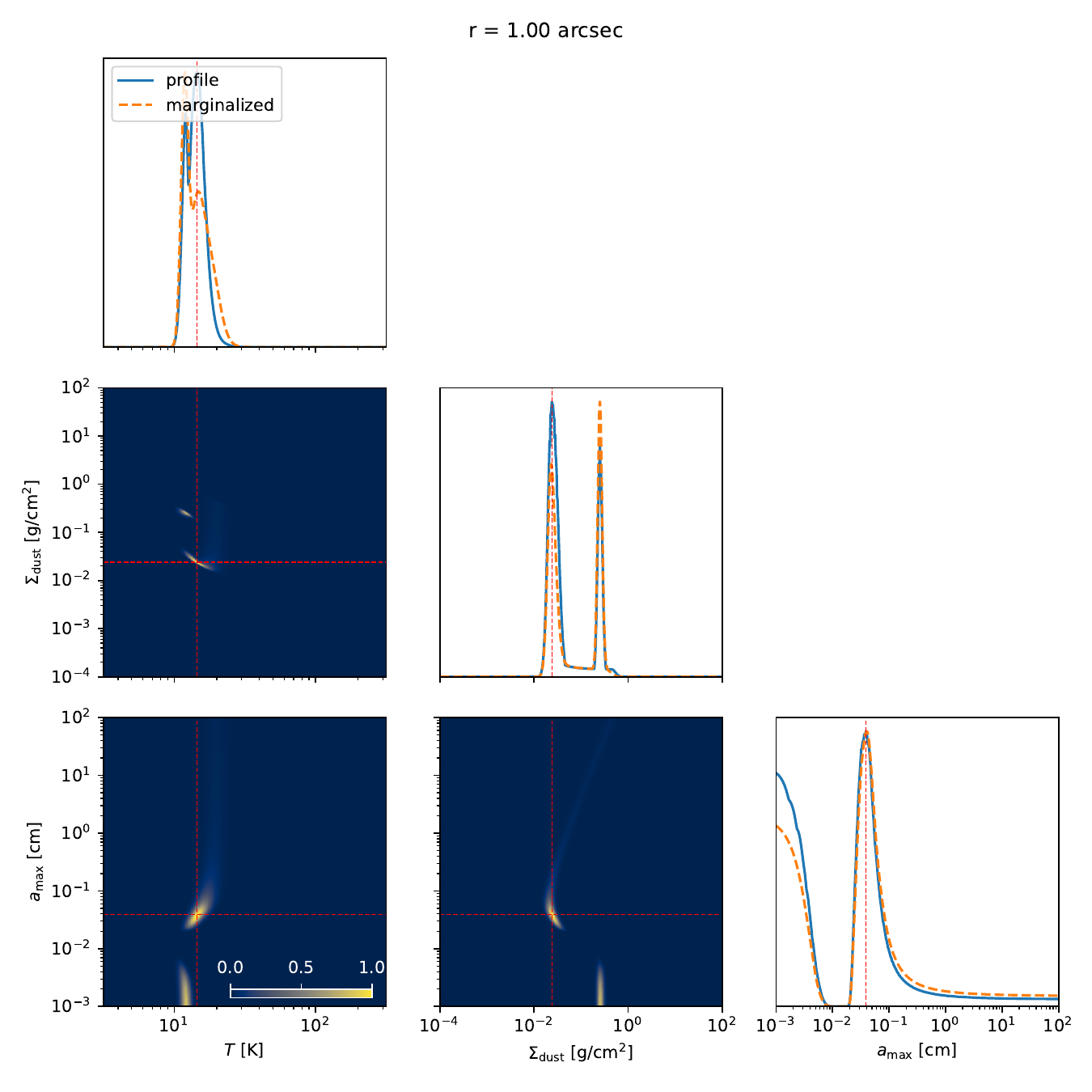}
    \end{center}
    \caption{
        2D corner plot of the profile likelihood for the dust properties fitted for three parameters: temperature, surface density, and maximum dust size in Section \ref{sec:SED_fitting}.
        Here is the result at a radius of $r=1\farcs0$ for the DSHARP default (compact) model (top) and the DSHARP Zubko (compact) model (bottom).
        The diagonal panels show the 1D profile likelihood (blue solid line) and marginalized distribution (orange dashed line).
        The vertical dashed red lines indicate the MLEs.
    }
    \label{fig:corner_DSHARP_default_100}
\end{figure}

Fig. \ref{fig:corner_DSHARP_default_100} shows corner plots for the DSHARP default (compact) and DSHARP Zubko (compact) models, respectively.
Each plot shows the 2D relative profile likelihood.
The diagonal panels show the 1D profile likelihood (blue solid line) and marginalized distribution (orange dashed line).
For the DSHARP Zubko model, the fitting shows two solutions with $a<100\ \mathrm{\mu m}$ and $a>100\ \mathrm{\mu m}$.

\begin{figure}[htbp]
    \begin{center}
        \includegraphics[width=9cm]{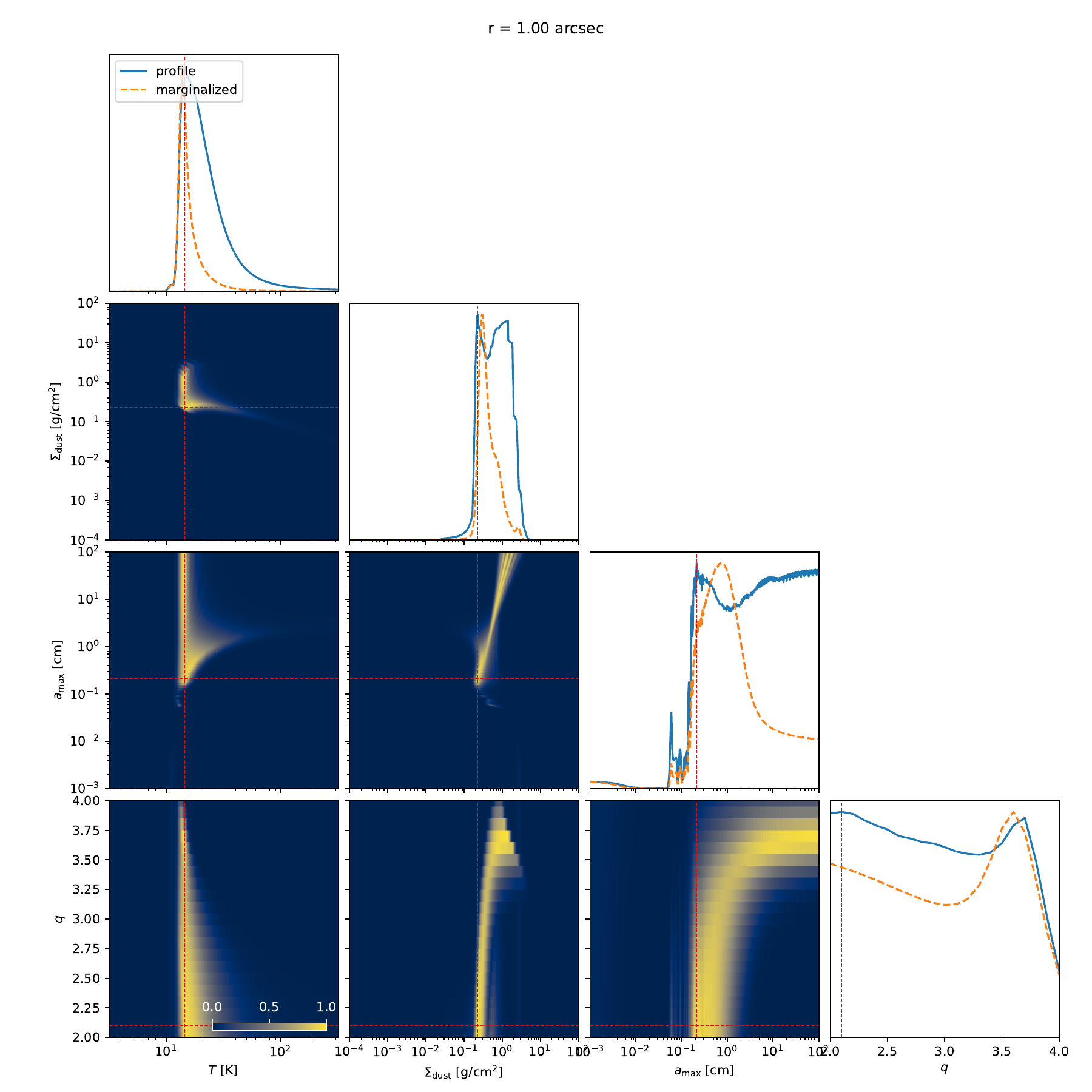}
    \end{center}
    \caption{
        2D corner plot of the profile likelihood as Fig. \ref{fig:corner_DSHARP_default_100} for the dust properties fitted for four parameters: temperature, surface density, maximum dust size, and power-law index $q$.
        Here is the result at a radius of $r=1\farcs0$ for the DSHARP default (compact) model.
    }
    \label{fig:corner_DSHARP_4param_100}
\end{figure}

Fig. \ref{fig:corner_DSHARP_4param_100} shows the corner plot for the case where we also fit the power-law index $q$ of the size distribution in addition to temperature, surface density, and dust size.
The maximum size and the power-law index show a strong degeneracy.

\end{appendix}

\end{document}